\documentclass[onecolumn,trackchanges]{aastex6}

\newcommand{\logg}{$\log\,g$}
\newcommand{\micro}{$\xi$}
\newcommand{\kms}{$km\,s^{-1}$}

\newcommand{\teff}{$T_{\!\mbox{\tiny\em eff}}$}
\newcommand{\teffq}{$T_{\!\mbox{\scriptsize \em eff}}^4$}
 
\newcommand{\lgf}{$\log\,g_{\!\mbox{\scriptsize \,\sc f}}$} 
\newcommand{\gf}{$g_{\!\mbox{\scriptsize \,\sc f}}$} 
\newcommand{\taur}{$\tau_{\!\mbox{\scriptsize \,\sc r}}$} 
\newcommand{\rv}{R$_{\!\mbox{\scriptsize \,\sc v}}$} 
\newcommand{\db}{D$_{\!\mbox{\scriptsize \,\sc b}}$~} 
\newcommand{\av}{A$_{\!\mbox{\scriptsize \,\sc v}}$}

\slugcomment{}

\shorttitle{FGLR LMC}
\shortauthors{Urbaneja et al.}

\begin{document}
\title{LMC Blue Supergiant Stars and the Calibration of the Flux-weighted Gravity--Luminosity Relationship.}

\author{M. A. Urbaneja}
s\affil{Institut f\"ur Astro- und Teilchenphysik, Universit\"at Innsbruck,
    Technikerstr. 25/8, 6020 Innsbruck, Austria}

\author{R.-P. Kudritzki}
\affil{Institute for Astronomy, University of Hawaii, 2680 Woodlawn Drive, Honolulu, HI 96822, USA}

\author{W. Gieren }
\affil{Departamento de Astronom\'{\i}a, Universidad de Concepci\'on, Casilla 160-C, Concepci\'on, Chile} 

\author{G. Pietrzy\'nski\altaffilmark{1}}
\affil{Departamento de Astronom\'{\i}a, Universidad de Concepci\'on, Casilla 160-C, Concepci\'on, Chile} 

\author{F. Bresolin}
\affil{Institute for Astronomy, University of Hawaii, 2680 Woodlawn Drive, Honolulu, HI 96822, USA}

\and

\author{N. Przybilla}
\affil{Institut f\"ur Astro- und Teilchenphysik, Universit\"at Innsbruck, Technikerstr. 25/8, 6020 Innsbruck, Austria}


\altaffiltext{1}{Nicolaus Copernicus Astronomical Center, Polish Academy of Sciences, ul. Bartycka 18, PL-00-716 Warszawa, Poland}

\begin{abstract}
High quality spectra of 90 blue supergiant stars in the Large Magellanic Cloud are analyzed with respect to effective temperature, 
gravity, metallicity, reddening, extinction and 
extinction law. An average metallicity, based on Fe and Mg abundances, relative to the Sun of [Z] = -0.35$\pm$0.09 dex is obtained. The reddening 
distribution peaks at E(B-V) = 0.08 mag, but significantly larger values are also encountered. A wide distribution of the ratio of extinction to 
reddening is found ranging from \rv = 2 to 6. The results are used to investigate the blue supergiant relationship between flux-weighted 
gravity, \gf $\equiv$ g/\teffq, and absolute
bolometric magnitude M$_{\rm bol}$. The existence of a tight relationship, the FGLR, is confirmed. However, in contrast to previous 
work the observations reveal that the FGLR is divided 
into two parts with a different slope. For flux-weighted gravities larger than 1.30 dex the slope is similar as found in previous work, 
but the relationship becomes significantly steeper
for smaller values of the flux-weighted gravity. A new calibration of the FGLR for extragalactic distance determinations is provided.   
\end{abstract}


\section{Introduction}

Blue supergiants of spectral type A and B are the brightest stars in the universe with absolute visual magnitudes up to $ M_V \simeq 
-9.5$ mag. They have
long been recognized as potentially valuable distance indicators in external galaxies \citep{hubble1936, petrie1965, crampton1979, 
tully1984, hill1986, humphreys1988}.
With the advent of very large telescopes of the 8 to 10m class equipped with efficient multi-object spectrographs and with the improvements of 
non-LTE model atmosphere techniques it has become 
possible to obtain high signal-to-noise spectra of individual supergiant stars in galaxies beyond the Local Group and to quantitatively 
analyze these spectra to determine stellar parameters
such as effective temperature \teff, gravity \logg~and metallicity with unprecedented accuracy and reliability. This work has led to the 
detection of a tight relationship between stellar 
absolute bolometric magnitude M$_{\rm bol}$ and the \textquotedblleft flux-weighted gravity\textquotedblright~\gf $\equiv$ g/\teffq~(\teff~in units 
of 10$^{4}$K)  of the form
\begin{equation}
 M_{\rm bol}\,=\,a (\mbox{\lgf}\,-\,1.5)\,+\,b, 
\end{equation}
the ``Flux-weighted Gravity - Luminosity Relationship (FGLR)'' \citep[see][]{kudritzki2003} . 

The simple form of the FGLR can be well understood as the result of massive star evolution 
from the main sequence to the red supergiant phase with constant luminosity and constant mass. Under these simplifying assumptions $g/$\teffq~remains constant during the 
evolution accross the Hertzsprung-Russell Diagram and the FGLR is then the result of a simple relationship between stellar luminosity 
$L$ and stellar mass $M$, $L \sim M^x$, where the exponent 
$x$ is a constant number of the order of 3 (see \citealt{kudritzki2008}, for details). In reality, the luminosity is not exactly constant during 
the evolution and also 
the exponent $x$ decreases with increasing stellar mass, but the FGLR still holds, as a recent detailed investigation using state-of-the-art stellar evolution calculations combined with 
population synthesis has shown \citep{meynet2015}.

After a first calibration \citep{kudritzki2008} the FGLR-technique has turned into a powerful tool to determine distances to a variety of 
galaxies in the Local Group (WLM -- \citealt{urbaneja2008}; M~\,33 -- \citealt{u2009}) and beyond 
(M\,81 -- \citealt{kudritzki2012}; NGC\,3109 -- \citealt{hosek2014}; NGC\,3621 -- \citealt{kudritzki2014}) with a great potential to map the 
local universe out to 10 Mpc with present day telescopes and to reach distances of 30 Mpc with the next generation of extremely large 
telescopes. The advantage of the technique is that it uses a 
spectroscopic method where the analysis of the spectrum yields the intrinsic stellar energy distribution allowing for an accurate correction of interstellar reddening and extinction, usually
a significant source of uncertainty for many other alternative distance determination methods. In addition, since stellar metallicity is also obtained from the spectra, effects of metallicity
on the calibration, if existent, could also be accounted for (we note, however, that the recent work by \citealt{meynet2015} indicates that metallicity effects of the FGLR-method are very small). 

A potential weakness of the present work applying the FGLR-technique lies in the calibration of the method. \citet{kudritzki2008} combined blue supergiant observations from eight galaxies, with 
distances obtained from a variety of other methods, such as cepheids, eclipsing binaries, stellar cluster membership, etc. A much cleaner way (in the absence of accurate distances for Milky Way 
blue supergiants) for the calibration is to use solely the Large Magellanic Cloud (LMC) and its blue supergiants. After the work by \citet{pietrzynski2013}, who carried out a comprehensive study
of late spectral type eclipsing binaries, the distance to the LMC is known with an accuracy of two percent. This provides an excellent basis for a new calibration of the FGLR.

In this paper, we carry out a first step towards such a new calibration. We present new spectroscopic observations of a sample of 32 
blue supergiants obtained with the Magellan Echellete spectrograph at the Las Campanas Clay Telescope. Our core sample is complemented 
with spectra of another 58 LMC supergiant stars collected from other sources.  The spectroscopic material is  
described in section 2, which also contains a discussion of the combined optical, near- and mid-IR photometry used for the 
FGLR in conjunction with spectroscopy. 
The spectra and photometry are then analyzed with respect to  effective temperature \teff, gravity \logg, metallicity, reddening E(B-V), 
extinction A$_V$ and extinction law characterized by \rv = A$_V$/E(B-V). The analysis method is introduced in section 3 and 
the results are presented in section 4. 
Section 5 finally discusses the observed FGLR of LMC blue supergiants
and introduces a new calibration. The paper finishes with the conclusions and important aspects of future work in section 6.
 
\section{Observational data}
\subsection{Spectroscopic data}

Optical spectra of 32 LMC blue supergiant stars were collected during the nights of October 29th 
and 30th, 2008,  and November 10th, 2010, using the Magellan Echellete spectrograph 
\citep[MagE,][]{marshall2008}, a medium resolution, cross-dispersed, optical spectrograph on the Clay Telescope at Las 
Campanas Observatory. We selected the 1 arcsec slit, providing a nominal resolving power 
of R$\sim$4100.  MagE has a plate scale of 0.3 arsec pixel$^{-1}$, and it covers a wide 
spectral range, $\sim$3100-10000 \AA. 

The conditions during the three nights were good, with excellent seeing in the first and third nights, between 0.6 
and 0.8 arcsec, with long spells of 0.6 arcsec. The seeing during the second night was somewhat worse, 
with a characteristic value of 0.8-0.9 arcsec and some periods slightly above 1.0 arcsec. We secured two to three 
consecutive exposures for each star, depending on its magnitude, followed 
by a ThAr lamp, for wavelength calibration purposes. Besides the program objects,   
spectrophotometric standards were observed several times during the night at different elevations. 
In all cases, the slit was placed following the parallactic angle. A number of standard calibration images 
were taken each night, including twilight flats for illumination correction. \\ 

This dataset was processed with the MagE Spectral Extractor pipeline 
\citep[MASE,][]{bochanski2009}.  Before processing the full sample with this 
pipeline, we convinced ourselves that the contribution by scattered light is at an irrelevant level, at 
least for the orders considered here (it could be of some importance for the very bluest orders, 
covering the range 3100--3400 \AA, because of the low flux level achieved during the 
observations). We achieved signal-to-noise ratios in the continuum of at least 200 for all the observed stars. Using the observations of the spectrophotometric standards
we also flux calibrated our spectra around the wavelength region of the Balmer discontinuity to measure the Balmer jump (see section 3).\\ 

Our primary MagE sample is supplemented with additional data taken with ESO facilities. First, we 
retrieved from the ESO archive spectra of 29 early OB-type supergiants obtained with the 
La Silla 2.2m and the FEROS spectrograph (program 074.D-0021(A), PI Evans, data  taken December 26, 27, 
29, 30, and 31, 2004). These spectra cover the range from 3500 to 9200 \AA~at a 
resolution of R = 48000. This dataset was fully processed using the ESO FEROS-DRS pipeline.

Fully reduced spectra of another 14 B-type supergiants in the clusters N11 and NGC\,2004, collected in the context of the 
VLT-FLAMES Survey of Massive Stars \citep{evans2006}, were already available to us. The spectral range is $
\sim3900-4750$ \AA~in the blue with a resolution between 20000 to 27500 and $\sim6300-6700$ \AA~in the red, at a 
somewhat lower resolution (R = 16700).  

Finally, reduced spectra of  another 16 early 
B-type supergiants in the Tarantula region, observed within the VLT--FLAMES Tarantula survey \citep{evans2011} 
were kindly made available to us by our colleagues Phillip Dufton and Chris Evans.  These cover the 3960--5050 \AA~ region at R$\sim$7000. 

Further information on these two last data sets  can be found in the aforementioned references. The characteristic SNR of the stars in 
the FLAMES, FEROS and VFTS samples ranges from 100 to 200, depending on the visual magnitude of the star considered.

In building our sample of supergiants for the FGLR calibration we selected only objects of luminosity class I with effective temperatures lower than 27000K
in order to avoid the main sequence phases of stellar evolution, where luminosity and flux-weighted gravity change significantly. Objects, for which
the spectral analysis resulted in a higher effective temperature, were dropped from the sample.

\subsection{Photometry}
Photometric data in multiple bands, from optical to IR, are available in the literature for most of the stars. Regarding the 
optical bands, because of the brightness of the stars, we mostly rely on early data taken with photometers. Only a handful 
of stars (the faint end of the sample) lacks this kind of photometric data. For these, we consider the CCD-based photometry 
(B and V passbands) provided by different works in the literature. The near-IR data were taken from the 2MASS All Sky Catalog of point 
sources \citep{skrutskie2006}. For a significant fraction of the sample, Spitzer/IRAC photometry is also available from the 
Spitzer survey of the Large Magellanic Cloud \citep{meixner2006}. The full list of targets along with some basic 
information is given in Tab.~\ref{table_info_ba} through Tab.~\ref{table_info_vfts}. For completeness, the 
photometric data used in this work are collected in Appendix 1, Tab.~\ref{table_photometry_optical} (optical), 
Tab.~\ref{table_photometry_2mass} (2MASS) and Tab.~\ref{table_photometry_irac} (Spitzer/IRAC). \\

The choice of photometric data is mainly driven by the requirement of minimising the possible sources of artificial scatter in the final FGLR. In the
case of the optical photometry, a comparison with the APASS survey \citep{henden2016} for the 72 stars studied in this work that are available in this survey
results  in a mean and standard deviation of the difference for V magnitudes of 0.005$\pm$0.05 mag, and of -0.008$\pm$0.025 mag for the 
B-V colour. Our conclusion from this comparison is that there is no effect that can be observed between the {\em old} photometry 
used in our work and the {\em modern} CCD photometry provided by this survey. 
On the other hand, a comparison for the 60 stars in common with \citet{zaritsky2004} is less favourable, with mean and standard deviation for the differences of 
0.02$\pm$0.14 mag (with a median absolute deviation of 0.11 mag) and -0.02$\pm$0.58 mag (with median absolute deviation 0.13 mag). Given the 
excellent agreement found with the APASS survey just discussed, these large differences might suggest that the bright stars in this survey could  
suffer from the same problems as indicated for example in \citet{massey2002}.

%
\begin{deluxetable}{ l l l c l l }
\tablecaption{LMC BA supergiants observed with MagE. \label{table_info_ba}}
\tablehead{
\colhead{Star} & \colhead{RA (J2000)} & \colhead{Dec (J2000)} & \colhead{V}     & \colhead{SpT/LC} & \colhead{Reference}  \\
\colhead{ }    & \colhead{}           & \colhead{}            & \colhead{(mag)} & \colhead{ }      & \colhead{ }  
}
\startdata
 NGC\,2004 8 &  05 30 40.106    &  -67 16 37.97  & 12.43 &  B9\,Ia     & E06   \\
 Sk-65 67    &  05 34 35.2302   &  -65 38 52.600 & 11.44 &  A2\,Ia     & tw    \\
 Sk-66 50    &  05 03 08.82368  &  -66 57 34.8883& 10.63 &  B8\,Ia$^+$ & F88   \\
 Sk-67 19    &  04 55 21.60684  &  -67 26 11.2579& 11.19 &  A3\,Ia     & tw    \\
 Sk-67 137   &  05 29 42.611    &  -67 20 47.71  & 11.93 &  B8\,Ia     & E06   \\
 Sk-67 140   &  05 29 56.507    &  -67 27 30.87  & 12.42 &  A1\,Ia     &  tw   \\
 Sk-67 141   &  05 30 01.246    &  -67 14 36.90  & 12.01 &  A2\,Iab    & E06   \\
 Sk-67 143   &  05 30 07.1165   &  -67 15 43.178 & 11.46 &  A0\,Ia     & tw    \\
 Sk-67 155   &  05 31 12.822    &  -67 15 07.98  & 11.60 &  A3\,Iab    & E06   \\
 Sk-67 157   &  05 31 27.933    &  -67 24 44.25  & 11.95 &  B9\,Ia     & E06   \\
 Sk-67 170   &  05 31 52.972    &  -67 12 15.32  & 12.48 &  A1\,Iab    & E06   \\
 Sk-67 171   &  05 32 00.791    &  -67 20 23.02  & 12.04 &  B8\,Ia     & E06   \\
 Sk-67 201   &  05 34 22.46749  &  -67 01 23.5699&  9.87 &  B9\,Ia     & E06   \\
 Sk-67 204   &  05 34 50.168    &  -67 21 12.50  & 10.87 &  B8\,Ia     & F88   \\
 Sk-69 2A    &  04 47 33.224    &  -69 14 32.87  & 12.47 &  A3\,Iab    & tw    \\
 Sk-69 24    &  04 53 59.519    &  -69 22 42.57  & 12.52 &  B9\,Ia     & A72   \\
 Sk-69 39A   &  04 55 40.447    &  -69 26 40.98  & 12.35 &  A0\,Iab    & tw    \\
 Sk-69 82    &  05 14 31.566    &  -69 13 53.39  & 10.92 &  B8\,Ia     & F88   \\
 Sk-69 113   &  05 21 22.399    &  -69 27 08.07  & 10.71 &  A1\,Ia     & F88   \\
 Sk-69 170   &  05 30 50.07611  &  -69 31 29.3838& 10.34 &  A0\,Ia     & tw    \\
 Sk-69 299   &  05 45 16.61696  &  -68 59 51.9691& 10.24 &  A2\,Ia$^+$ & tw    \\
 Sk-70 45    &  05 02 17.804    &  -70 26 56.74  & 12.53 &  A0\,Iab    & S76   \\
 Sk-71 14    &  05 09 38.91592  &  -71 24 02.2387& 10.62 &  A0\,Ia     & F88   \\
\enddata
\tablecomments{References for the spectral classification: tw--this work; A72--\cite{ardeberg1972}; E06--\cite{evans2006}; 
F88--\cite{fitzpatrick1988}; S76--\cite{stock1976}. }
\end{deluxetable}

%
\begin{deluxetable}{ l l l c l l c c }
\tablecaption{LMC OB supergiants (MagE and FLAMES\,I) \label{table_info_ob}}
\tablehead{
\colhead{Star} & \colhead{RA (J2000)} & \colhead{Dec (J2000)} & \colhead{V} & \colhead{SpT/LC} & \colhead{Reference} \\
\colhead{ }    & \colhead{}           & \colhead{}            & \colhead{(mag)} & \colhead{ } & \colhead{ } 
}
\startdata
N11 24       &  04 55 32.931    &  -66 25 27.90  & 13.45   & B1\,Ib       & E06     \\
N11 36       &  04 57 41.015    &  -66 29 56.54  & 13.72   & B0.5\,Ib     & E06     \\
N11 54       &  04 57 18.332    &  -66 25 59.59  & 14.10   & B1\,Ib       & E06     \\
NGC\,2004 12 &  05 30 37.491    &  -67 16 53.57  & 13.39   & B1.5\,Iab    & E06     \\ 
NGC\,2004 21 &  05 30 42.043    &  -67 21 41.58  & 13.67   & B1.5\,Ib     & E06     \\
NGC\,2004 22 &  05 30 47.362    &  -67 17 23.40  & 13.77   & B1.5\,Ib     & E06     \\
Sk-66 1      &  04 52 19.053    &  -66 43 53.23  & 11.61   & B1.5\,Ia     & F88     \\
Sk-66 15     &  04 55 22.355    &  -66 28 19.26  & 12.81   & B0.5\,Ia     & E06     \\ 
Sk-66 23     &  04 56 17.555    &  -66 18 18.96  & 13.09   & B2.5\,Iab    & E06     \\
Sk-66 26     &  04 56 20.575    &  -66 27 13.83  & 12.91   & B1\,Ib       & E06     \\
Sk-66 27     &  04 56 23.476    &  -66 29 52.04  & 11.82   & B3\,Ia       & E06     \\
Sk-66 36     &  04 57 08.824    &  -66 23 25.15  & 11.35   & B2\,Ia       & E06     \\
Sk-66 37     &  04 57 22.078    &  -66 24 27.67  & 12.98   & B0.7\,Ib     & E06     \\
Sk-66 166    &  05 36 04.149    &  -66 13 43.51  & 11.71   & B2.5\,Ia     & F88     \\
Sk-67 14     &  04 54 31.8907   &  -67 15 24.664 & 11.52   & B1.5\,Ia     & W02     \\ 
Sk-67 36     &  05 01 22.590    &  -67 20 10.02  & 12.01   & B2.5\,Ia     & F88     \\
Sk-67 133    &  05 29 21.684    &  -67 20 11.34  & 12.59   & B2.5\,Iab    & E06     \\
Sk-67 154    &  05 31 03.776    &  -67 21 20.46  & 12.61   & B1.5\,Ia     & E06     \\
Sk-67 228    &  05 37 41.012    &  -67 43 16.56  & 11.49   & B2\,Ia       & F88     \\
Sk-68 40     &  05 05 15.203    &  -68 02 14.15  & 11.71   & B2.5\,Ia     & F88     \\
Sk-68 92     &  05 28 16.157    &  -68 51 45.58  & 11.71   & B1\,Ia       & F88     \\
Sk-68 171    &  05 50 22.979    &  -68 11 24.62  & 12.01   & B0.7\,Ia     & F91     \\
Sk-69 43     &  04 56 10.457    &  -69 15 38.21  & 11.94   & B1\,Ia       & A72     \\
\enddata
\tablecomments{References for the spectral classification: A72--\cite{ardeberg1972}; E06--\cite{evans2006}; 
F88--\cite{fitzpatrick1988}; F91--\cite{fitzpatrick1991}; W02--\cite{walborn2002}. }
\end{deluxetable}

\begin{deluxetable}{ l l l c l l c c }
\tablecaption{LMC OB supergiants (FEROS) \label{table_info_lennon}}
\tablehead{
\colhead{Star} & \colhead{RA (J2000)} & \colhead{Dec (J2000)} & \colhead{V} & \colhead{SpT/LC} & \colhead{Reference} \\
\colhead{ }    & \colhead{}           & \colhead{}            & \colhead{(mag)} & \colhead{ } & \colhead{ } 
}
\startdata
Sk-66 5     &  04 53 30.029     & -66 55 28.24  & 10.73 & B2.5\,Ia   & F88      \\
Sk-66 35    &  04 57 04.440     & -66 34 38.45  & 11.60 & BC1\,Ia    & F88      \\
Sk-66 106   &  05 29 00.981     & -66 38 27.69  & 11.72 & B1.5\,Ia   & F88      \\
Sk-66 118   &  05 30 51.912     & -66 54 09.12  & 11.81 & B2\,Ia     & F88      \\
Sk-67 2     &  04 47 04.44925   & -67 06 53.1076& 11.26 & B1\,Ia$^+$ & F88      \\
Sk-67 28    &  04 58 39.23      & -67 11 18.7   & 12.28 & B0.7\,Ia   & F88      \\
Sk-67 78    &  05 20 19.07797   & -67 18 05.6893& 11.26 & B3\,Ia     & F88      \\
Sk-67 90    &  05 23 00.670     & -67 11 21.93  & 11.29 & B1\,Ia     & F88      \\
Sk-67 112   &  05 26 56.471     & -67 39 34.98  & 11.90 & B0.5\,Ia   & F88      \\
Sk-67 150   &  05 30 31.690     & -67 00 53.41  & 12.24 & B0.7\,Ia   & F88      \\
Sk-67 169   &  05 31 51.588     & -67 02 22.29  & 12.18 & B1\,Ia     & W02      \\
Sk-67 172   &  05 32 07.336     & -67 29 13.94  & 11.88 & B2.5\,Ia   & F88      \\
Sk-67 173   &  05 32 10.772     & -67 40 25.00  & 12.04 & B0\,Ia     & M95      \\
Sk-67 206   &  05 34 55.109     & -67 02 37.45  & 12.00 & B0.5\,Ia   & F88      \\
Sk-67 256   &  05 44 25.036     & -67 13 49.61  & 11.90 & B1\,Ia     & F88      \\
Sk-68 26    &  05 01 32.239     & -68 10 42.92  & 11.63 & BC2\,Ia    & F88      \\
Sk-68 41    &  05 05 27.131     & -68 10 02.61  & 12.01 & B0.5\,Ia   & F87      \\
Sk-68 45    &  05 06 07.295     & -68 07 06.11  & 12.03 & B0\,Ia     & F88      \\
Sk-68 111   &  05 31 00.842     & -68 53 57.17  & 12.01 & B0.5\,Ia   & M10      \\
Sk-69 89    &  05 17 17.5791    & -69 46 44.165 & 11.39 & B2.5\,Ia   & F88      \\
Sk-69 214   &  05 36 16.436     & -69 31 27.09  & 12.19 & B0.7\,Ia   & F88      \\
Sk-69 228   &  05 37 09.218     & -69 20 19.54  & 12.12 & BC1.5\,Ia  & F88      \\
Sk-69 237   &  05 38 01.3126    & -69 22 14.079 & 12.08 & B1\,Ia     & F88      \\
Sk-69 270   &  05 41 20.408     & -69 05 07.36  & 11.27 & B2.5\,Ia   & F88      \\
Sk-69 274   &  05 41 27.679     & -69 48 03.70  & 11.21 & B2.5\,Ia   & F88      \\
Sk-70 78    &  05 06 16.035     & -70 29 35.76  & 11.29 & B0.7\,Ia   & F88      \\
Sk-70 111   &  05 41 36.790     & -70 00 52.65  & 11.85 & B0.5\,Ia   & F88      \\
Sk-70 120   &  05 51 20.78027   & -70 17 09.3281& 11.59 & B1\,Ia     & F88      \\
Sk-71 42    &  05 30 47.77615   & -71 04 02.2976& 11.17 & B2\,Ia     & F91      \\
\enddata     
\tablecomments{References for the spectral classification: F87--\cite{fitzpatrick1987}; F88--\cite{fitzpatrick1988}; 
F91--\cite{fitzpatrick1991}; M95--\cite{massey1995}; M10--\cite{massey2010}; 
W02--\cite{walborn2002}. }
\end{deluxetable}

\begin{deluxetable}{ l l l c l }
\tablecaption{LMC VFTS OB Supergiants \label{table_info_vfts}}
\tablehead{
\colhead{Star} & \colhead{RA (J2000)} & \colhead{Dec (J2000)} & \colhead{V}   & \colhead{SpT/LC} \\
\colhead{ }    & \colhead{}           & \colhead{}            & \colhead{(mag)} & \colhead{ } 
}
\startdata
VFTS 3    &  05 36 55.17     &  -69 11 37.7   & 11.59 & B1\,Ia$^+$          \\
VFTS 28   &  05 37 17.856    &  -69 09 46.23  & 13.40 & B0.7\,Ia Nwk        \\
VFTS 69   &  05 37 33.76     &  -69 08 13.2   & 13.59 & B0.7\,Ib-Iab        \\
VFTS 82   &  05 37 36.082    &  -69 06 44.90  & 13.61 & B0.5\,Ib-Iab        \\
VFTS 232  &  05 38 04.774    &  -69 09 05.53  & 14.52 & B3\,Ia              \\
VFTS 270  &  05 38 15.140    &  -69 04 01.18  & 14.35 & B3\,Ib              \\
VFTS 302  &  05 38 19.003    &  -69 11 12.80  & 15.53 & B1.5\,Ib            \\
VFTS 315  &  05 38 20.600    &  -69 15 37.88  & 14.81 & B1\,Ib              \\
VFTS 431  &  05 38 36.971    &  -69 05 07.75  & 12.05 & B1.5\,Ia Nstr       \\
VFTS 533  &  05 38 42.742    &  -69 05 42.44  & 11.82 & B1.5\,Ia+p Nwk      \\
VFTS 590  &  05 38 45.5859   &  -69 05 47.772 & 12.49 & B0.7\,Iab           \\
VFTS 696  &  05 38 57.18     &  -68 56 53.1   & 12.74 & B0.7\,Ib-Iab Nwk    \\
VFTS 732  &  05 39 04.777    &  -69 04 09.96  & 13.00 & B1.5\,Iap Nwk       \\
VFTS 831  &  05 39 39.870    &  -69 12 04.34  & 13.04 & B5\,Ia              \\
VFTS 867  &  05 40 01.325    &  -69 07 59.60  & 14.63 & B1\,Ib Nwk          \\
\enddata
\tablecomments{Reference for the spectral classification: \cite{mcevoy2015} }
\end{deluxetable}

\section{Spectral Analysis}
As in our previous works \citep[see for example][]{kudritzki2008, kudritzki2012}, the stars are separated in two 
groups, according to their spectral types. We use the term {\em OB supergiants} for stars of Ib to Ia$^+$
luminosity classes with spectral types B5 or earlier (note that the
earliest star in our sample is of spectral type B0.5 so no O-type 
supergiants are included), whilst the 
{\em BA supergiants} group is comprised of stars with spectral type B8--A5. For each group, we utilized a different 
set of model atmosphere and line formation codes suitable 
for the physical conditions in the atmospheres of these objects. The next two subsections describe the codes employed 
for the two different groups and a third subsection summarises the 
methodology of the spectral analysis.

\subsection{OB supergiants}
For the B0--B5 range we use the model atmosphere/line formation code {\sc fastwind} \citep{puls2005}. 
{\sc fastwind} includes the effects of stellar winds which are important for the atmospheres of these objects. It 
solves the radiative transfer problem in the co-moving frame of the expanding atmosphere of an early type 
star, for a spherically symmetric geometry, considering the constraints of statistical equilibrium and energy 
conservation, along with chemical homogeneity and steady state. The density stratification is set by the 
velocity field (pseudo static photosphere plus a wind following a $\beta$-law) via the equation of continuity. A 
smooth transition between the photosphere, described by a depth-dependent pressure scale-height, and the 
wind is enforced \citep[see][]{santolaya1997}.\\
 
Each model is then specified by a set of parameters: effective temperature, surface gravity and stellar radius (all defined at Rosseland optical depth \taur\,$=2/3$), the $\beta$ exponent of the wind-velocity law, the steady mass-loss rate, the wind terminal velocity, the micro-turbulence velocity, and the chemical abundances. In order to reduce the number of parameters to be explored with the model grid, we follow the ideas presented in \cite{urbaneja2005}, with the exception being a fixed micro-turbulence velocity for the calculation of the atmospheric structure. The reader is referred to this last 
reference for specific information on the model atoms used in our calculations.\\

The grid of {\sc fastwind} models utilised in this work explores a total of 10 parameters: effective temperature 
(\teff), surface gravity (\logg), micro-turbulence (\micro, fixed to 10 \kms~for the calculation of the atmospheric 
density and temperature stratification but variable for the formal solution), the outflow velocity field parameter 
$\beta$, an optical depth invariant parameter $Q\,=\dot{M}\left(R_\star\,v_\infty\right)^{-2/3}$ that characterizes 
the strength of the wind \citep[assumed to be homogeneous, see][]{urbaneja2011}, and the abundances of He, 
C, N, O, Mg and Si. For all other species, we assume the solar abundance pattern scaled down to 0.4\,Z$_\odot$. 
We adopt a scaling relation between the wind terminal velocity and the 
escape velocity as prescribed by \citet[][]{kudritzki2000}.

\subsection{BA supergiants}
In the case of BA supergiants, we follow closely the procedures introduced by \cite{przybilla2006}. 
Since the effects of stellar winds are less important in the photospheres of these objects, the atmospheric 
structure is provided by plane-parallel, hydrostatic model atmospheres calculated assuming local 
thermodynamic 
equilibrium (LTE) with the ATLAS9 code \citep{kurucz1993}, whilst the occupation numbers and line formation 
calculation are 
performed with the DETAIL/SURFACE codes \citep{giddings1981, butler1985}, assuming either LTE or non-LTE 
depending on the species. For the analyses 
presented in the present work, we consider a model grid defined by a set of 6 parameters for each model: 
\teff, \logg, 
\micro, and the abundances of He, Mg and Fe (all these 3 species, plus H, are treated in non-LTE in all 
cases). The reader 
is referred to \cite{przybilla2006} for details on the applicability limits, and the model atoms used in the non-
LTE calculations. 

\subsection{Methodology}
The quantitative spectroscopic analysis of B- and A-type supergiants is well established and the methods 
have been carefully tested. The reader 
is referred to the papers cited in the previous paragraphs for a detailed description. Briefly, 
the combined diagnostic information of ionisation equilibria (Si~{\sc ii/iii/iv} supplemented by He~{\sc i/ii} for 
OB stars; 
Mg~{\sc i/ii} and Fe~{\sc i/ii} in the case of BA stars) and of the higher Balmer lines (starting usually at H$
\gamma$) are used 
to determine the fundamental atmospheric parameters (effective temperature, gravity, helium abundance, 
microturbulence, metal abundances), 
requiring at the same time that any line is well reproduced. In the case of OB 
supergiants we additionally utilise H$\alpha$ and H$\beta$  to constrain the strengths and influence of stellar winds, 
i.e. the $\beta$ and Q parameters.\\
 
The methodology just described can be implemented in different ways. In our case, a simultaneous solution for all the 
relevant parameters is found through a procedure that minimizes the differences between the observation and the model. 
For each group (OB or BA), a set of key diagnostic lines is defined. The model that minimizes the residuals in these lines 
then defines the parameters and abundances of the observed star. Tab.~\ref{table_diagnostic_lines} shows 
the set of lines that we have considered for this procedure. We note that all the metal lines used are 
fairly {\em isolated} (unblended) at the spectral resolution of our observational dataset.\\

\begin{deluxetable}{ l l }
\tablecaption{Diagnostic lines used for the spectral analysis. \label{table_diagnostic_lines}}
\tablehead{ \colhead{Species} & \colhead{Lines} }
\tablewidth{0pt}
\startdata
\multicolumn{2}{c}{OB Supergiants} \\
\hline
H    & H$\alpha$, H$\beta$, H$\gamma$, H$\delta$, H$\zeta$ \\
He   & He~{\sc i} 4026, 4387, 4713, 4922, 5015, 5876 \\
     & He~{\sc ii} 4541, 5411 \\
Mg   & Mg~{\sc ii} 4481 \\
Si   & Si~{\sc ii} 4128, 4130 \\
     & Si~{\sc iii} 4553, 4567, 4574, 5740 \\
     & Si~{\sc iv} 4116 \\
\hline
\hline
\multicolumn{2}{c}{BA Supergiants} \\
\hline
H    & H$\gamma$, H$\delta$, H$\zeta$, H$\eta$ \\
He   & He~{\sc i} 5876, 6678  \\
Mg   & Mg~{\sc i} 5172, 5183 \\
     & Mg~{\sc ii} 4390, 4481 \\
Fe   & Fe~{\sc i} 3859, 4045, 4063, 4404 \\
     & Fe~{\sc ii} 4233, 4303, 4490, 4505, 4554, 4576 \\
     & Fe~{\sc ii} 4731, 5325, 6145, 6416  \\
\enddata
\end{deluxetable}

In the case of BA stars, an extra powerful  constraint on \teff~is provided by the Balmer jump, complementing 
and sometimes substituting the use of ionisation equilibria, IE  \citep[see for example][]{kudritzki2008}. We 
have conducted separate analyses, considering on the one hand an IE and the Balmer jump on another, of those 
stars in the BA group for which we could achieve a reliable flux calibration (only stars observed during our 
2008 run, the 2010 data turned out to be of insufficient quality in the region of the Balmer jump for a reliable 
flux calibration), in order to evaluate the consistency between both \teff-indicators. This will be discussed in 
the coming sections. 

\section{Results}

\begin{figure*}
\includegraphics[scale=0.75,angle=180]{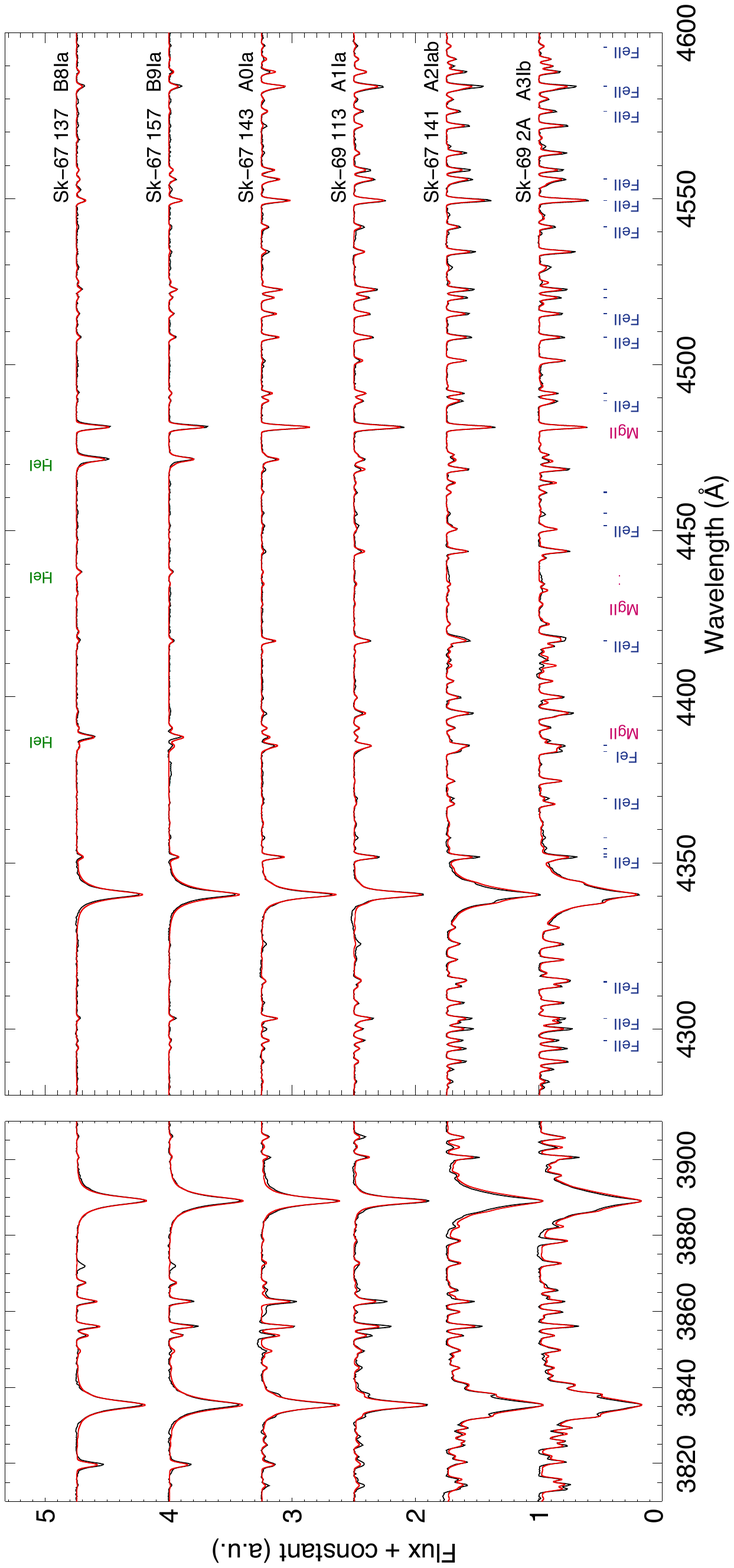}
\caption{\label{spec_abfit} Example fits for some BA supergiants of our sample as a function of spectral type. Iron lines, as well as 
some other prominent features, are identified.}
\end{figure*}

\begin{figure*}
\epsscale{0.95}
\plotone{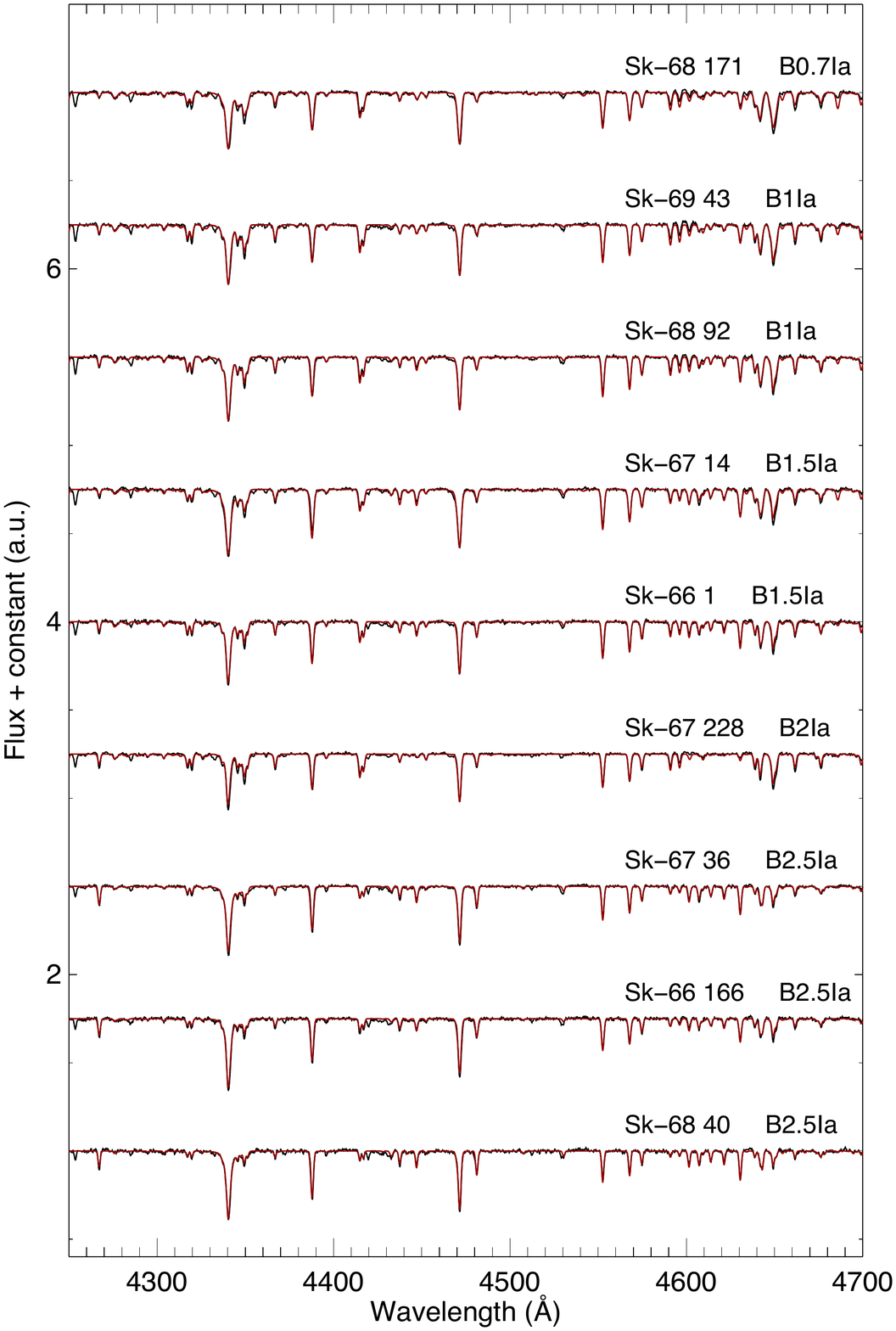}
\caption{\label{spec_obfit} Example fits for some OB supergiants of our sample as a function of spectral type.}
\end{figure*}

\begin{figure*}
\includegraphics[scale=0.6]{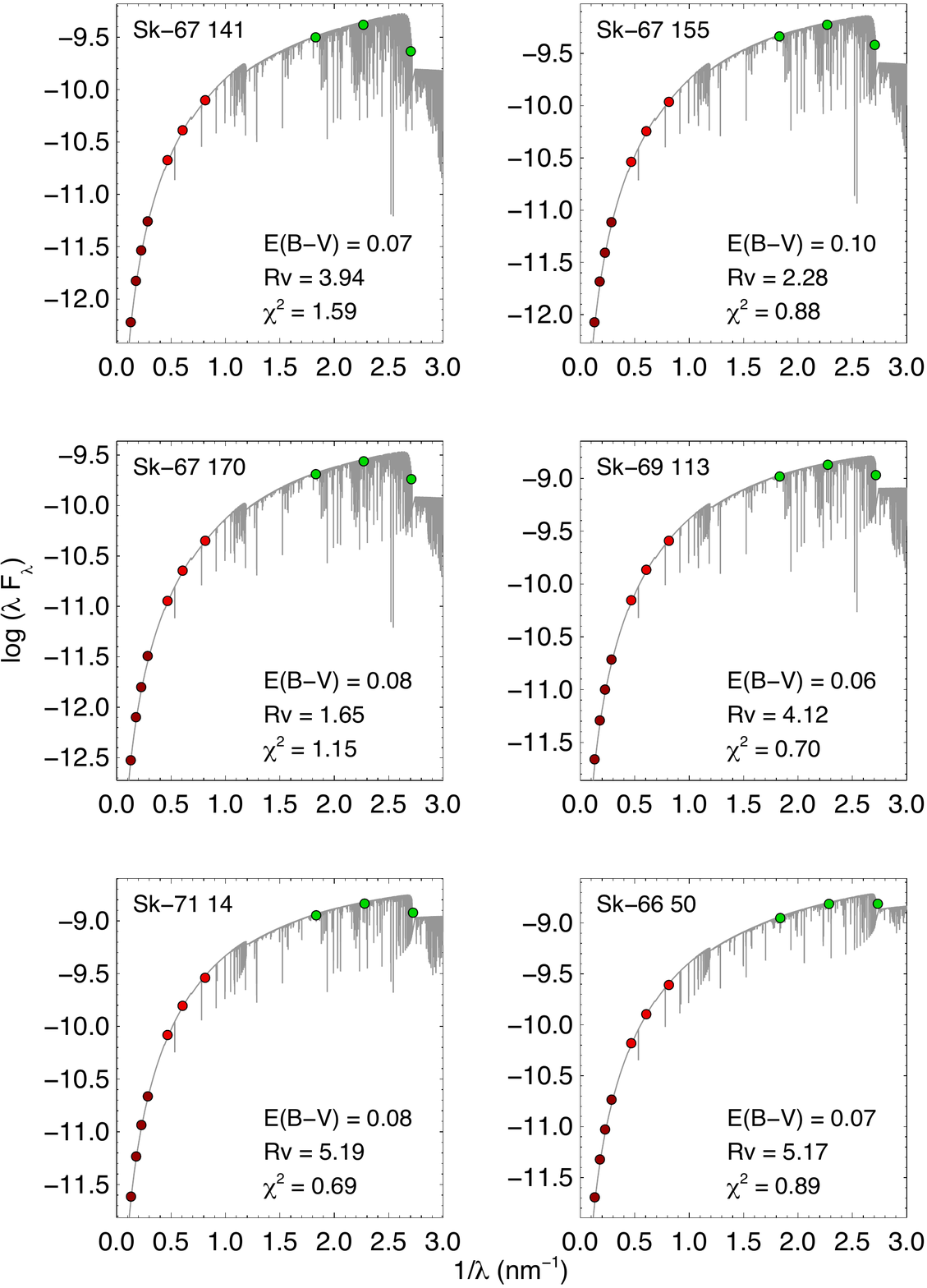}
\caption{\label{sed_abfit} Spectral energy distributions of some BA supergiant stars in our sample. Photometric 
measurements (filled circles) correspond to Johnson (U, B and V
--green circles), 2MASS (J, H, and Ks -- light-red circles) and IRAC
(3.6, 4.5, 5.8 and 8.0 microns channels -- dark-red circles) measurements. Whilst this figure, and the following one, are shown in flux units 
for illustration purposes, the derivation of E(B-V) and \rv~is based on colours. The reader is referred to the text for a detailed explanation.  }
\end{figure*}

\begin{figure*}
\epsscale{0.95}
\includegraphics[scale=0.6]{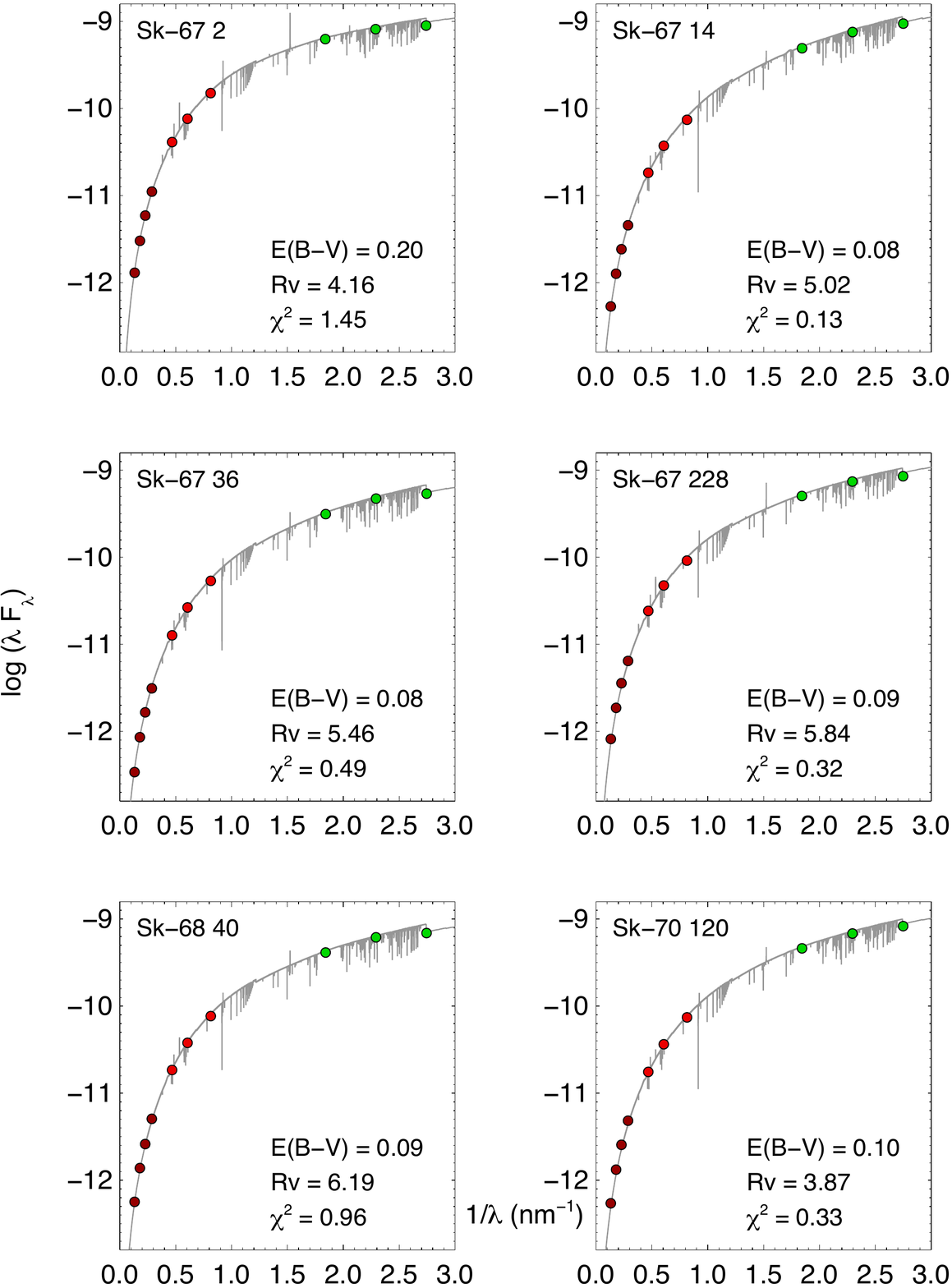}
\caption{\label{sed_obfit} Energy distribution fit of some OB supergiants in the sample. Photometric measurements are in Johnson 
(U, B and V-- green circles), 2MASS (J, H, and Ks -- light-red circles) and IRAC (3.6, 4.5, 5.8 and 8.0 microns 
channels -- dark-red circles) measurements. See comments in the caption of the previous figure.}
\end{figure*}

The results of the analysis of our spectra are summarized in Tables~\ref{table_results_ba},~\ref{table_results_ob}, 
~\ref{table_results_lennon} and ~\ref{table_results_vfts}. 
Fig.~\ref{spec_abfit} and~\ref{spec_obfit} give an impression of the quality of the fits and demonstrate that the model
spectra calculated with the final parameters of our target stars reproduce the observed spectra very well. 
Fig.~\ref{sed_abfit} and Fig.~\ref{sed_obfit} show 
example fits of the observed energy distributions, which are used to determine E(B-V), A$_V$ and 
\rv~(see subsection 4.4). Note that while we display monochromatic fluxes in these figures for a better illustration
of the results, the determination of the reddening parameters is based on filter integrated colours (see below). In the 
following, we discuss details of our results with respect to stellar 
evolution (\S4.1), metallicity (\S4.2), low resolution spectroscopy as applied in the case of extragalactic studies 
beyond the Local Group (\S4.3), and reddening, extinction and extinction law (\S4.4).

\begin{deluxetable}{ l c c c c c c c c c }
\tablecaption{LMC BA supergiants results. \label{table_results_ba}}
\tablehead{
\colhead{Star} & \colhead{\teff} & \colhead{\lgf} & \colhead{Mg} & \colhead{Fe} & \colhead{BC} & \colhead{E(B-V)} & \colhead{Rv} & \colhead{Av} & \colhead{$\chi^2$}  \\
\colhead{} & \colhead{($\times 10^4$ K)} & \colhead{ (dex) } & \colhead{(dex)} & \colhead{ (dex) }  & \colhead{ (mag) } & \colhead{ (mag) } & \colhead{} & \colhead{mag}   & \colhead{}
}
\tablewidth{0pt}
\startdata
NGC\,2004 8& 1.1485$^{+0.0300}_{-0.0275}$   &   1.69$^{+0.05}_{-0.06}$   & 7.39$^{+0.20}_{-0.13}$  &  7.06$^{+0.14}_{-0.16}$  &  -0.59$\pm$0.04   & 0.06$^{+0.02}_{-0.03}$ &  6.02$^{+1.88}_{-2.27}$ &  0.36$^{+0.03}_{-0.03}$  &  0.97   \\
Sk-65 67   & 0.9440$^{+0.0050}_{-0.0050}$   &   1.57$^{+0.02}_{-0.01}$   & 7.26$^{+0.18}_{-0.18}$  &  7.05$^{+0.13}_{-0.14}$  &  -0.21$\pm$0.02   & 0.06$^{+0.02}_{-0.03}$ &  4.50$^{+1.69}_{-2.73}$ &  0.25$^{+0.03}_{-0.03}$  &  2.12   \\
Sk-66 50   & 1.1900$^{+0.0420}_{-0.0315}$   &   1.19$^{+0.02}_{-0.08}$   & 7.38$^{+0.21}_{-0.14}$  &  7.09$^{+0.18}_{-0.18}$  &  -0.72$\pm$0.03   & 0.07$^{+0.02}_{-0.03}$ &  5.17$^{+1.60}_{-2.32}$ &  0.39$^{+0.03}_{-0.03}$  &  1.10   \\
Sk-67 19   & 0.8860$^{+0.0075}_{-0.0125}$   &   1.59$^{+0.05}_{-0.02}$   & 7.37$^{+0.17}_{-0.16}$  &  7.31$^{+0.18}_{-0.14}$  &  -0.10$\pm$0.06   & 0.14$^{+0.04}_{-0.04}$ &  2.83$^{+0.70}_{-1.16}$ &  0.40$^{+0.03}_{-0.03}$  &  1.76   \\
Sk-67 137  & 1.2060$^{+0.0365}_{-0.0390}$   &   1.53$^{+0.06}_{-0.07}$   & 7.41$^{+0.17}_{-0.11}$  &  7.06$^{+0.14}_{-0.17}$  &  -0.69$\pm$0.05   & 0.11$^{+0.03}_{-0.03}$ &  1.27$^{+0.46}_{-0.80}$ &  0.13$^{+0.04}_{-0.04}$  &  0.90   \\
Sk-67 140  & 0.9675$^{+0.0045}_{-0.0045}$   &   1.83$^{+0.02}_{-0.01}$   & 7.33$^{+0.17}_{-0.18}$  &  7.11$^{+0.12}_{-0.11}$  &  -0.24$\pm$0.02   & 0.11$^{+0.03}_{-0.03}$ &  3.22$^{+0.91}_{-1.53}$ &  0.34$^{+0.03}_{-0.03}$  &  1.03   \\
Sk-67 141  & 0.9040$^{+0.0045}_{-0.0045}$   &   1.95$^{+0.02}_{-0.01}$   & 7.32$^{+0.10}_{-0.12}$  &  7.29$^{+0.10}_{-0.09}$  &  -0.12$\pm$0.03   & 0.07$^{+0.02}_{-0.03}$ &  3.94$^{+1.18}_{-2.02}$ &  0.29$^{+0.04}_{-0.04}$  &  1.45   \\
Sk-67 143  & 1.0135$^{+0.0095}_{-0.0090}$   &   1.51$^{+0.04}_{-0.02}$   & 7.37$^{+0.04}_{-0.07}$  &  7.28$^{+0.08}_{-0.05}$  &  -0.34$\pm$0.04   & 0.07$^{+0.03}_{-0.03}$ &  3.63$^{+1.36}_{-2.58}$ &  0.24$^{+0.03}_{-0.03}$  &  1.09   \\
Sk-67 155  & 0.9295$^{+0.0035}_{-0.0030}$   &   1.74$^{+0.02}_{-0.01}$   & 7.38$^{+0.15}_{-0.14}$  &  7.31$^{+0.10}_{-0.10}$  &  -0.17$\pm$0.02   & 0.10$^{+0.03}_{-0.03}$ &  2.28$^{+0.71}_{-1.33}$ &  0.23$^{+0.03}_{-0.03}$  &  1.10   \\
Sk-67 157  & 1.1420$^{+0.0205}_{-0.0200}$   &   1.66$^{+0.04}_{-0.04}$   & 7.44$^{+0.13}_{-0.09}$  &  7.09$^{+0.13}_{-0.14}$  &  -0.58$\pm$0.03   & 0.07$^{+0.03}_{-0.03}$ &  3.65$^{+1.31}_{-2.41}$ &  0.27$^{+0.04}_{-0.03}$  &  3.04   \\
Sk-67 170  & 0.9615$^{+0.0070}_{-0.0055}$   &   1.89$^{+0.02}_{-0.02}$   & 7.30$^{+0.17}_{-0.19}$  &  7.14$^{+0.13}_{-0.14}$  &  -0.22$\pm$0.02   & 0.08$^{+0.03}_{-0.03}$ &  1.65$^{+0.69}_{-1.50}$ &  0.13$^{+0.04}_{-0.04}$  &  1.06   \\
Sk-67 171  & 1.2080$^{+0.0225}_{-0.0215}$   &   1.63$^{+0.04}_{-0.04}$   & 7.39$^{+0.20}_{-0.13}$  &  7.01$^{+0.11}_{-0.14}$  &  -0.71$\pm$0.03   & 0.06$^{+0.02}_{-0.03}$ &  4.33$^{+1.67}_{-2.75}$ &  0.24$^{+0.03}_{-0.03}$  &  1.76   \\
Sk-67 201  & 1.0120$^{+0.0090}_{-0.0160}$   &   1.20$^{+0.03}_{-0.02}$   & 7.49$^{+0.12}_{-0.07}$  &  7.25$^{+0.11}_{-0.08}$  &  -0.36$\pm$0.03   & 0.06$^{+0.02}_{-0.03}$ &  5.36$^{+1.77}_{-2.48}$ &  0.33$^{+0.03}_{-0.03}$  &  1.03   \\
Sk-67 204  & 1.0730$^{+0.0365}_{-0.0400}$   &   1.26$^{+0.06}_{-0.07}$   & 7.36$^{+0.21}_{-0.14}$  &  7.06$^{+0.17}_{-0.17}$  &  -0.47$\pm$0.05   & 0.06$^{+0.02}_{-0.03}$ &  3.96$^{+1.52}_{-2.76}$ &  0.23$^{+0.03}_{-0.03}$  &  1.88   \\
Sk-69 2A   & 0.8465$^{+0.0045}_{-0.0040}$   &   1.92$^{+0.03}_{-0.01}$   & 7.34$^{+0.13}_{-0.14}$  &  7.20$^{+0.09}_{-0.09}$  &  -0.01$\pm$0.02   & 0.22$^{+0.03}_{-0.03}$ &  3.31$^{+0.45}_{-0.57}$ &  0.71$^{+0.03}_{-0.03}$  &  2.37   \\
Sk-69 24   & 1.0660$^{+0.0140}_{-0.0155}$   &   1.77$^{+0.03}_{-0.03}$   & 7.42$^{+0.15}_{-0.11}$  &  7.14$^{+0.11}_{-0.11}$  &  -0.41$\pm$0.03   & 0.16$^{+0.03}_{-0.03}$ &  2.79$^{+0.58}_{-0.84}$ &  0.43$^{+0.03}_{-0.03}$  &  1.70   \\
Sk-69 39A  & 0.9140$^{+0.0055}_{-0.0055}$   &   1.89$^{+0.03}_{-0.01}$   & 7.34$^{+0.14}_{-0.15}$  &  7.25$^{+0.13}_{-0.12}$  &  -0.13$\pm$0.03   & 0.13$^{+0.03}_{-0.03}$ &  4.13$^{+0.93}_{-1.50}$ &  0.54$^{+0.03}_{-0.03}$  &  2.22   \\
Sk-69 82   & 1.0935$^{+0.0400}_{-0.0440}$   &   1.24$^{+0.05}_{-0.07}$   & 7.35$^{+0.22}_{-0.15}$  &  7.06$^{+0.17}_{-0.18}$  &  -0.53$\pm$0.05   & 0.07$^{+0.02}_{-0.03}$ &  4.58$^{+1.49}_{-2.28}$ &  0.33$^{+0.03}_{-0.03}$  &  1.57   \\
Sk-69 113  & 0.9245$^{+0.0085}_{-0.0115}$   &   1.33$^{+0.03}_{-0.02}$   & 7.35$^{+0.14}_{-0.14}$  &  7.05$^{+0.10}_{-0.11}$  &  -0.19$\pm$0.05   & 0.06$^{+0.02}_{-0.03}$ &  4.12$^{+1.50}_{-2.74}$ &  0.26$^{+0.04}_{-0.03}$  &  1.02   \\
Sk-69 170  & 1.0140$^{+0.0105}_{-0.0175}$   &   1.24$^{+0.06}_{-0.02}$   & 7.48$^{+0.14}_{-0.07}$  &  7.28$^{+0.12}_{-0.07}$  &  -0.33$\pm$0.04   & 0.12$^{+0.03}_{-0.03}$ &  4.42$^{+1.03}_{-1.61}$ &  0.53$^{+0.03}_{-0.03}$  &  1.10   \\
Sk-69 299  & 0.8850$^{+0.0040}_{-0.0040}$   &   1.24$^{+0.02}_{-0.01}$   & 7.39$^{+0.14}_{-0.14}$  &  7.19$^{+0.11}_{-0.11}$  &  -0.13$\pm$0.02   & 0.17$^{+0.03}_{-0.03}$ &  3.09$^{+0.58}_{-0.79}$ &  0.52$^{+0.03}_{-0.03}$  &  1.52   \\
Sk-70 45   & 1.1200$^{+0.0225}_{-0.0200}$   &   1.79$^{+0.03}_{-0.05}$   & 7.39$^{+0.17}_{-0.13}$  &  7.05$^{+0.12}_{-0.12}$  &  -0.54$\pm$0.03   & 0.08$^{+0.03}_{-0.03}$ &  2.76$^{+0.96}_{-1.89}$ &  0.23$^{+0.04}_{-0.04}$  &  1.45   \\
Sk-71 14   & 0.9900$^{+0.0045}_{-0.0040}$   &   1.27$^{+0.01}_{-0.01}$   & 7.29$^{+0.27}_{-0.21}$  &  7.29$^{+0.21}_{-0.18}$  &  -0.33$\pm$0.01   & 0.08$^{+0.02}_{-0.03}$ &  5.19$^{+1.46}_{-2.27}$ &  0.43$^{+0.03}_{-0.03}$  &  1.01   \\
\enddata
\tablecomments{The values in the last column correspond to the final
reduced-$\chi^2$ value resulting in the \rv--E(B-V) analysis.}
\end{deluxetable}

\begin{deluxetable}{ l c c c c c c c c}
\tablecaption{LMC OB supergiants results (MagE and FLAMES\,I). \label{table_results_ob}}
\tablehead{
\colhead{Star} & \colhead{\teff}  & \colhead{\lgf} & \colhead{Mg}  & \colhead{BC}  & \colhead{E(B-V)}
& \colhead{Rv} & \colhead{Av}   & \colhead{$\chi^2$} \\
\colhead{       } & \colhead{($\times 10^4$ K)} & \colhead{ (dex) } &
\colhead{(dex)} & \colhead{ (mag) } & \colhead{ (mag) } & \colhead{ } &
\colhead{ (mag) }& \colhead{} 
}
\tablewidth{0pt}
\startdata
N11 24       &  2.3360 $^{+0.0285}_{-0.0290}$  &  1.41 $^{+0.03}_{-0.03}$& 7.21$^{+0.10}_{-0.11}$  &   -2.31$\pm$0.01 & 0.14$^{+0.03}_{-0.03}$ & 4.20$^{+0.81}_{-1.25}$ & 0.57$^{+0.04}_{-0.03}$   &  1.73    \\
N11 36       &  2.5450 $^{+0.0290}_{-0.0260}$  &  1.53 $^{+0.03}_{-0.03}$& 7.17$^{+0.12}_{-0.12}$  &   -2.52$\pm$0.01 & 0.07$^{+0.01}_{-0.02}$ & 7.07$^{+1.75}_{-1.75}$ & 0.52$^{+0.04}_{-0.04}$   &  0.22    \\
N11 54       &  2.5550 $^{+0.0310}_{-0.0290}$  &  1.52 $^{+0.03}_{-0.03}$& 7.14$^{+0.12}_{-0.12}$  &   -2.52$\pm$0.01 & 0.14$^{+0.03}_{-0.03}$ & 5.89$^{+1.03}_{-1.44}$ & 0.85$^{+0.04}_{-0.04}$   &  0.77    \\
NGC\,2004 12 &  2.4110 $^{+0.0330}_{-0.0315}$  &  1.52 $^{+0.03}_{-0.03}$& 7.20$^{+0.11}_{-0.12}$  &   -2.40$\pm$0.01 & 0.03$^{+0.01}_{-0.02}$ & 5.20$^{+2.37}_{-2.82}$ & 0.13$^{+0.04}_{-0.04}$   &  1.02    \\
NGC\,2004 21 &  2.3620 $^{+0.0320}_{-0.0310}$  &  1.63 $^{+0.02}_{-0.02}$& 7.18$^{+0.12}_{-0.11}$  &   -2.29$\pm$0.01 & 0.05$^{+0.02}_{-0.03}$ & 3.38$^{+1.57}_{-2.98}$ & 0.16$^{+0.05}_{-0.05}$   &  0.50    \\
NGC\,2004 22 &  2.3880 $^{+0.0300}_{-0.0310}$  &  1.69 $^{+0.03}_{-0.03}$& 7.09$^{+0.10}_{-0.11}$  &   -2.35$\pm$0.01 & 0.08$^{+0.01}_{-0.01}$ & 8.52$^{+1.42}_{-0.97}$ & 0.65$^{+0.05}_{-0.05}$   &  0.66    \\
Sk-66 1      &  2.0665 $^{+0.0300}_{-0.0250}$  &  1.18 $^{+0.01}_{-0.01}$& 7.28$^{+0.06}_{-0.07}$  &   -1.98$\pm$0.01 & 0.09$^{+0.02}_{-0.03}$ & 5.99$^{+1.41}_{-1.93}$ & 0.57$^{+0.03}_{-0.03}$   &  0.25    \\
Sk-66 15     &  2.6065 $^{+0.0255}_{-0.0270}$  &  1.28 $^{+0.03}_{-0.03}$& 7.14$^{+0.12}_{-0.13}$  &   -2.59$\pm$0.01 & 0.08$^{+0.02}_{-0.03}$ & 5.79$^{+1.50}_{-2.04}$ & 0.49$^{+0.03}_{-0.03}$   &  1.21    \\
Sk-66 23     &  1.9700 $^{+0.0315}_{-0.0310}$  &  1.37 $^{+0.02}_{-0.02}$& 7.15$^{+0.09}_{-0.10}$  &   -1.86$\pm$0.01 & 0.23$^{+0.03}_{-0.03}$ & 4.41$^{+0.53}_{-0.70}$ & 1.02$^{+0.04}_{-0.04}$   &  0.70    \\
Sk-66 26     &  2.3085 $^{+0.0390}_{-0.0350}$  &  1.32 $^{+0.02}_{-0.02}$& 7.07$^{+0.08}_{-0.08}$  &   -2.23$\pm$0.01 & 0.13$^{+0.03}_{-0.03}$ & 3.87$^{+0.83}_{-1.27}$ & 0.49$^{+0.04}_{-0.03}$   &  0.62    \\
Sk-66 27     &  1.7765 $^{+0.0345}_{-0.0300}$  &  1.24 $^{+0.02}_{-0.02}$& 7.25$^{+0.10}_{-0.15}$  &   -1.62$\pm$0.01 & 0.15$^{+0.03}_{-0.03}$ & 4.82$^{+0.87}_{-1.23}$ & 0.71$^{+0.03}_{-0.03}$   &  1.32    \\
Sk-66 36     &  1.9075 $^{+0.0325}_{-0.0330}$  &  1.16 $^{+0.02}_{-0.02}$& 7.17$^{+0.09}_{-0.09}$  &   -1.80$\pm$0.01 & 0.19$^{+0.03}_{-0.03}$ & 4.95$^{+0.70}_{-0.95}$ & 0.94$^{+0.03}_{-0.03}$   &  1.60    \\
Sk-66 37     &  2.5400 $^{+0.0370}_{-0.0330}$  &  1.34 $^{+0.02}_{-0.02}$& 7.05$^{+0.11}_{-0.13}$  &   -2.47$\pm$0.01 & 0.11$^{+0.02}_{-0.03}$ & 6.26$^{+1.35}_{-1.76}$ & 0.69$^{+0.03}_{-0.03}$   &  0.08    \\
Sk-66 166    &  1.9195 $^{+0.0135}_{-0.0120}$  &  1.21 $^{+0.01}_{-0.01}$& 7.21$^{+0.06}_{-0.07}$  &   -1.87$\pm$0.01 & 0.10$^{+0.03}_{-0.03}$ & 5.41$^{+1.26}_{-1.89}$ & 0.55$^{+0.03}_{-0.04}$   &  0.76    \\
Sk-67 14     &  2.2890 $^{+0.0120}_{-0.0125}$  &  1.24 $^{+0.01}_{-0.01}$& 7.19$^{+0.10}_{-0.11}$  &   -2.22$\pm$0.01 & 0.08$^{+0.02}_{-0.03}$ & 5.02$^{+1.47}_{-2.30}$ & 0.38$^{+0.03}_{-0.03}$   &  0.13    \\
Sk-67 36     &  1.9395 $^{+0.0165}_{-0.0175}$  &  1.26 $^{+0.01}_{-0.01}$& 7.26$^{+0.09}_{-0.10}$  &   -1.88$\pm$0.01 & 0.08$^{+0.02}_{-0.03}$ & 5.46$^{+1.55}_{-2.14}$ & 0.42$^{+0.04}_{-0.03}$   &  0.49    \\
Sk-67 133    &  1.9300 $^{+0.0270}_{-0.0280}$  &  1.46 $^{+0.02}_{-0.02}$& 7.22$^{+0.09}_{-0.11}$  &   -1.87$\pm$0.01 & 0.11$^{+0.03}_{-0.03}$ & 3.80$^{+0.89}_{-1.52}$ & 0.40$^{+0.04}_{-0.03}$   &  0.42    \\
Sk-67 154    &  2.3280 $^{+0.0300}_{-0.0300}$  &  1.35 $^{+0.03}_{-0.03}$& 7.15$^{+0.11}_{-0.11}$  &   -2.29$\pm$0.01 & 0.11$^{+0.03}_{-0.03}$ & 1.63$^{+0.50}_{-0.82}$ & 0.18$^{+0.04}_{-0.04}$   &  0.98    \\
Sk-67 228    &  2.0465 $^{+0.0215}_{-0.0215}$  &  1.15 $^{+0.02}_{-0.02}$& 7.21$^{+0.09}_{-0.08}$  &   -2.05$\pm$0.01 & 0.09$^{+0.02}_{-0.03}$ & 5.84$^{+1.47}_{-1.99}$ & 0.52$^{+0.03}_{-0.03}$   &  0.32    \\
Sk-68 40     &  1.9125 $^{+0.0250}_{-0.0230}$  &  1.25 $^{+0.02}_{-0.02}$& 7.33$^{+0.06}_{-0.06}$  &   -1.84$\pm$0.01 & 0.08$^{+0.02}_{-0.03}$ & 6.19$^{+1.55}_{-1.98}$ & 0.53$^{+0.03}_{-0.03}$   &  0.96    \\
Sk-68 92     &  2.2090 $^{+0.0210}_{-0.0195}$  &  1.19 $^{+0.02}_{-0.02}$& 7.21$^{+0.08}_{-0.08}$  &   -2.13$\pm$0.01 & 0.09$^{+0.03}_{-0.03}$ & 4.83$^{+1.23}_{-1.99}$ & 0.46$^{+0.03}_{-0.03}$   &  0.34    \\
Sk-68 171    &  2.4405 $^{+0.0275}_{-0.0265}$  &  1.18 $^{+0.03}_{-0.03}$& 7.31$^{+0.14}_{-0.14}$  &   -2.43$\pm$0.01 & 0.09$^{+0.03}_{-0.03}$ & 4.94$^{+1.31}_{-2.00}$ & 0.45$^{+0.04}_{-0.04}$   &  0.80    \\
Sk-69 43     &  2.2845 $^{+0.0245}_{-0.0220}$  &  1.18 $^{+0.02}_{-0.02}$& 7.22$^{+0.11}_{-0.11}$  &   -2.25$\pm$0.01 & 0.10$^{+0.03}_{-0.03}$ & 4.33$^{+1.11}_{-1.81}$ & 0.42$^{+0.04}_{-0.04}$   &  0.48    \\
\enddata
\tablecomments{The values in the last column correspond to the final
reduced-$\chi^2$ value resulting in the \rv--E(B-V) analysis.}
\end{deluxetable}

\begin{deluxetable}{ l c c c c c c c c}
\tablecaption{LMC OB supergiants results (FEROS). \label{table_results_lennon}}
\tablehead{
\colhead{Star} & \colhead{\teff}                        &
\colhead{\lgf} & \colhead{Mg} &  \colhead{BC} & \colhead{E(B-V)} & \colhead{Rv}
& \colhead{Av} & \colhead{$\chi^2$} \\
\colhead{       } & \colhead{($\times 10^4$ K)} & \colhead{ (dex) } &
\colhead{(dex)} &\colhead{ (mag) } & \colhead{ (mag) } & \colhead{  }   &
\colhead{ (mag) } & \colhead{}
}
\tablewidth{0pt}
\startdata
Sk-66 5   &  1.7185$^{+0.0340}_{-0.0310}$  &   1.10$^{+0.02}_{-0.02}$  & 7.16$^{+0.04}_{-0.08}$ &  -1.54 $\pm$0.01 & 0.08$^{+0.02}_{-0.03}$ &  4.56$^{+1.37}_{-2.26}$ &  0.35$^{+0.03}_{-0.03}$  &  0.63     \\
Sk-66 35  &  2.2000$^{+0.0320}_{-0.0310}$  &   1.23$^{+0.02}_{-0.02}$  & 7.06$^{+0.10}_{-0.10}$ &  -2.20 $\pm$0.01 & 0.09$^{+0.02}_{-0.03}$ &  5.48$^{+1.39}_{-2.02}$ &  0.51$^{+0.03}_{-0.03}$  &  0.66     \\
Sk-66 106 &  2.2550$^{+0.0216}_{-0.0225}$  &   1.25$^{+0.03}_{-0.03}$  & 7.13$^{+0.10}_{-0.09}$ &  -2.20 $\pm$0.01 & 0.09$^{+0.03}_{-0.03}$ &  4.78$^{+1.18}_{-1.92}$ &  0.45$^{+0.03}_{-0.04}$  &  1.88     \\
Sk-66 118 &  2.1355$^{+0.0273}_{-0.0170}$  &   1.25$^{+0.01}_{-0.01}$  & 7.21$^{+0.09}_{-0.07}$ &  -2.04 $\pm$0.01 & 0.11$^{+0.03}_{-0.03}$ &  3.99$^{+0.96}_{-1.49}$ &  0.43$^{+0.03}_{-0.03}$  &  0.56     \\
Sk-67 2   &  1.9910$^{+0.0358}_{-0.0390}$  &   1.12$^{+0.03}_{-0.02}$  & 7.11$^{+0.10}_{-0.10}$ &  -2.00 $\pm$0.01 & 0.20$^{+0.03}_{-0.03}$ &  4.15$^{+0.58}_{-0.74}$ &  0.83$^{+0.03}_{-0.03}$  &  1.45     \\
Sk-67 28  &  2.4900$^{+0.0233}_{-0.0255}$  &   1.15$^{+0.02}_{-0.02}$  & 7.08$^{+0.12}_{-0.12}$ &  -2.46 $\pm$0.01 & 0.05$^{+0.01}_{-0.02}$ &  6.10$^{+1.92}_{-2.25}$ &  0.33$^{+0.04}_{-0.04}$  &  0.36     \\
Sk-67 78  &  1.6150$^{+0.0098}_{-0.0175}$  &   1.20$^{+0.01}_{-0.02}$  & 7.28$^{+0.10}_{-0.05}$ &  -1.47 $\pm$0.01 & 0.06$^{+0.02}_{-0.03}$ &  4.05$^{+1.47}_{-2.51}$ &  0.24$^{+0.04}_{-0.04}$  &  1.82     \\
Sk-67 90  &  2.2240$^{+0.0200}_{-0.0215}$  &   1.19$^{+0.02}_{-0.02}$  & 6.92$^{+0.08}_{-0.08}$ &  -2.16 $\pm$0.01 & 0.08$^{+0.02}_{-0.03}$ &  5.60$^{+1.56}_{-2.19}$ &  0.43$^{+0.03}_{-0.03}$  &  0.08     \\
Sk-67 112 &  2.5725$^{+0.0710}_{-0.0670}$  &   1.19$^{+0.05}_{-0.05}$  & 7.14$^{+0.23}_{-0.25}$ &  -2.55 $\pm$0.01 & 0.06$^{+0.02}_{-0.03}$ &  3.25$^{+1.30}_{-2.44}$ &  0.19$^{+0.04}_{-0.04}$  &  0.51     \\
Sk-67 150 &  2.3715$^{+0.0230}_{-0.0245}$  &   1.16$^{+0.03}_{-0.03}$  & 7.35$^{+0.15}_{-0.17}$ &  -2.33 $\pm$0.01 & 0.08$^{+0.02}_{-0.03}$ &  5.76$^{+1.62}_{-2.17}$ &  0.44$^{+0.04}_{-0.04}$  &  1.46     \\
Sk-67 169 &  2.3660$^{+0.0260}_{-0.0235}$  &   1.20$^{+0.02}_{-0.02}$  & 6.99$^{+0.11}_{-0.10}$ &  -2.30 $\pm$0.01 & 0.07$^{+0.02}_{-0.02}$ &  6.16$^{+1.72}_{-2.12}$ &  0.43$^{+0.03}_{-0.03}$  &  1.16     \\
Sk-67 172 &  1.9800$^{+0.0455}_{-0.0390}$  &   1.29$^{+0.02}_{-0.02}$  & 7.10$^{+0.08}_{-0.05}$ &  -1.92 $\pm$0.01 & 0.09$^{+0.03}_{-0.03}$ &  4.81$^{+1.32}_{-2.12}$ &  0.41$^{+0.04}_{-0.03}$  &  2.15     \\
Sk-67 173 &  2.7000$^{+0.0360}_{-0.0345}$  &   1.20$^{+0.03}_{-0.03}$  & 6.99$^{+0.14}_{-0.16}$ &  -2.66 $\pm$0.01 & 0.08$^{+0.02}_{-0.03}$ &  5.14$^{+1.42}_{-2.17}$ &  0.42$^{+0.04}_{-0.04}$  &  1.51     \\
Sk-67 206 &  2.5230$^{+0.0235}_{-0.0180}$  &   1.14$^{+0.01}_{-0.01}$  & 7.21$^{+0.15}_{-0.16}$ &  -2.50 $\pm$0.01 & 0.08$^{+0.03}_{-0.03}$ &  3.85$^{+1.13}_{-2.06}$ &  0.31$^{+0.03}_{-0.03}$  &  0.87     \\
Sk-67 256 &  2.2400$^{+0.0570}_{-0.0530}$  &   1.23$^{+0.04}_{-0.04}$  & 7.00$^{+0.14}_{-0.18}$ &  -2.21 $\pm$0.01 & 0.10$^{+0.02}_{-0.03}$ &  6.15$^{+1.47}_{-1.91}$ &  0.59$^{+0.03}_{-0.03}$  &  2.43     \\
Sk-68 26  &  1.8160$^{+0.0210}_{-0.0200}$  &   1.15$^{+0.02}_{-0.02}$  & 7.09$^{+0.09}_{-0.09}$ &  -1.74 $\pm$0.01 & 0.22$^{+0.03}_{-0.03}$ &  4.25$^{+0.55}_{-0.66}$ &  0.94$^{+0.03}_{-0.04}$  &  0.13     \\
Sk-68 41  &  2.4970$^{+0.0350}_{-0.0300}$  &   1.14$^{+0.02}_{-0.02}$  & 7.15$^{+0.15}_{-0.15}$ &  -2.48 $\pm$0.01 & 0.06$^{+0.01}_{-0.02}$ &  7.07$^{+1.91}_{-1.77}$ &  0.42$^{+0.03}_{-0.03}$  &  1.98     \\
Sk-68 45  &  2.6675$^{+0.0445}_{-0.0410}$  &   1.13$^{+0.02}_{-0.02}$  & 7.07$^{+0.14}_{-0.14}$ &  -2.63 $\pm$0.01 & 0.07$^{+0.03}_{-0.03}$ &  3.75$^{+1.21}_{-2.21}$ &  0.27$^{+0.04}_{-0.04}$  &  0.36     \\
Sk-68 111 &  2.3645$^{+0.0700}_{-0.0700}$  &   1.25$^{+0.08}_{-0.09}$  & 7.07$^{+0.20}_{-0.27}$ &  -2.37 $\pm$0.01 & 0.10$^{+0.03}_{-0.03}$ &  3.94$^{+0.97}_{-1.67}$ &  0.41$^{+0.04}_{-0.04}$  &  2.57     \\
Sk-69 89  &  1.8255$^{+0.0222}_{-0.0302}$  &   1.21$^{+0.01}_{-0.01}$  & 7.24$^{+0.07}_{-0.07}$ &  -1.70 $\pm$0.01 & 0.08$^{+0.02}_{-0.03}$ &  6.05$^{+1.59}_{-2.05}$ &  0.50$^{+0.03}_{-0.03}$  &  0.52     \\
Sk-69 214 &  2.4870$^{+0.0215}_{-0.0230}$  &   1.28$^{+0.03}_{-0.03}$  & 7.10$^{+0.11}_{-0.11}$ &  -2.44 $\pm$0.01 & 0.19$^{+0.03}_{-0.03}$ &  3.92$^{+0.58}_{-0.76}$ &  0.75$^{+0.04}_{-0.04}$  &  1.36     \\
Sk-69 228 &  2.0615$^{+0.0570}_{-0.0405}$  &   1.22$^{+0.02}_{-0.02}$  & 6.97$^{+0.08}_{-0.08}$ &  -2.01 $\pm$0.01 & 0.20$^{+0.03}_{-0.03}$ &  3.92$^{+0.58}_{-0.73}$ &  0.77$^{+0.04}_{-0.04}$  &  0.06     \\
Sk-69 237 &  2.4625$^{+0.0430}_{-0.0405}$  &   1.24$^{+0.06}_{-0.06}$  & 7.02$^{+0.16}_{-0.20}$ &  -2.41 $\pm$0.01 & 0.15$^{+0.03}_{-0.03}$ &  4.20$^{+0.76}_{-1.09}$ &  0.62$^{+0.03}_{-0.04}$  &  0.23     \\
Sk-69 270 &  1.6825$^{+0.0335}_{-0.0300}$  &   1.12$^{+0.02}_{-0.02}$  & 7.28$^{+0.09}_{-0.09}$ &  -1.51 $\pm$0.01 & 0.22$^{+0.03}_{-0.03}$ &  3.84$^{+0.47}_{-0.61}$ &  0.87$^{+0.04}_{-0.04}$  &  1.01     \\
Sk-69 274 &  1.8020$^{+0.0170}_{-0.0180}$  &   1.17$^{+0.02}_{-0.02}$  & 7.43$^{+0.05}_{-0.04}$ &  -1.66 $\pm$0.01 & 0.15$^{+0.03}_{-0.03}$ &  4.40$^{+0.75}_{-1.13}$ &  0.68$^{+0.03}_{-0.03}$  &  2.88     \\
Sk-70 78  &  2.3430$^{+0.0415}_{-0.0415}$  &   1.24$^{+0.02}_{-0.02}$  & 7.05$^{+0.11}_{-0.10}$ &  -2.29 $\pm$0.01 & 0.10$^{+0.03}_{-0.03}$ &  3.88$^{+1.02}_{-1.73}$ &  0.37$^{+0.03}_{-0.04}$  &  0.95     \\
Sk-70 111 &  2.5650$^{+0.0416}_{-0.0365}$  &   1.25$^{+0.02}_{-0.02}$  & 7.03$^{+0.11}_{-0.10}$ &  -2.56 $\pm$0.01 & 0.13$^{+0.03}_{-0.03}$ &  4.47$^{+0.93}_{-1.39}$ &  0.56$^{+0.03}_{-0.03}$  &  0.42     \\
Sk-70 120 &  2.2875$^{+0.0570}_{-0.0580}$  &   1.12$^{+0.02}_{-0.02}$  & 7.08$^{+0.11}_{-0.10}$ &  -2.22 $\pm$0.01 & 0.10$^{+0.03}_{-0.03}$ &  3.87$^{+1.01}_{-1.65}$ &  0.39$^{+0.03}_{-0.04}$  &  0.33     \\
Sk-71 42  &  2.0120$^{+0.0782}_{-0.0755}$  &   1.15$^{+0.05}_{-0.04}$  & 7.33$^{+0.18}_{-0.17}$ &  -1.95 $\pm$0.01 & 0.11$^{+0.03}_{-0.03}$ &  5.12$^{+1.21}_{-1.89}$ &  0.54$^{+0.03}_{-0.03}$  &  2.11     \\
\enddata
\tablecomments{The values in the last column correspond to the final
reduced-$\chi^2$ value resulting in the \rv--E(B-V) analysis.}
\end{deluxetable}

\begin{deluxetable}{ l c c c c c c c c}
\tablecaption{LMC VFTS OB Supergians results. \label{table_results_vfts}}
\tablehead{
\colhead{Star} & \colhead{\teff} &\colhead{\lgf} & \colhead{Mg}
&\colhead{BC}        &\colhead{E(B-V)} &  \colhead{Rv}& \colhead{Av}& \colhead{$\chi^2$} \\
\colhead{       } & \colhead{($\times 10^4$ K)} & \colhead{ (dex) }
&\colhead{(dex)} &\colhead{ (mag) } & \colhead{ (mag) } & \colhead{  }
& \colhead{ (mag) }   & \colhead{ }
}
\tablewidth{0pt}
\startdata
VFTS 3   & 2.2590$^{+0.0480}_{-0.0470}$ &   1.23$^{+0.04}_{-0.04}$ &  7.06$^{+0.16}_{-0.14}$ &  -2.19$\pm$0.01 & 0.25$^{+0.03}_{-0.03}$ &  4.01$^{+0.45}_{-0.58}$ &  1.01$^{+0.03}_{-0.03}$  &  0.64   \\
VFTS 28  & 2.4585$^{+0.0425}_{-0.0500}$ &   1.11$^{+0.04}_{-0.05}$ &  7.07$^{+0.15}_{-0.15}$ &  -2.41$\pm$0.01 & 0.44$^{+0.03}_{-0.03}$ &  4.58$^{+0.31}_{-0.34}$ &  2.01$^{+0.03}_{-0.04}$  &  0.41   \\
VFTS 69  & 2.4995$^{+0.0480}_{-0.0400}$ &   1.23$^{+0.06}_{-0.06}$ &  7.04$^{+0.15}_{-0.15}$ &  -2.42$\pm$0.01 & 0.34$^{+0.03}_{-0.03}$ &  5.86$^{+0.49}_{-0.60}$ &  1.98$^{+0.04}_{-0.03}$  &  0.32   \\
VFTS 82  & 2.6775$^{+0.0460}_{-0.0540}$ &   1.32$^{+0.05}_{-0.06}$ &  7.05$^{+0.15}_{-0.14}$ &  -2.60$\pm$0.01 & 0.15$^{+0.02}_{-0.03}$ &  7.21$^{+1.17}_{-1.45}$ &  1.06$^{+0.03}_{-0.03}$  &  1.37   \\
VFTS 232 & 1.7300$^{+0.0300}_{-0.0300}$ &   1.53$^{+0.03}_{-0.03}$ &  7.14$^{+0.10}_{-0.09}$ &  -1.55$\pm$0.01 & 0.64$^{+0.03}_{-0.03}$ &  3.53$^{+0.19}_{-0.21}$ &  2.25$^{+0.05}_{-0.04}$  &  0.50   \\
VFTS 270 & 2.0100$^{+0.0390}_{-0.0510}$ &   1.95$^{+0.04}_{-0.04}$ &  7.27$^{+0.17}_{-0.16}$ &  -2.01$\pm$0.01 & 0.11$^{+0.02}_{-0.02}$ &  7.57$^{+1.26}_{-1.36}$ &  0.80$^{+0.03}_{-0.03}$  &  2.53   \\
VFTS 302 & 2.2990$^{+0.0635}_{-0.0590}$ &   1.64$^{+0.05}_{-0.03}$ &  7.21$^{+0.12}_{-0.12}$ &  -2.25$\pm$0.01 & 0.47$^{+0.03}_{-0.03}$ &  4.90$^{+0.31}_{-0.34}$ &  2.31$^{+0.04}_{-0.04}$  &  0.35   \\
VFTS 315 & 2.4785$^{+0.0655}_{-0.0600}$ &   1.68$^{+0.02}_{-0.01}$ &  7.15$^{+0.12}_{-0.12}$ &  -2.41$\pm$0.01 & 0.21$^{+0.03}_{-0.03}$ &  4.49$^{+0.60}_{-0.76}$ &  0.96$^{+0.05}_{-0.04}$  &  0.81   \\
VFTS 431 & 2.0760$^{+0.0550}_{-0.0565}$ &   1.16$^{+0.06}_{-0.06}$ &  7.25$^{+0.15}_{-0.15}$ &  -2.00$\pm$0.01 & 0.24$^{+0.03}_{-0.03}$ &  5.50$^{+0.66}_{-0.80}$ &  1.32$^{+0.05}_{-0.05}$  &  0.37   \\
VFTS 533 & 1.9275$^{+0.0500}_{-0.0515}$ &   1.19$^{+0.05}_{-0.05}$ &  7.06$^{+0.16}_{-0.18}$ &  -1.90$\pm$0.01 & 0.39$^{+0.03}_{-0.03}$ &  3.72$^{+0.31}_{-0.33}$ &  1.45$^{+0.05}_{-0.05}$  &  0.29   \\
VFTS 590 & 2.5340$^{+0.0385}_{-0.0385}$ &   1.11$^{+0.04}_{-0.04}$ &  7.08$^{+0.15}_{-0.15}$ &  -2.42$\pm$0.01 & 0.37$^{+0.03}_{-0.03}$ &  4.91$^{+0.39}_{-0.46}$ &  1.81$^{+0.06}_{-0.06}$  &  0.50   \\
VFTS 696 & 2.4700$^{+0.0500}_{-0.0500}$ &   1.19$^{+0.05}_{-0.05}$ &  7.00$^{+0.14}_{-0.17}$ &  -2.41$\pm$0.01 & 0.25$^{+0.03}_{-0.03}$ &  4.53$^{+0.51}_{-0.60}$ &  1.15$^{+0.04}_{-0.04}$  &  0.36   \\
VFTS 732 & 2.2150$^{+0.0810}_{-0.0750}$ &   1.29$^{+0.06}_{-0.06}$ &  7.03$^{+0.16}_{-0.17}$ &  -2.12$\pm$0.01 & 0.34$^{+0.03}_{-0.03}$ &  5.04$^{+0.43}_{-0.50}$ &  1.70$^{+0.04}_{-0.04}$  &  0.17   \\
VFTS 831 & 1.6250$^{+0.0160}_{-0.0190}$ &   1.33$^{+0.03}_{-0.03}$ &  7.38$^{+0.15}_{-0.14}$ &  -1.41$\pm$0.01 & 0.38$^{+0.03}_{-0.03}$ &  4.31$^{+0.35}_{-0.39}$ &  1.62$^{+0.04}_{-0.04}$  &  1.30   \\
VFTS 867 & 2.6165$^{+0.0360}_{-0.0360}$ &   1.66$^{+0.04}_{-0.02}$ &  7.04$^{+0.12}_{-0.13}$ &  -2.58$\pm$0.01 & 0.32$^{+0.03}_{-0.03}$ &  4.35$^{+0.39}_{-0.48}$ &  1.39$^{+0.04}_{-0.04}$  &  0.21   \\
\enddata
\tablecomments{The values in the last column correspond to the final
reduced-$\chi^2$ value resulting in the \rv--E(B-V) analysis.}
\end{deluxetable}

\subsection{Evolutionary status}

\begin{figure*}
\includegraphics[scale=0.55,angle=90]{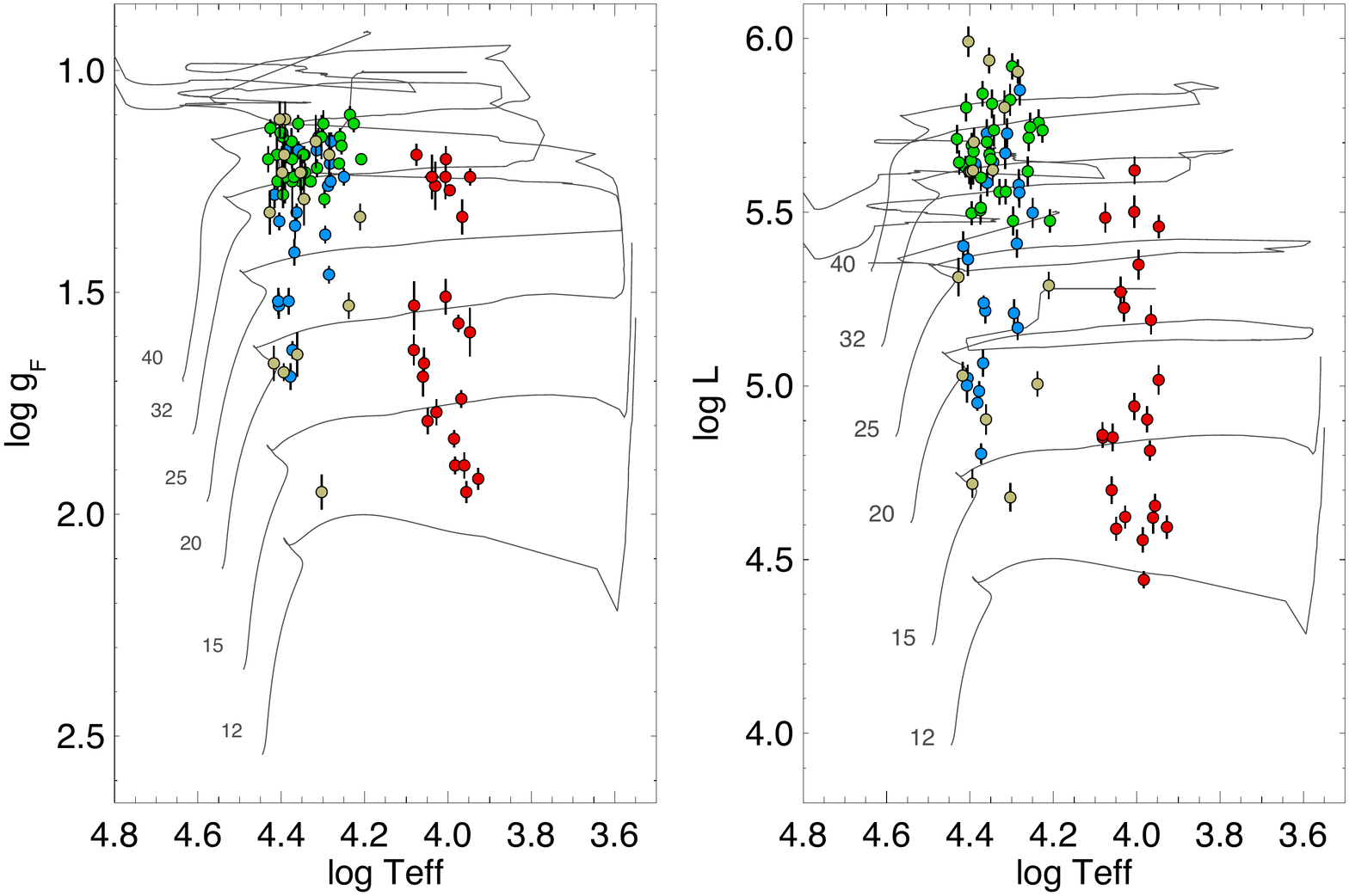}
\caption{\label{hrd} {Hertzsprung-Russell diagram of the supergiant stars of this study compared with evolutionary tracks 
from \cite{eckstroem2012}. Left: spectroscopic HRD; Right: classic HRD. The different samples are shown in different colours: blue--MagE, green--FEROS, 
gold--VFTS.} }
\end{figure*}

Fig.~\ref{hrd}, right-hand side, shows the Hertzsprung-Russell diagram (HRD) of the supergiants investigated by our 
spectroscopic study. We have used \av~and bolometric correction BC as given in the tables 
and a distance modulus m-M = 18.494 mag \citep{pietrzynski2013} to obtain absolute bolometric magnitudes. The theoretical 
evolutionary tracks overplotted are from {\color{green} \citet{eckstroem2012}} and calculated for solar metallicity and with a rotation rate on the
zero age main sequence of 40\% of the critical break-up velocity. The evolved evolutionary status of our targets is obvious from the comparison with the tracks.

An alternative way to discuss stellar evolution, which has been introduced by \citet{langer2014}, is the use of the ``spectroscopic'' 
HRD (sHRD), which replaces the absolute bolometric magnitude by 
the logarithm of the flux-weighted gravity \lgf. The sHRD is particularly useful in cases where the stellar distances are 
uncertain. In addition, in situations where the distance is well know,
the differential comparison of the position of supergiants relative to evolutionary tracks in the HRD versus the sHRD can reveal 
extreme outliers for which the stellar masses are not consistent
with stellar evolution and the location in the HRD (see \citealt{langer2014}). We, therefore, show the sHRD of our targets in the 
left-hand side panel of Fig.~\ref{hrd}. While for some of our objects one might infer different masses from  the inspection of the 
two diagrams, we do not see a reason to exclude objects from the sample.

\subsection{Metallicity}

The major focus of our work is the recalibration of the FGLR. Thus, we postpone the detailed investigation of the chemical composition 
of our sample to a forthcoming publication. However, since
we have adopted a metallicity a factor of 0.4 lower than the Sun for our model atmosphere calculations and for the fit of the spectra, we 
need a straightforward check whether this assumption
is basically correct. Therefore, we use two elements, Mg and Fe, as proxies for metallicity. Strong Mg lines are present in the spectra of 
both the BA and OB supergiants and, thus, we determined
Mg abundances for the whole sample. The spectra of the BA show multiple Fe~{\sc ii} features, as well as weaker Fe~{\sc i} lines, that 
can 
be individually resolved at the resolution and SNR of our sample. On the other hand, the OB supergiants show only a handful of 
extremely weak Fe~{\sc iii} lines in their optical spectra, which makes 
the determination of iron abundances for these objects uncertain. We, thus, determine iron abundances only for the BA supergiant 
subsample.

\subsubsection{Iron abundances from BA supergiants} 

Individual Fe abundances derived from our sample of BA supergiant stars are collected in 
Tab.~\ref{table_results_ba}. These are based on the set of spectral lines defined in 
Tab.~\ref{table_diagnostic_lines}. The mean and standard deviation of the BA sample is $\big\langle Fe/H \big
\rangle\,=\,7.16\pm0.11$ dex.  There are no significant outliers; the median absolute deviation is 0.09 dex, 
indicating chemical homogeneity for the BA stars analysed in this work.  In terms of relative values with 
respect to the Sun and the solar neighbourhood, our value corresponds to $\left[Fe/H\right]\,=\,-0.34,\,-0.36$ 
dex, depending on the reference value used for Fe: $7.50\pm0.04$ dex for photospheric abundance of Fe in 
the Sun \citep{asplund2009} or $7.52\pm0.04$ dex for the mean Fe abundance of a sample of B-type 
dwarf/giant stars in the solar neighbourhood \citep{nieva2012}. 

The derived Fe abundances from our BA sample agree well with the values obtained for classic Cepheids by \cite{luck1998}, $\left[Fe/H\right]\,=\,-0.35\pm0.15$ dex, and more recently by \cite{romaniello2008}, $\left[Fe/H\right]\,=\,-0.33\pm0.10$ dex. This is very encouraging, giving the different nature of the objects, analysis techniques and models that are employed in these studies, when compared to ours. This would give support to the idea that, for more distant galaxies, where metallicity studies of Cepheids are currently not possible, blue supergiant stars (in this particular case, BA-type supergiants) can be used as good proxies for the iron content of Cepheids, even though these objects belong to a somewhat older population.

\subsubsection{Magnesium abundance throughout the sample}

Individual Mg abundances derived for all the stars in the sample are presented in Tab.~\ref{table_results_ba} (BA supergiants) 
and Tab.~\ref{table_results_ob},~\ref{table_results_lennon} and ~\ref{table_results_vfts}  (OB supergiants). Before discussing the 
results, there are a few aspects to consider. First, Mg 
abundances are not affected by stellar evolution, hence we would expect for both samples to show a high 
degree of homogeneity. Secondly, the set of model atmosphere/line formation codes are different for the two 
groups (see previous sections). But, more importantly, whilst treated in non-LTE, the Mg model atom used is 
also not the same for both groups. Hence, slight differences in the zero point of both sub-samples are 
possible. But at the same time, each group should be highly homogeneous internally. 

The mean and standard deviation of the BA subsample is $\big\langle Mg/H $\big\rangle\,=\,7.33$\pm0.07$ 
dex. For the OB subsample, we obtained $\big\langle Mg/H $\big\rangle
\,=\,7.20$\pm0.12$ dex, a value which agrees well with the LMC B-supergiant studies by 
\cite{dufton2006} and \cite{trundle2007}.  As expected, both groups are 
individually highly homogeneous. Also, not totally unexpected given the different atomic data employed for 
each group, there is a zero point difference of 0.13~dex. Disregarding this difference, the value for the 
combined sample is $\big\langle Mg/H $\big\rangle\,=\,7.24$\pm0.14$ dex. There are arguments supporting the 
idea that the value from the BA sample should be preferred (for 
once, there are more lines included in the analysis; only Mg~{\sc ii} 4481 is observed in the OB supergiants). 
However, for the time being, we will adopt the mean value from the full sample as the representative value for 
the present-day Mg abundance in the LMC. Comparing our value with the Sun and solar neighbourhood 
(see references in previous section), we obtain 
$\left[Mg/H\right]\,=\,-0.36,\,-0.32$ dex (Sun: 7.60$\pm$0.04 dex; solar neighbourhood: 7.56$\pm$0.05).

Whilst a detailed discussion on the chemical abundances is deferred to a dedicated paper, 
we would like to mention here that our Mg abundance (an $\alpha$-element), regardless of the solar reference 
considered, agrees very well with the corresponding oxygen value obtained by \cite{bresolin2011} for a 
sample of LMC H~{\sc ii} regions, $12+\log\,(O/H)\,=\,8.36\pm\,0.10$ dex, or  $\left[O/H\right]\,=\,-0.33\pm0.10$ dex, reinforcing
the concept that blue supergiants' and H~{\sc ii} region abundances (based on the electronic temperature of the gas, the 
so-called {\em direct} method) yield consistent results at metallicities below solar.\\

Combining the Mg abundances with the previously discussed Fe abundances from the BA supergiants, it seems clear that 
our sample is chemically homogeneous, and that it shows a \textquotedblleft metallicity\textquotedblright~that 
corresponds to $\left[Z\right]\,=\,-0.35\pm0.09$ dex, where {\em Z} represents the characteristic metallicity 
based on Fe (iron-group) and Mg ($\alpha$-element) abundances (weighted mean and sigma of their 
relative values with respect to the solar values). One can compare this result with the one obtained recently
by \cite{davies2015}, based on the analysis of mid resolution IR spectra of a sample of 9 LMC red supergiant stars. Using LTE hydrostatic model atmospheres and detailed non-LTE calculations of Si~{\sc i}, Ti~{\sc i} and Fe~{\sc i} lines, these authors find 
$\left[Z\right]\,=\,-0.37\pm0.14$ dex. We want to stress out once more that this agreement is remarkable, given the differences in the analysis techniques, model atmospheres, atomic data and nature of the objects analysed in this work. \\

\subsection{Effects of spectral resolution}
The Flux-weighted Gravity as defined by \citet{kudritzki2003},  \gf $\equiv$ g/\teffq, requires the 
determination of surface gravity and effective temperature of the star. These fundamental parameters can in 
principle be easily determined from the quantitative analysis of the optical spectrum of B- and A-type 
supergiant stars, as shown in previous sections. But, whilst the surface gravity diagnostic lines (the hydrogen 
Balmer lines) are strong enough so that they remain accessible even at the low spectral resolutions required for 
the extragalactic work beyond the Local Group, this is not the case of the ionisation equilibria required for the 
determination of the effective temperature of BA-type supergiants, based on weak lines from neutral species (chiefly Mg~{\sc i} and Fe~{\sc i}). 
For OB supergiants, ionization
equilibria (Si~{\sc ii/iii/iv} and He~{\sc i/ii}, see section 3.3) can still be used at low resolution, since the lines are strong enough and fairly isolated 
\citep{urbaneja2003}.\\

To circumvent the issue of the lack of ionization equilibria for BA-supergiants at low spectral resolution \citet{kudritzki2008} utilised 
the Balmer jump (Balmer discontinuity, located at $\sim$3650 
\AA) as an alternative diagnostic. This technique was then used in many of the extragalactic FGLR applications cited in the 
introduction. Our LMC observations of BA supergiants provide
an excellent opportunity to compare the two alternative diagnostic methods. This comparison is carried out in the following 
subsections.

We also note that more recently \citet{hosek2014} and \citet{kudritzki2014} have introduced a technique, which determines 
\teff~from information encoded in the spectra: the strength of He~{\sc i} 
lines and/or the relative strength of Ti~{\sc ii} and Fe~{\sc ii} lines. This alternative method has already been thoroughly tested by 
\citet{hosek2014} and we, thus, refrain from an additional test here.

\subsubsection{Consistency between Balmer jump and ionization equilibrium based temperatures}
The relative flux calibration of the data collected during the two nights in October 2014 is regarded as very good, as is 
shown by the high consistency between the Balmer jump measurements for the different standard stars observed during 
these nights. It was not possible however to achieve the same satisfactory relative calibration for the data collected during 
the third night. Hence, the results shown in this section refer only to a subsample of our stars, 17 BA 
supergiant stars (see Tab.~\ref{table_results_db}).

\begin{figure*}
\includegraphics[scale=0.7]{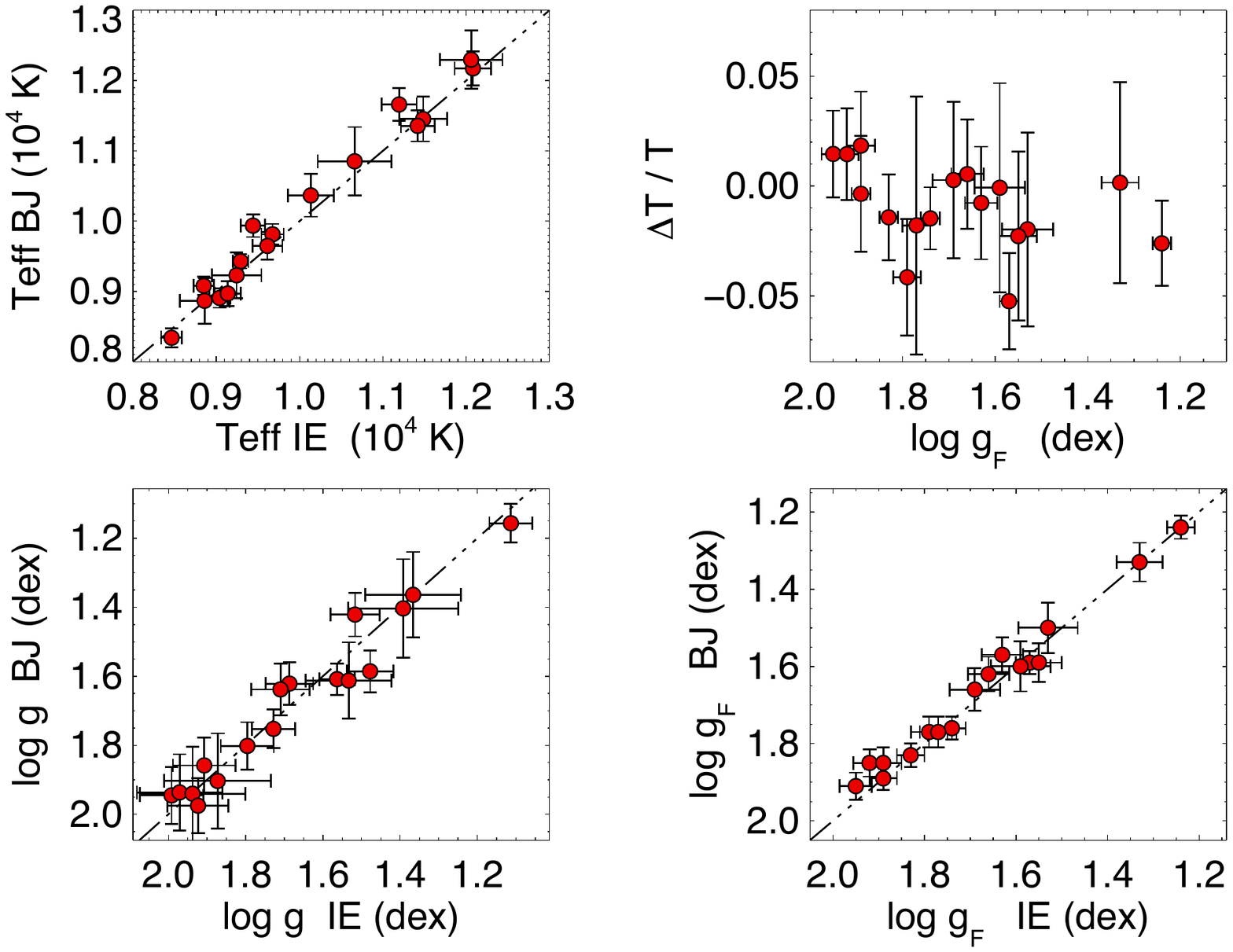}
\caption{ \label{fig_teff_ie_bjump} Comparison of Balmer jump based and IE based results. Clockwise: effective temperatures; relative difference in Teff against flux--weighted gravity; flux weighted gravities; surface gravities. See text for a discussion.}
\end{figure*}

First, we conducted two separated analyses, one considering the ionisation equilibrium of Mg~{\sc i/ii} and 
Fe~{\sc i/ii}, and a second one using the \db index as \teff~indicator. The results are summarized in 
Fig.~\ref{fig_teff_ie_bjump}, and the measurements of the Balmer jump \db as defined in \citet{kudritzki2008} 
are given in Tab~\ref{table_results_db}. Within the uncertainties of the analyses, the results show a high consistency 
between the 
temperatures derived via the application of an ionisation equilibrium and the ones obtained through the \db 
index. As for the latter, the measurement of \db could be potentially affected by the particular form of the
extinction law modifying the intrinsic spectral energy distribution of each star individually. However, as shown
in 
Fig.~\ref{fig_effect_redlaw_on_db}, this is not the case. The left hand side of this figure illustrates the 
difference between the true \db index measured from a synthetic spectral energy distribution of a typical 
A0 supergiant star, reddened by the amount indicated in the x-axis, when considering different forms of the 
reddening law (MW--\citealt{cardelli1989}, LMC--\citealt{misselt1999}, SMC--\citealt{gordon2003}), and the 
{\em measured} \db value under the 
assumption that the form of the extinction law is given by \cite{cardelli1989}, with $\mbox{\rv}/E\left(B-V\right)=3.1$. 
An error in \db of 0.02 dex, comparable to the accuracy achievable in external galaxies \citep[i.e.][]
{kudritzki2008, urbaneja2008}, would require the combination of a significantly different form of the extinction 
law (SMC versus MW) and a high reddening value, $E\left(B-V\right)\sim0.9$ mag. It is very unlikely that a 
star showing such high reddening (hence extinction) would be selected for spectroscopic observations, since it 
will appear as a faint, reddish target.

The right-hand side of Fig.~\ref{fig_effect_redlaw_on_db} illustrates the effects of assuming a fix \rv$\,=\,$3.1 
value (for an extinction curve given by Cardelli et al. parametrisation) when our typical A0 star is 
reddened with different \rv~values (x-axis). It is also considered in this case that the observed photometric 
colour is V-I, hence there is a contribution from the translation from $E\left(V-I\right)$ to $E\left(B-V\right)$, 
which is \rv~dependent, also present. The lower right-hand side panel shows the difference in $E\left(B-V\right)$ 
(real minus recovered) for different values of \rv, whilst the upper panel displays the (negligible) 
effect on \db.\\

We conclude that the Balmer jump is in fact an excellent \teff~discriminant in the B8--A3 spectral range, 
rivalling in accuracy with the standard techniques used at high 
spectral resolution. We note that this conclusions also agrees with the results obtained by \cite{firnstein2012}, who studied 
a large sample of Milky Way BA supergiants. Fig.~\ref{fig_teff_ie_bjump} also demonstrates that flux-weighted gravities are 
not systematically affected by the choice of the method used for the determination of \teff.

\subsubsection{Consistency between low and high spectral resolution results for the flux-weighted gravity.}
In the preceding subsection we used our high spectral resolution observations to compare the results of the two alternative methods, the 
Balmer jump or ionization 
equilibria, as a constraint for the determination of effective temperature. However, in the extragalactic application of the FGLR-method for 
distance determinations 
spectra of moderate resolution of only 4.5 or 5~\AA~are used because of the faintess of the targets at large distances. The transition from 
high to moderate 
resolution could introduce systematic effects affecting temperatures and, most importantly, flux-weighted gravities. 

In order to investigate these effects we degraded the resolution of the spectra of the BA supergiants in Tab.~\ref{table_results_db} to 5~\AA~and repeated the 
spectral analysis again using the Balmer jump for temperature (note that we adopted a general uncertainty of 0.02 dex for the Balmer jump to be consistent with the 
extragalactic work. This is larger than in the previous section and leads to larger \teff~errors).

\begin{figure*}
\includegraphics[scale=0.7]{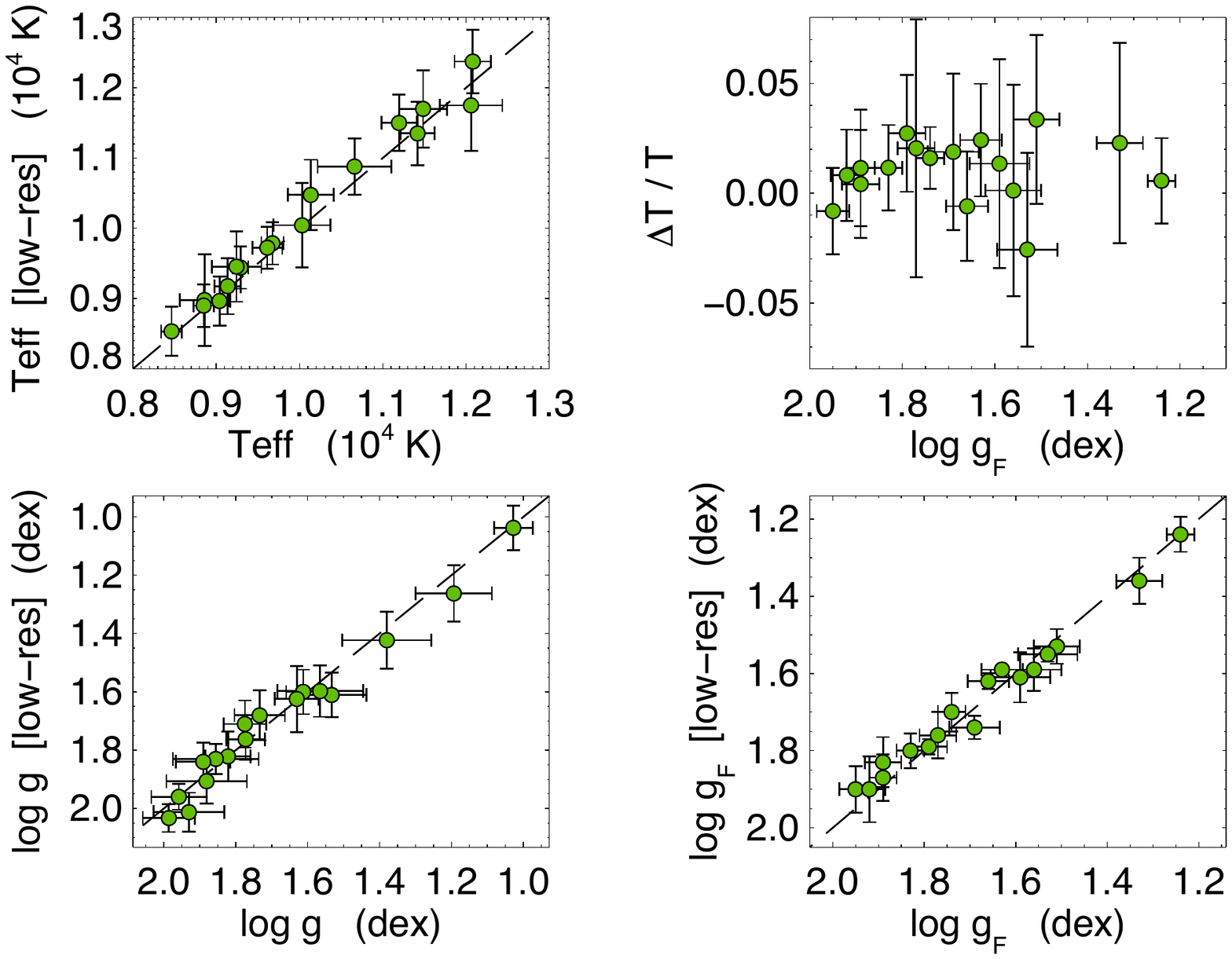}
\caption{ \label{fig_gf_low_res} Comparison of flux-weighted gravities and temperatures derived from high- and low-resolution optical spectra. Clockwise: effective 
temperatures (based on IE at high-res, based on the Balmer jump at low-res); relative difference in Teff against flux--weighted gravity; flux weighted gravities; 
surface gravities. See text for a discussion.}
\end{figure*}

\begin{figure*}
\plottwo{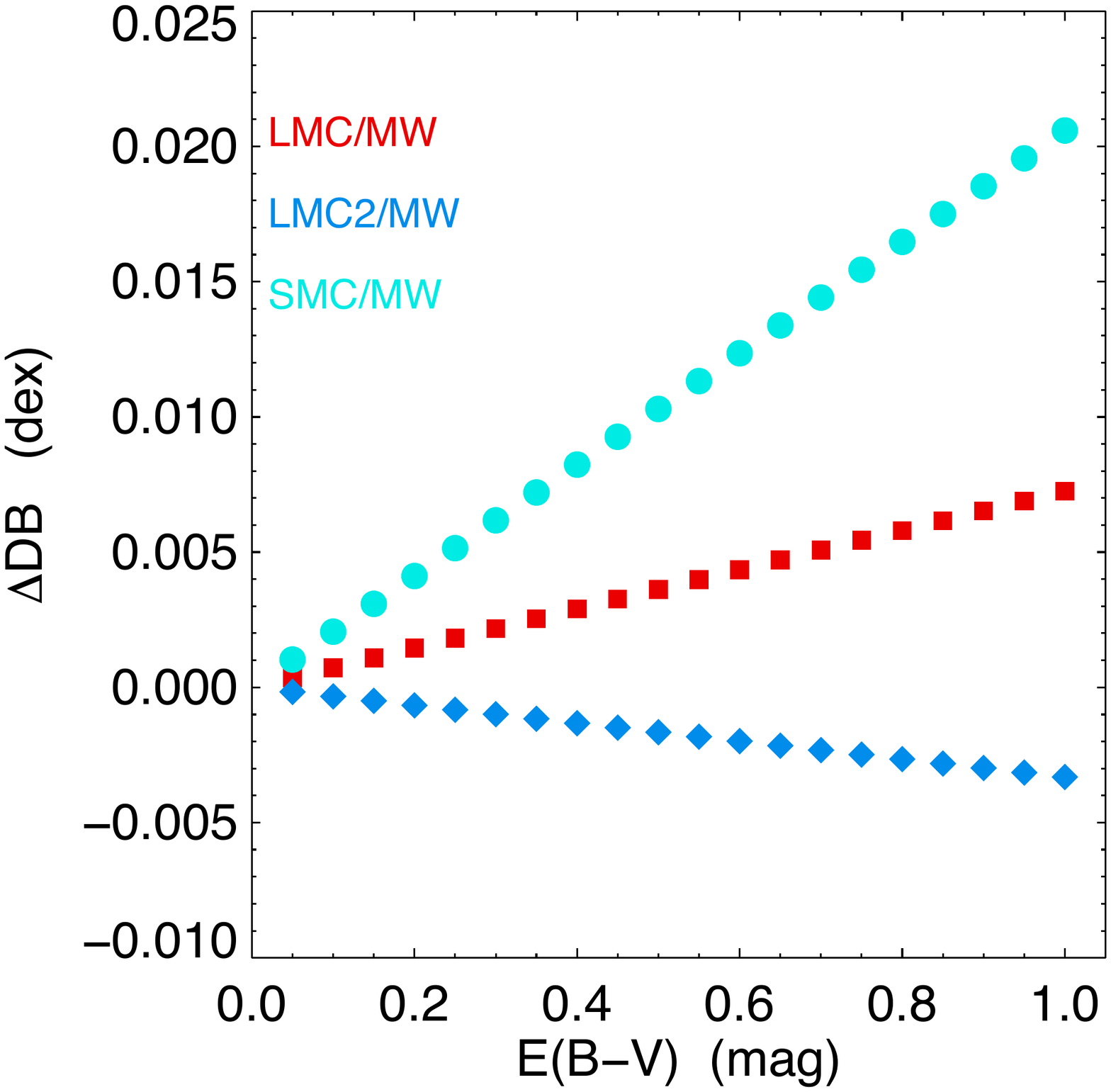}{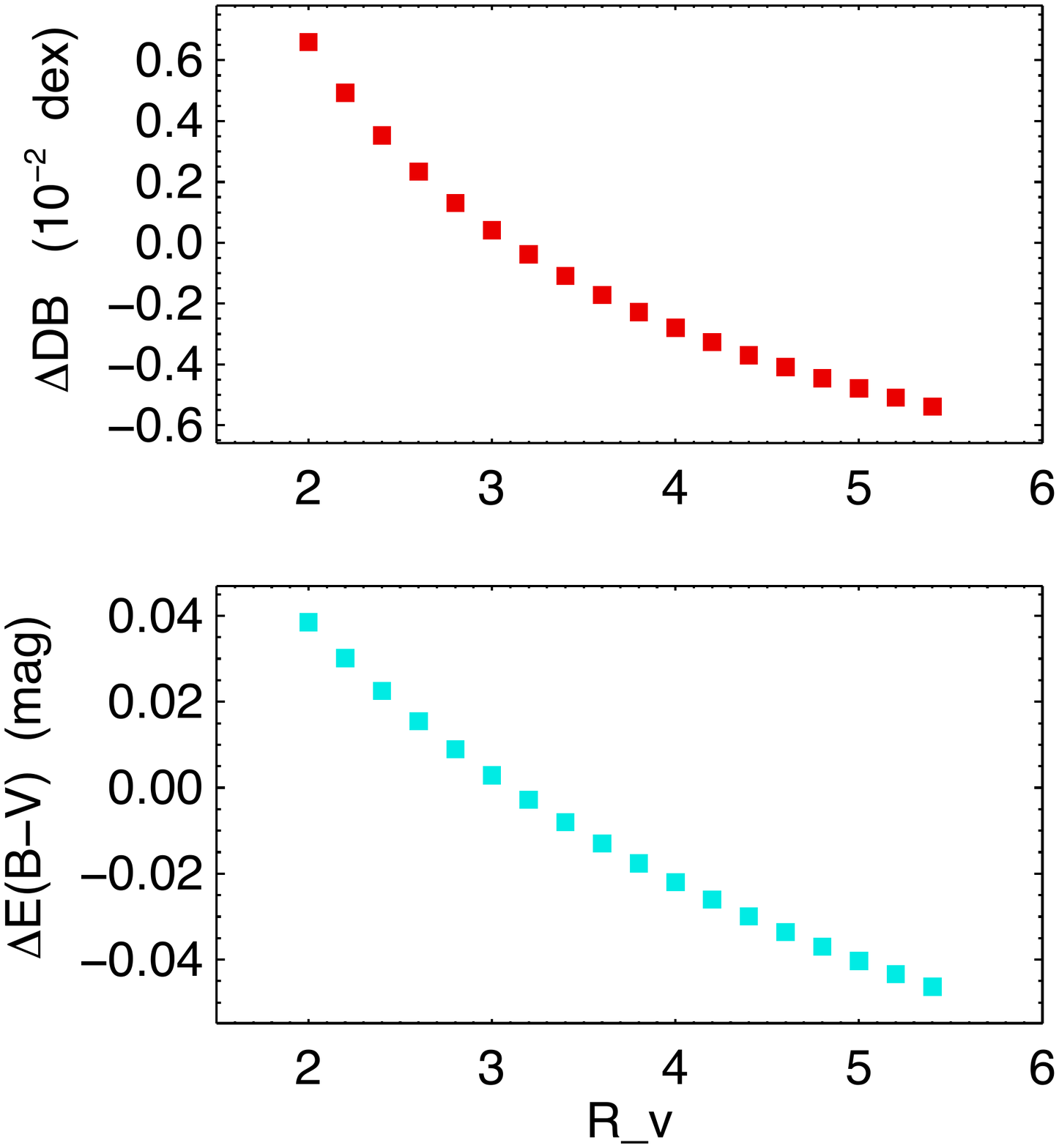}
\caption{\label{fig_effect_redlaw_on_db} Effects of the reddening law on the Balmer jump. Different reddening laws are considered: MW--\citealt{cardelli1989}, LMC and LMC2--\citealt{misselt1999}, SMC--\citealt{gordon2003}. See text for discussion.}
\end{figure*}

Fig.~\ref{fig_gf_low_res} shows the comparison between temperatures and gravities obtained in this way with the values resulting from 
the ionization equilibrium analysis of the high resolution spectra. Within the error margins no systematic effects are found.

\begin{deluxetable}{ l l l l l l  }
\tablecaption{BA Supergians results based on the Balmer jump as \teff~indicator, at nominal and degraded spectral resolution. \label{table_results_db} }
\tablehead{
\colhead{Star}  & \colhead{\db } & \colhead{\teff} & \colhead{$\mbox{\lgf}$} & \colhead{\teff} & \colhead{$\mbox{\lgf}$}  \\
\colhead{      }  & \colhead{ (dex) } & \colhead{($\times 10^4$ K)} & \colhead{(dex)}       & \colhead{($\times 10^4$ K)} & \colhead{(dex)}   
}
\startdata
 NGC\,2004 8 &  0.21  &  1.1455$\pm$ 0.0290 &  1.66$\pm$ 0.06  &  1.1700$\pm$ 0.0248 &  1.74$\pm$ 0.03  \\
 Sk-65 67     &  0.29  &  0.9935$\pm$ 0.0075 &  1.59$\pm$ 0.03  &  1.0045$\pm$ 0.0192 &  1.59$\pm$ 0.05  \\
 Sk-67 19     &  0.43  &  0.8865$\pm$ 0.0150 &  1.60$\pm$ 0.07  &  0.8980$\pm$ 0.0167 &  1.61$\pm$ 0.06  \\
 Sk-67 137   &  0.16  &  1.2300$\pm$ 0.0375 &  1.50$\pm$ 0.07  &  1.1750$\pm$ 0.0215 &  1.55$\pm$ 0.02 \\
 Sk-67 140   &  0.36  &  0.9815$\pm$ 0.0070 &  1.83$\pm$ 0.03  &  0.9790$\pm$ 0.0138 &  1.80$\pm$ 0.05  \\
 Sk-67 141   &  0.49  &  0.8910$\pm$ 0.0065 &  1.91$\pm$ 0.04  &  0.8965$\pm$ 0.0108 &  1.90$\pm$ 0.06  \\
 Sk-67 143   &  0.29  &  1.0365$\pm$ 0.0140 &  1.59$\pm$ 0.05  &  1.0475$\pm$ 0.0189 &  1.53$\pm$ 0.05  \\
 Sk-67 155   &  0.38  &  0.9430$\pm$ 0.0045 &  1.76$\pm$ 0.03  &  0.9440$\pm$ 0.0144 &  1.70$\pm$ 0.05  \\
 Sk-67 157   &  0.19  &  1.1355$\pm$ 0.0200 &  1.62$\pm$ 0.05  &  1.1350$\pm$ 0.0301 &  1.62$\pm$ 0.02  \\
 Sk-67 170   &  0.39  &  0.9648$\pm$ 0.0090 &  1.89$\pm$ 0.03  &  0.9725$\pm$ 0.0143 &  1.87$\pm$ 0.06  \\
 Sk-67 171   &  0.17  &  1.2175$\pm$ 0.0220 &  1.57$\pm$ 0.05  &  1.2375$\pm$ 0.0245 &  1.59$\pm$ 0.01  \\
 Sk-69 2A     &  0.62  &  0.8340$\pm$ 0.0065 &  1.85$\pm$ 0.04  &  0.8535$\pm$ 0.0140 &  1.90$\pm$ 0.09  \\
 Sk-69 24     &  0.26  &  1.0855$\pm$ 0.0220 &  1.77$\pm$ 0.04  &  1.0880$\pm$ 0.0103 &  1.76$\pm$ 0.06  \\
 Sk-69 39A   &  0.47  &  0.8970$\pm$ 0.0080 &  1.85$\pm$ 0.04  &  0.9175$\pm$ 0.0109 &  1.83$\pm$ 0.06  \\
 Sk-69 113   &  0.25  &  0.9230$\pm$ 0.0150 &  1.33$\pm$ 0.05  &  0.9455$\pm$ 0.0196 &  1.36$\pm$ 0.06  \\
 Sk-69 299   &  0.26  &  0.9080$\pm$ 0.0065 &  1.24$\pm$ 0.03  &  0.8900$\pm$ 0.0158 &  1.24$\pm$ 0.05  \\
 Sk-70 45     &  0.23  &  1.1660$\pm$ 0.0210 &  1.77$\pm$ 0.04  &  1.1500$\pm$ 0.0185 &  1.79$\pm$ 0.02  \\
 \enddata
 \tablecomments{First (second) \teff--$\log\,g_F$ pair corresponds to the values derived at nominal 
 (degraded) spectral resolution. The uncertainty in the measured \db index at the nominal resolution is in all 
 cases smaller than 0.01~dex. See text for details.}
\end{deluxetable}

\subsection{Individual reddenings and total to selective extinction ratios, E(B-V) and \rv.}
Once the fundamental parameters of each star are known, we calculate a tailored model for each object. By comparing the 
predicted spectral energy distribution with the observed one, individual filter pass-band integrated
reddening values E(B-V) and \rv = A$_V$/E(B-V) can be derived for each
object. Our procedure to determine the reddening, E(B-V) and \rv, is
based entirely on the use of observed photometric colours and
magnitudes, which are then compared to colours obtained from the
spectral energy distribution of our model atmospheres. We do not use
monochromatic fluxes for the determination of E(B-V) and \rv.

\begin{figure*}
\includegraphics[scale=0.7]{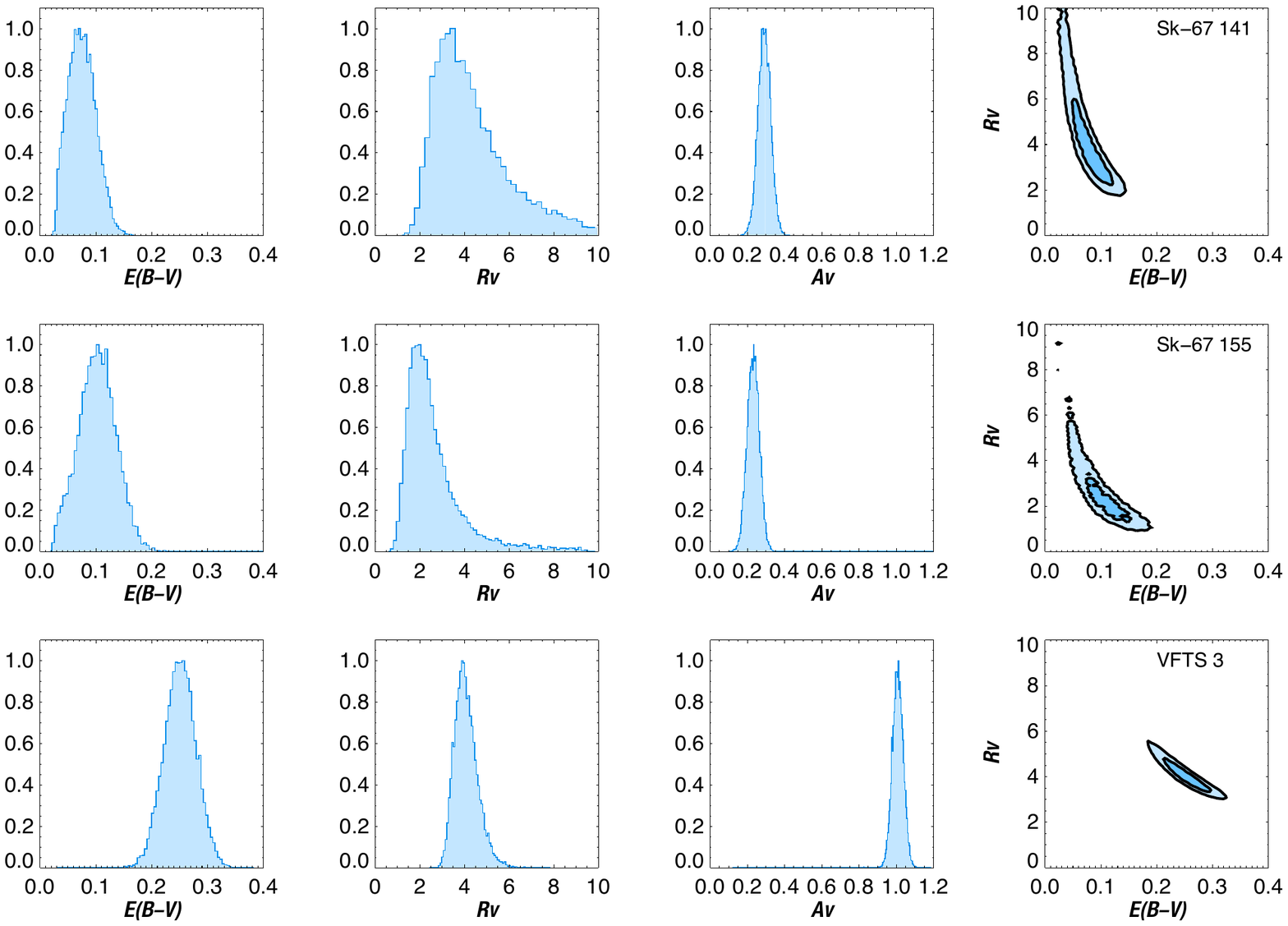}
\caption{\label{ebv_rv_av_fig} Monte Carlo Marko Chain determination
of E(B-V),\rv~and \av~for the stars Sk-67 141, Sk-67 155 and VFTS 3.
For each target posterior probability distribution functions for
each variable are displayed together with the conditional
E(B-V)--\rv~probability function, showing the isocontours enclosing
67\%and 95\% of the solutions.}
\end{figure*}

\begin{figure*}
\gridline{\fig{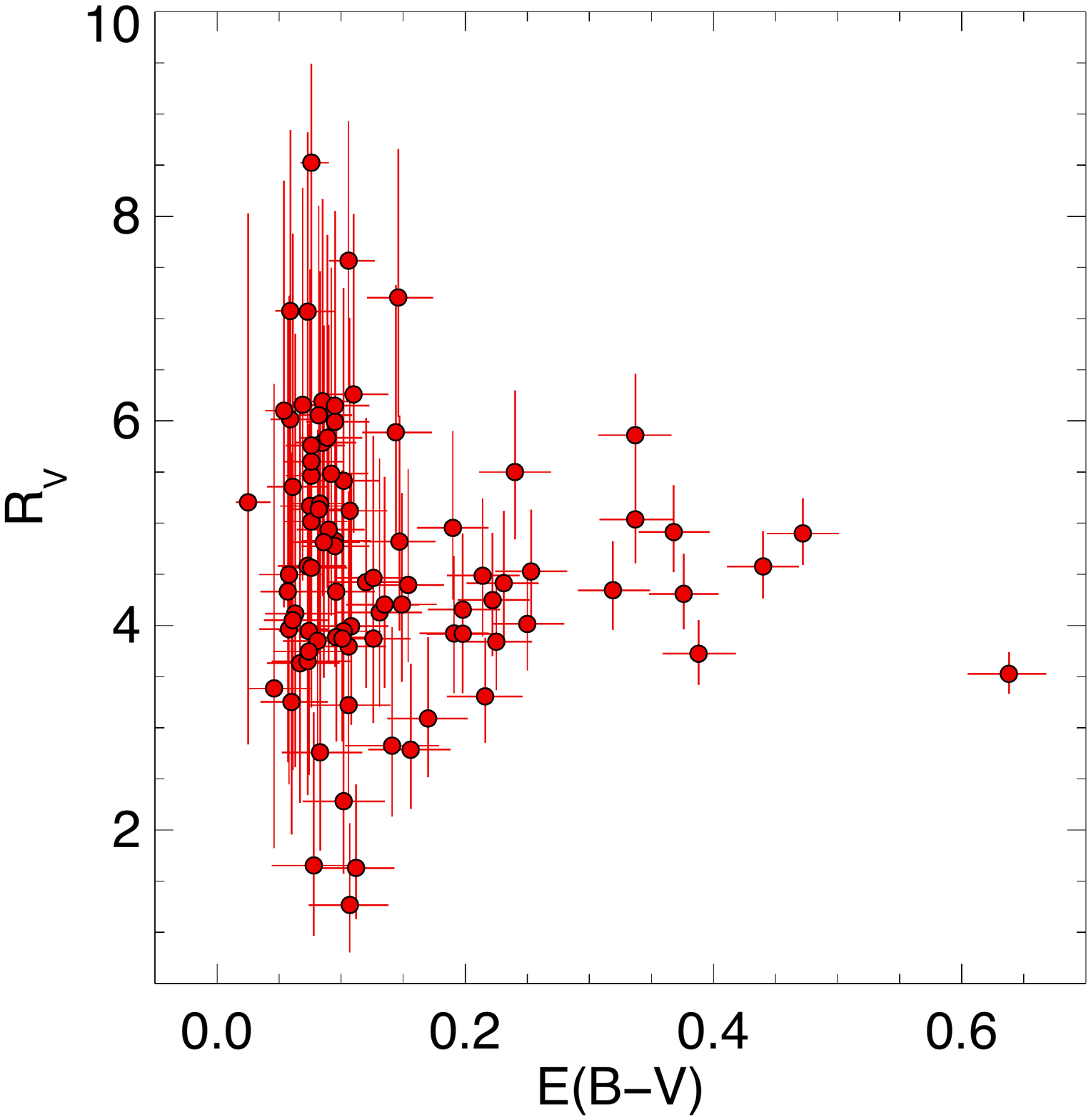}{0.3\textwidth}{(a)}
              \fig{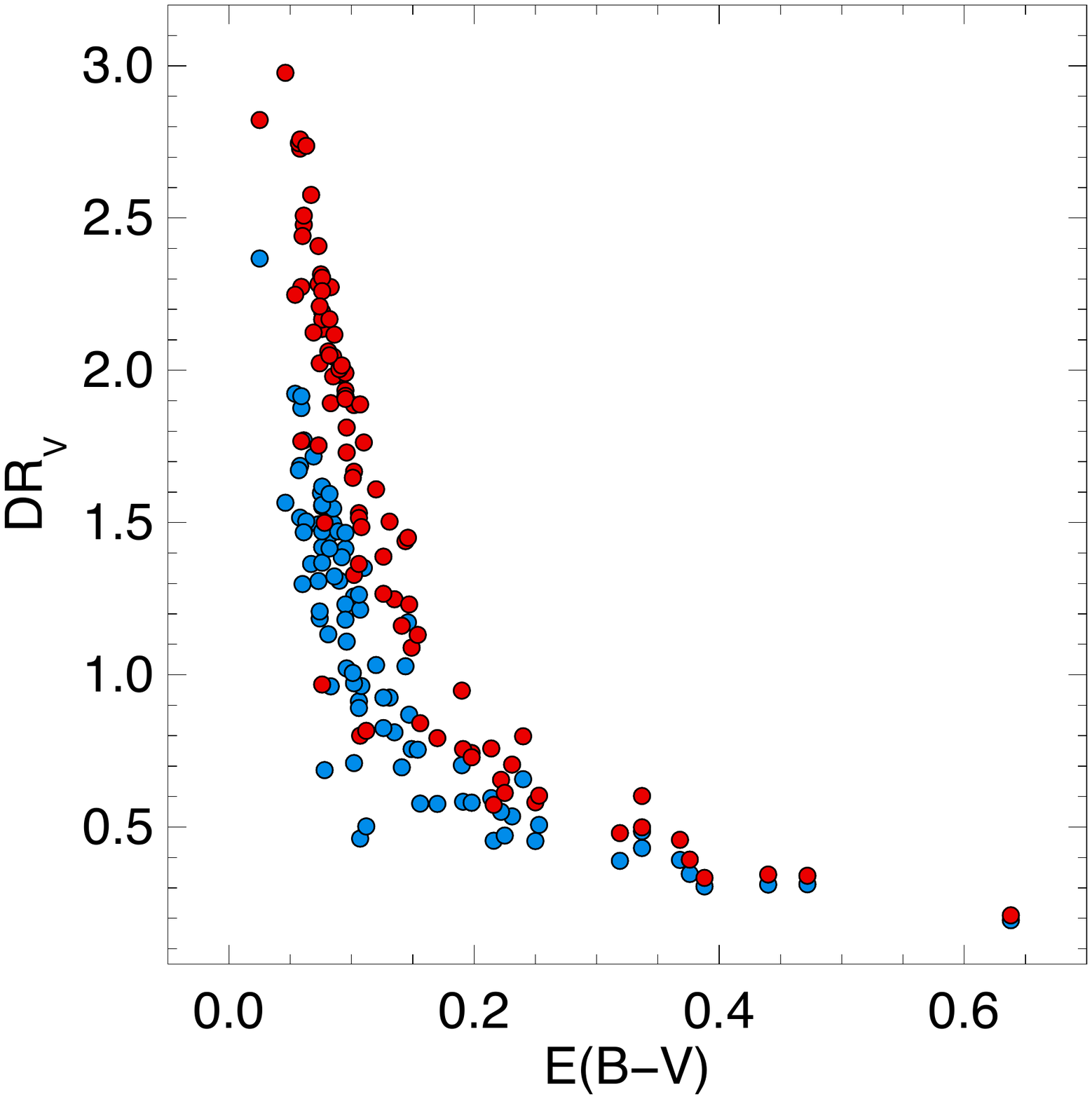}{0.3\textwidth}{(b)}
              \fig{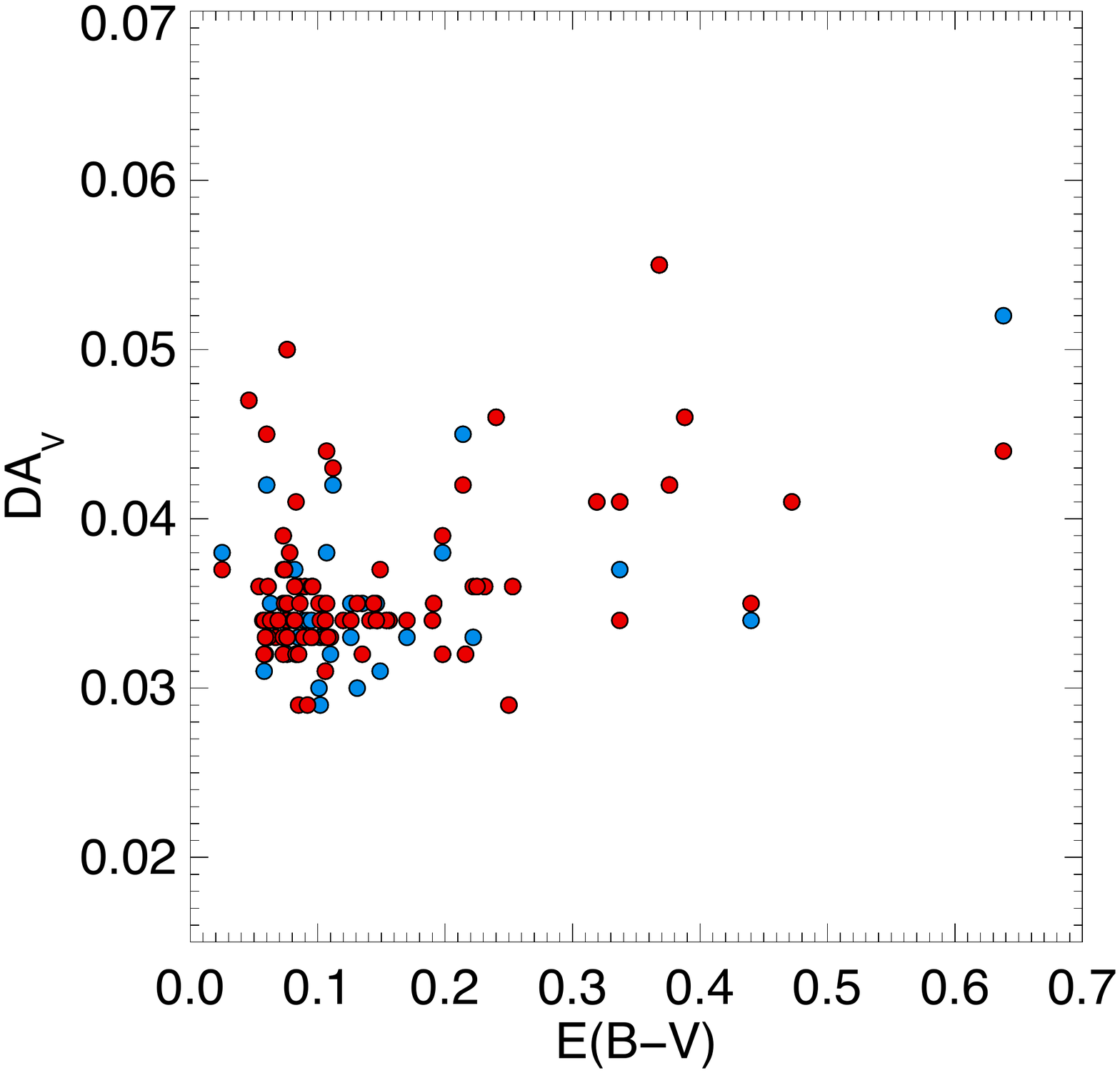}{0.3\textwidth}{(c)} 
             }
\caption{\label{erros_ebv_rv_av} E(B-V), \rv, \av~and
uncertainties for the full sample. (a) \rv~vs. E(B-V). (b)
Possitive (blue) and negative (red) uncertatinties in \rv~vs. E(B-V).
(c) Possitive (blue) and negative (red) uncertainties in
\av~vs. E(B-V).  }
\end{figure*}

We proceed as follows: (1) a pair of monochromatic E(4405-5495) and
R$_{5495}\,=\,$A$_{5495}$/E(4405-5495) values are
drawn. This is done because the reddening laws, for which we
have applied our procedure (see below), are defined as monochromatic
laws and the key parameter characterizing them is R$_{5495}$. (2) the synthetic spectral 
energy distribution, SED, predicted by the tailored model calculated with the results obtained in our spectroscopic analysis, is then 
reddened using the \cite{cardelli1989} monochromatic reddening law (the use of other reddening laws has no significant effect on the 
final FGLR, see below); (3) the corresponding B-V, V-J, V-H and V-Ks model colours are calculated from the reddened SED; (4) the differences 
between model and observed colours are calculated and enter the calculation of the cost function. Errors of the colour differences account 
for the observed values and a contribution from the synthetic photometry (assumed to be a constant 0.01 mag); (5) steps 1 to 4 are repeated 
in a Monte Carlo Markov Chain (MCMC) procedure, minimizing the cost function (a $\chi^2$ with 2 degrees of freedom, since we are using four colours 
two solve simultaneously for two variables).
Fig.~\ref{ebv_rv_av_fig}
displays the results of our MCMC procedure for three of the targets in
our sample. Posterior probability distribution functions for
E(B-V),\rv~and \av~are shown, as well as the conditional
E(B-V)--\rv~distribution function, with the isocontours enclosing 67\%
and 95\% of the solutions obtained. We determine mean
values and (asymmetric) uncertainties from the posterior probability
distribution functions. Note that R$_{5495}$ and E(4405-5495) are converted to the 
corresponding filter integrated quantities using the final model atmospheres SEDs. These values of \rv~and E(B-V) are then used to 
calculate the extinction A$_{\!\mbox{\scriptsize \,\sc v}}$.
Fig.~\ref{erros_ebv_rv_av} summarizes the results for the full
sample. Whilst the uncertainties of \rv~become large for small values of
reddenig, the extinction \av~remains well constrained with relatively
small errors. The reason is explained by the isocontours shown in
Fig.~\ref{ebv_rv_av_fig}, which demonstrate that errors in \rv~and
E(B-V) are anticorrelated.

Note that whilst IRAC/Spitzer photometry is available for most of the sample, we used it just 
as a posterior validation of the result, and not in the derivation of \rv. 

We decided to work with line-of-sight values, i.e. we do not consider separated contributions for the Milky 
Way and the LMC. The values derived for the full sample are collected in Tab.~\ref{table_results_ba} through Tab.~\ref{table_results_vfts}. 
There are 4 stars in common with the work by \cite{gordon2003}: Sk-69 270, Sk-67 2, Sk-68 26 and VFTS\,696 (aka Sk-68 140). 
Note that \cite{gordon2003} implicitly \textquotedblleft corrected\textquotedblright~for Galactic extinction, since they use {\em lightly 
reddened} LMC stars as comparisons to estimate these values (whilst we use model atmospheres
instead). Nonetheless, our E(B-V) and \rv~values are
in agreement with theirs, when considering our larger uncertainties for \rv; our derived E(B-V) values 
are $\sim$0.05--0.08 mag higher, which can be easily understood as the combined 
Galactic and LMC contributions to reddening of the stars used as reference by these authors \citep[see Tab. 2]{gordon2003}.
Unfortunately a comparison with the recent study by \cite{maiz2014} is not possible since we do not have stars in common 
with this work. \\

\begin{figure*}
\includegraphics[scale=0.45,angle=90]{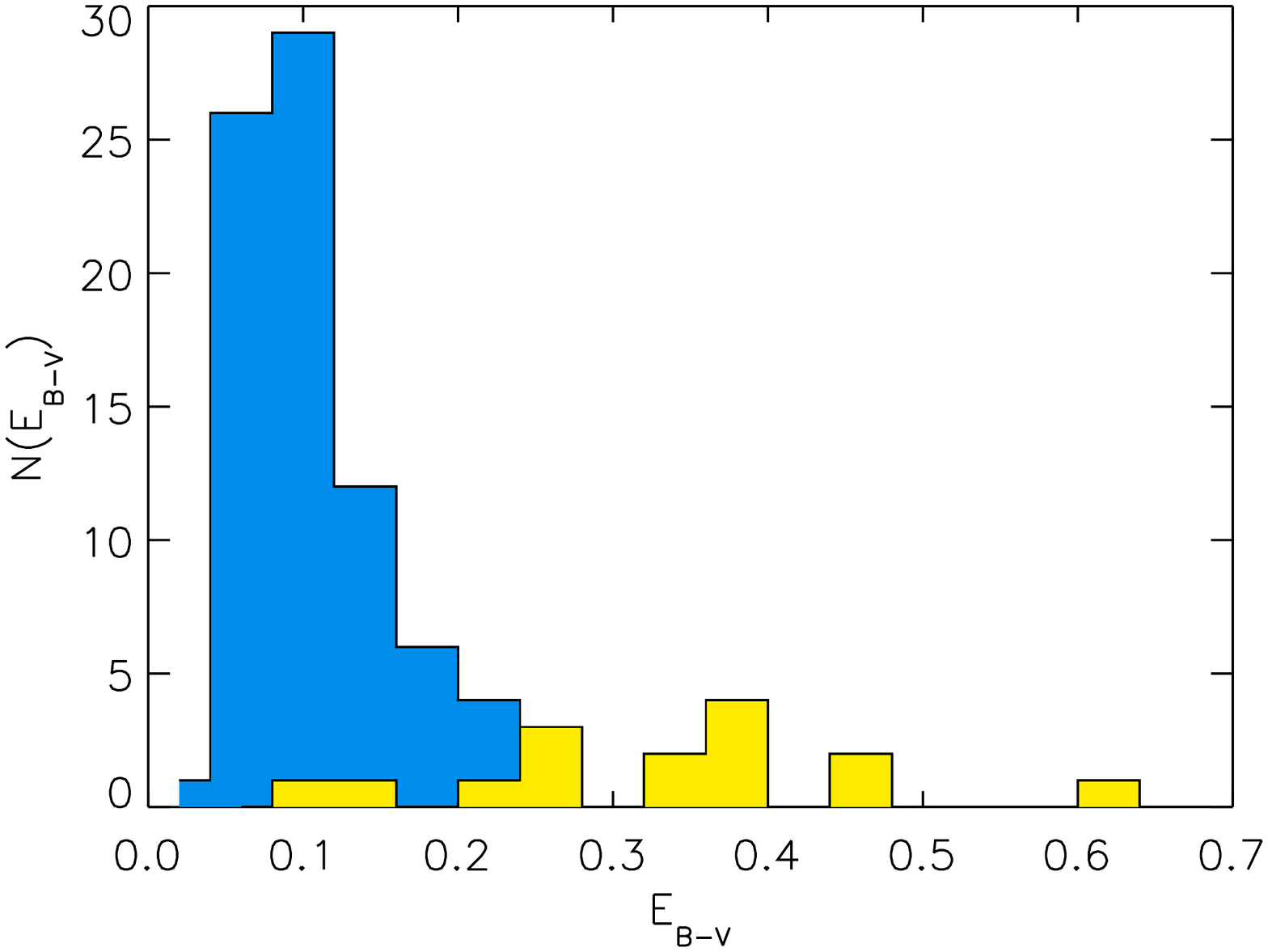}
\caption{\label{ebv_rv}  Histogram of observed E(B-V) values. The yellow histograms correspond to 
the subsample of
stars in the Tarantula region (VFTS sample). }
\end{figure*}

\begin{figure*}
\includegraphics[scale=0.45]{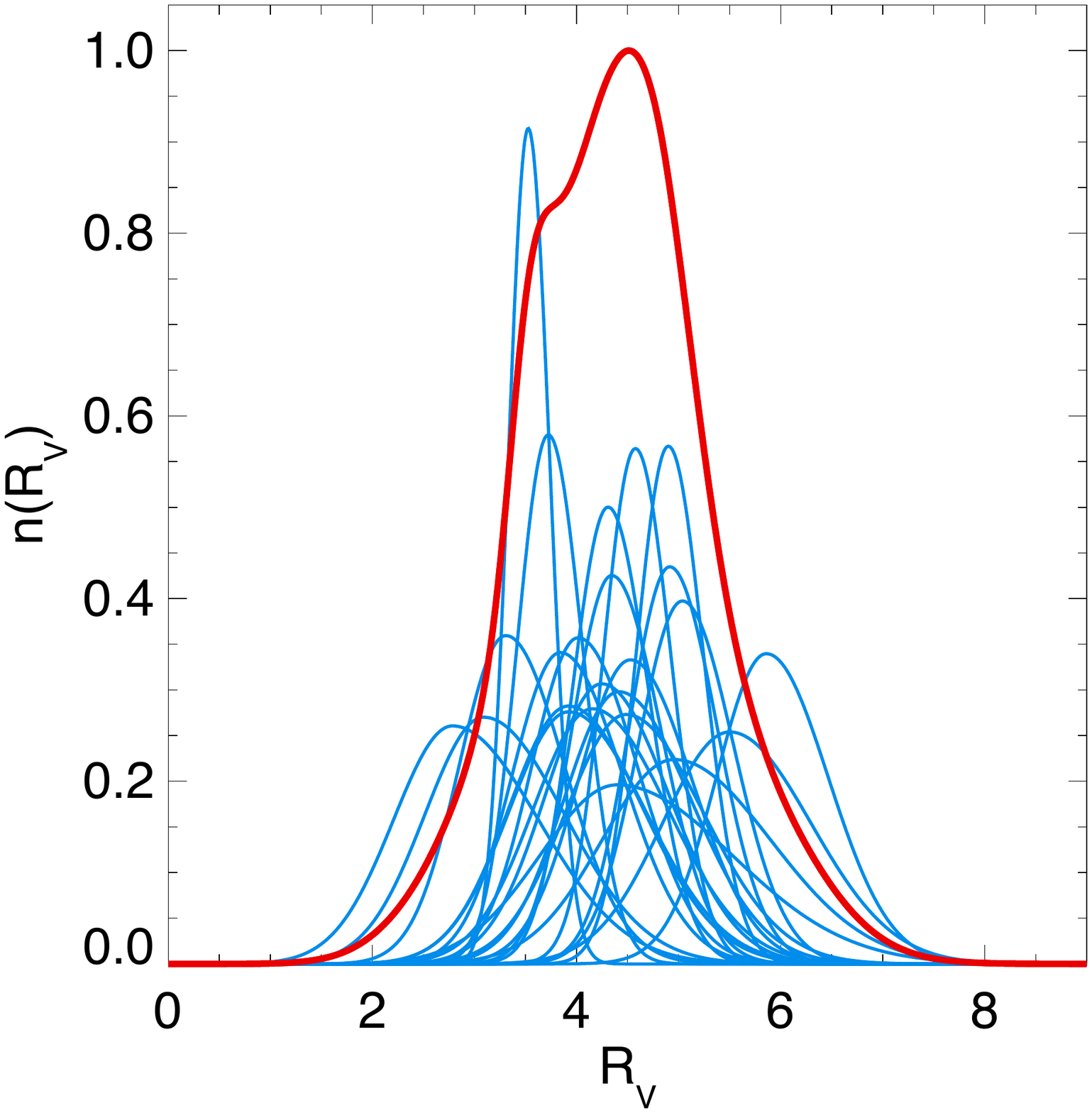}
\includegraphics[scale=0.45]{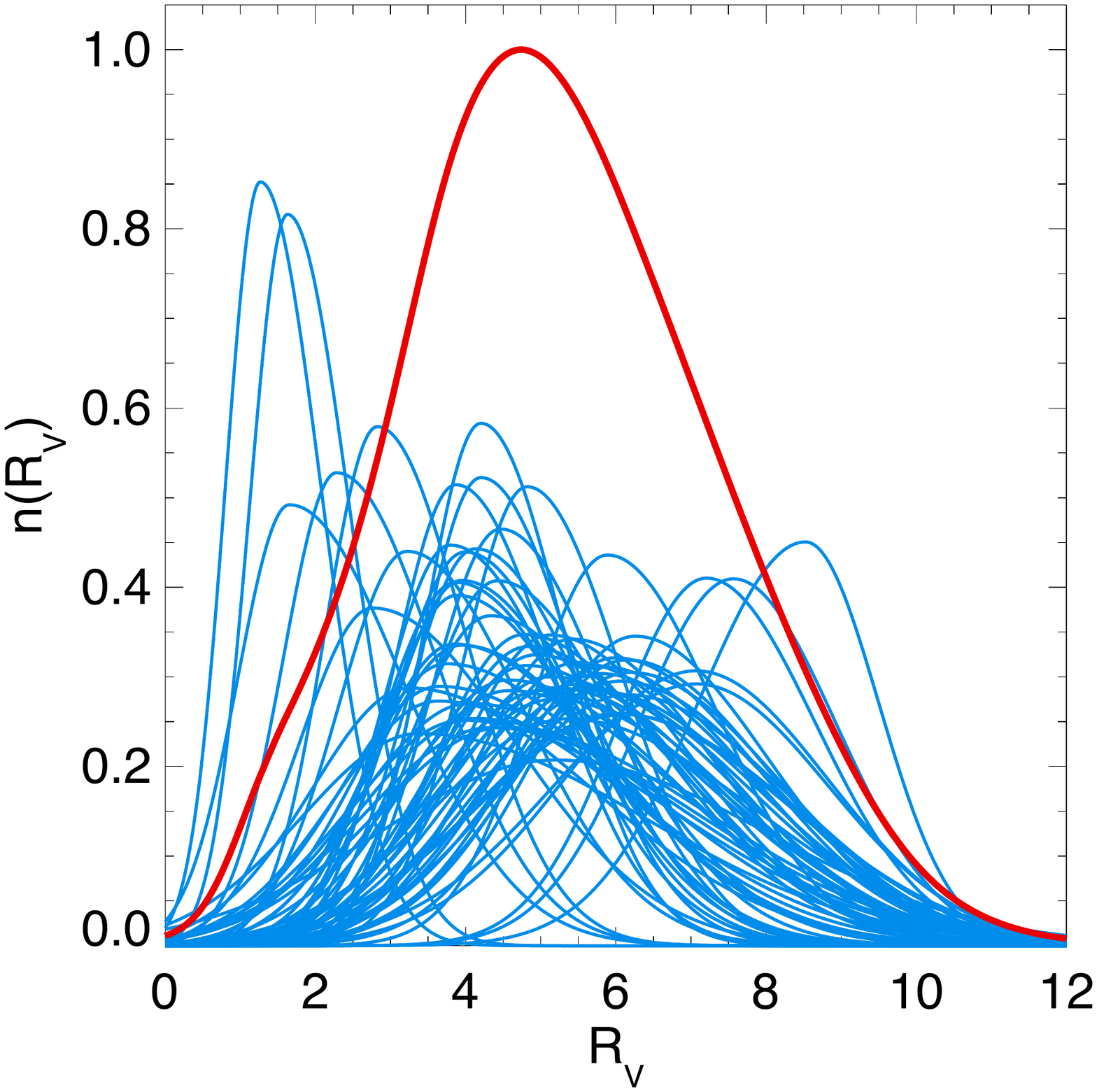}
\caption{\label{dist_rv} Probability distribution N(\rv) for all
targets with E(B-V)$\geq$0.15 mag (left,red), and E(B-V)$<$0.15
mag. The distributions are obtained by adding the assymetric Gaussian
distributions of all individual targets (plotted in blue) and
subsequent re-normalization.} 
\end{figure*}

The E(B-V) distribution is shown in Fig.~\ref{ebv_rv}. For the sample of targets
observed with MagE and FEROS the distribution peaks strongly at
E(B-V)\,=\,0.08 mag. However, there are also line-of-sights with 
significantly higher reddening, in particular, for targets located in the Tarantula Nebula shown as a yellow histogram in the plot. 
Of particular, interest is the distribution of \rv-values. However, because of the relatively
large errors of \rv~a simple histogram is not useful. Instead, we
construct a probability distribution for the total sample by assuming
asymmetric Gaussian distributions for each individual star, which are
them summed up to obtain the total distribution. The results are shown
in Fig.~\ref{dist_rv} for target subsamples with E(B-V) larger and smaller
than 0.15 mag respectively. It is evident that we encounter a
relatively wide range of \rv~values between 3 and 6,in agreement with
previous work \citep[see for example ][]{maiz2017}. 
Also worth mentioning are the three cases with very
low values of \rv~(namely Sk\,-67 137, Sk\,-67 154 and 
Sk\,-67 170). Interestingly, all these
three objects are members of the cluster NGC\,2004, but they are not spatially 
concentrated (see Fig. 16 in \citealt{evans2006}). These putatively low values are not modified 
when using any other available photometry. In this regard, we would like to point out that there are indications of low values 
of \rv~in lines of sight of SN\,Ia, 
ranging from \rv = 1.0 to \rv = 2.5 \citep{cikota2016}.\\ 

Finally, we note that the standard assumption for extragalactic distance determinations using stellar distance indicators
such as cepheids or blue supergiants is to use a standard value of \rv\,=\,3.1 or 3.3. However, Fig.~\ref{ebv_rv} indicates that this 
may introduce systematic errors, 
in particular, with respect to the ambitious goal to increase the precision of the Hubble constant to a few percent \citep[see, for instance][]{riess2011}.\\ 

\section{Flux-weighted Gravity--Luminosity Relationship of LMC supergiants}

\begin{figure*}
\plottwo{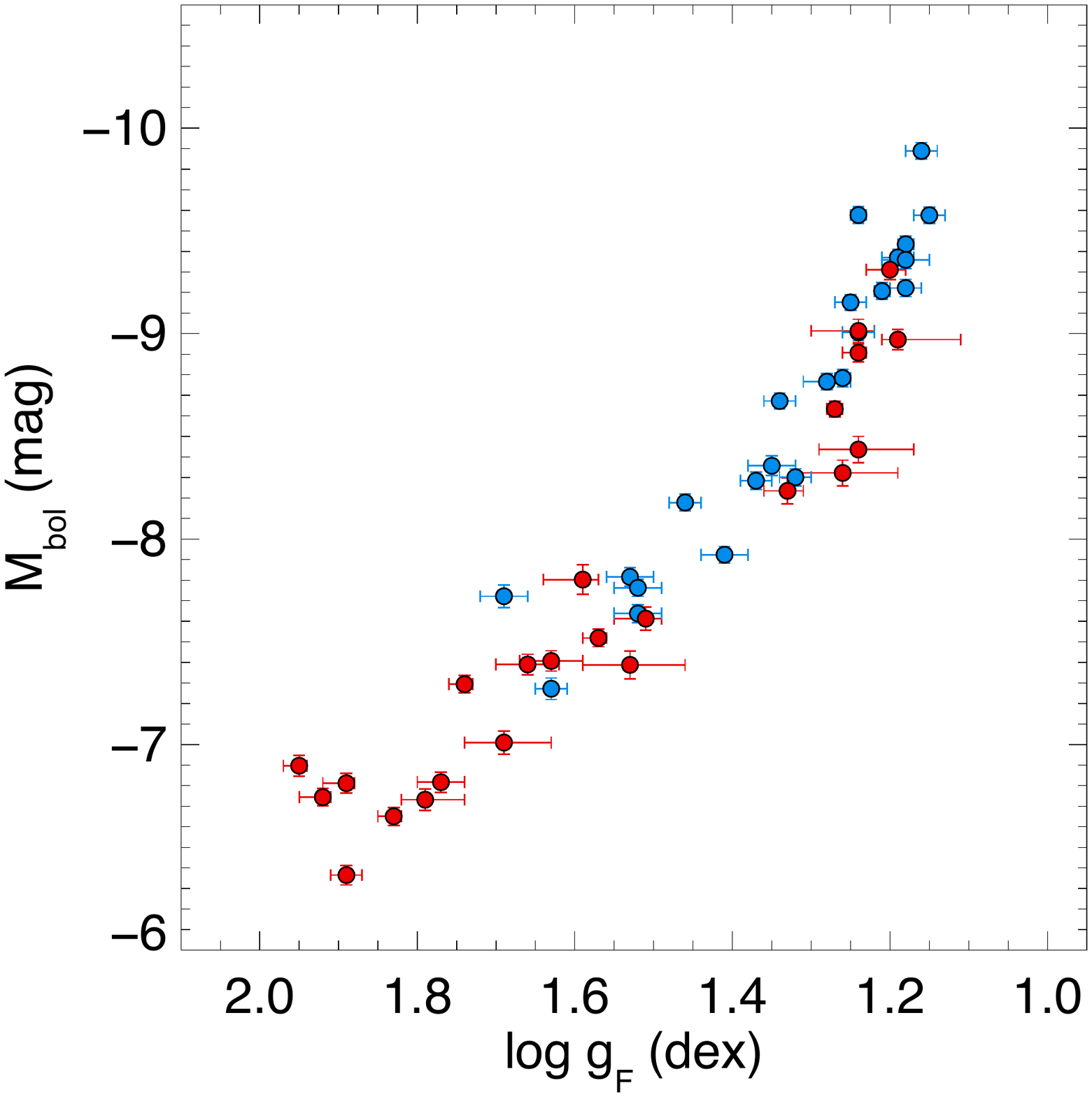}{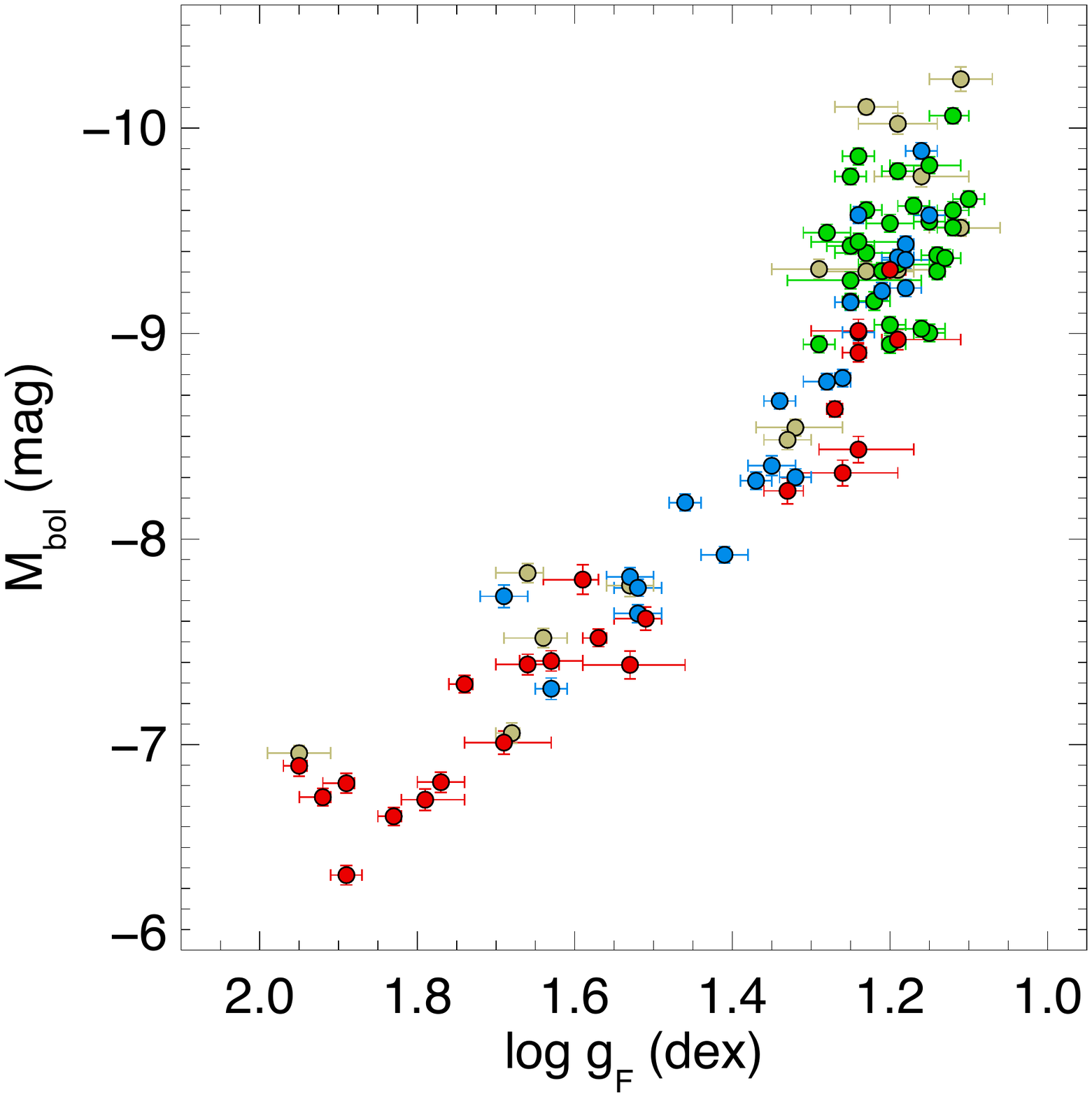}
\caption{\label{fglr_obs}  Observed FGLR of LMC supergiants. Left: Sample of supergiants observed with MagE (BA-supergiants: red, OB-supergiants: blue).
Right: The full sample including targets observed with FEROS (green) and VFTS (gold).}
\end{figure*}

Fig.~\ref{fglr_obs} shows a plot of absolute bolometric magnitudes versus flux-weighted gravities. As mentioned before, we used a distance modulus to the LMC of 
m-M = 18.494 mag \citep{pietrzynski2013} together with the values of E(B-V), \rv~and bolometric correction as given in the tables to obtain M$_{bol}$. The left hand side shows the
the subsample of supergiants observed with the MagE spectrograph only. This was the starting point of our investigation to accomplish a new calibration of the FGLR. Realizing that the data indicate
a change in the slope of the FGLR for \lgf~values smaller than 1.30 dex we decided to supplement our MagE sample with 
the FEROS and VLT/FLAMES spectra to enlarge the sample. The full sample is then given on the right 
hand side and the change in the slope is now very apparent.

A straightforward fit to the data is to introduce a flux-weighted gravity $\log\,g_{\!\mbox{\scriptsize \,\sc f}}^{\!\mbox{\scriptsize \,break}}$, 
where the slope of the FGLR changes. For 
flux-weighted gravities larger than or equal to this value
we adopt a FGLR of the form of eq. (1) and determine the coefficients $a$ and $b$ by a linear regression that accounts for errors both in \lgf~and M$_{\rm bol}$. For flux-weighted gravities smaller
than $\log\,g_{\!\mbox{\scriptsize \,\sc f}}^{\!\mbox{\scriptsize \,break}}$ we than adopt an FGLR with a different slope, which meets 
the high gravity FGLR at $\log\,g_{\!\mbox{\scriptsize \,\sc f}}^{\!\mbox{\scriptsize \,break}}$. 
This latter slope is obtained by $\chi^2$-minimalization again 
accounting for errors both in \lgf~and M$_{\rm bol}$. In this way we have
\begin{equation}
 \mbox{\lgf}\, \geq \,\log\,g_{\!\mbox{\scriptsize \,\sc f}}^{\!\mbox{\scriptsize \,break}}:\,   M_{\rm bol}\,=\,a (\mbox{\lgf}\,-\,1.5)\,+\,b
\end{equation}
and 
\begin{equation}
\mbox{\lgf} \leq \,\log\,g_{\!\mbox{\scriptsize \,\sc f}}^{\!\mbox{\scriptsize \,break}} \,:\,     M_{\rm bol}\,=\,a_{\rm low} (\mbox{\lgf}\,-\,\log\,g_{\!\mbox{\scriptsize \,\sc f}}^{\!\mbox{\scriptsize \,break}})\,+\,b_{\!\mbox{\scriptsize \,break}}  
\end{equation}
with
\begin{equation}
 b_{\!\mbox{\scriptsize \,break}}\,=\,a (\log\,g_{\!\mbox{\scriptsize \,\sc f}}^{\!\mbox{\scriptsize \,break}}\,-\,1.5)\,+\,b.
\end{equation}

\begin{figure*}
\includegraphics[scale=0.7]{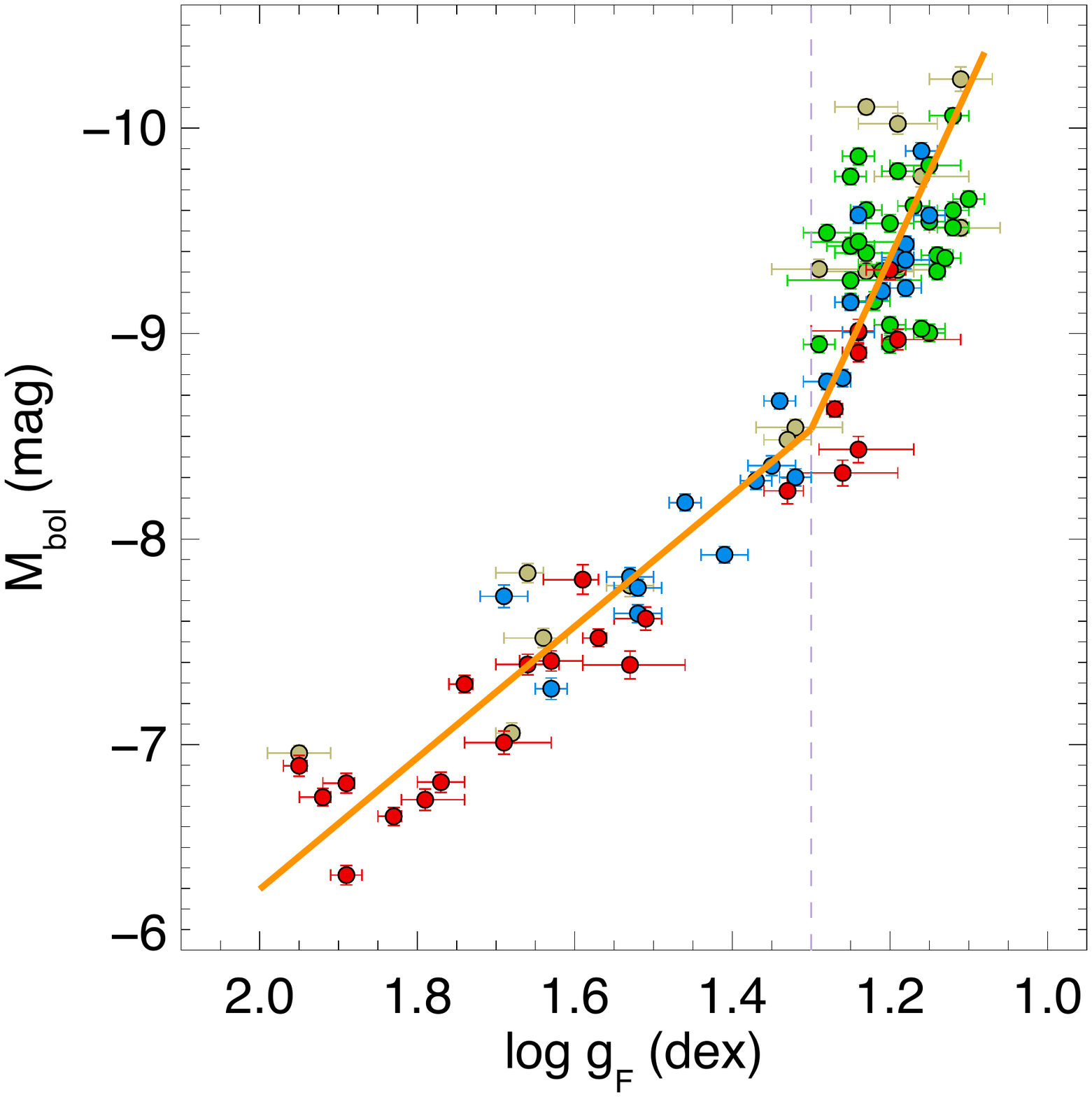}
\caption{\label{fglr_fit} Linear regression fits (orange) to the observed FGLR of Fig.~\ref{fglr_obs} for the flux-weighted gravity intervals log g$_F$ larger or 
smaller than 1.29\,dex (see text). The grey vertical dashed line indicates the the flux-weighted gravity log g$_F^{break}$ = 1.29\,dex, where the 
slopes of the regressions change. } 
\end{figure*}

We select $\log\,g_{\!\mbox{\scriptsize \,\sc f}}^{\!\mbox{\scriptsize \,break}} = 1.30$\,dex and obtain 
a = 3.20$\pm$0.08, b = -7.90$\pm$0.02 mag and a$_{\rm low}$ = 8.34$\pm$0.25. 
The standard deviations from the fit are $\sigma\,=\,0.24$ mag for $\log\,g_{\!\mbox{\scriptsize \,\sc f}} \ge 1.30$ dex
and 0.42 mag for $\log\,g_{\!\mbox{\scriptsize \,\sc f}} < 1.30$ dex. 
Fig.~\ref{fglr_fit} shows the corresponding fit to the data. We note that
for gravities above $\log\,g_{\!\mbox{\scriptsize \,\sc f}}^{\!\mbox{\scriptsize \,break}}$ the slope obtained for our LMC sample is slightly lower than the slope found in the original calibration by 
\citet{kudritzki2008}, while the intercept corresponds
to absolute magnitudes 0.12 mag fainter ($a_{\rm old}$ = 3.41$\pm$0.16, $b_{\rm old}$ = -8.02$\pm$0.04 mag).

The detection of an obvious break in the slope of the FGLR is a surprise. The previous data sets leading to the detection of the FGLR 
\citep{kudritzki2003} and used for the first calibration
\citep{kudritzki2008} contained only a few objects with flux-weighted gravities lower than $\log\,g_{\!\mbox{\scriptsize \,\sc f}}^{\!\mbox{\scriptsize \,break}}$ and did not provide any 
hints that an FGLR with two slopes would be needed to fit 
the data. On the other hand, as discussed in these early papers, the theory of stellar evolution predicts a mildly curved FGLR with 
a steeper slope at the high luminosity end,
in particular at lower metallicity. Observationally, a first indication in this regard was recognized by \citet{hosek2014} in the FGLR observations of NGC\,3109 and a combined sample of blue 
supergiants of all FGLR-studied galaxies so far.

As first noticed by \citet{kudritzki2008}, in the simple interpretation of the FGLR as a result of stellar evolution across the HRD with constant luminosity and mass the slope of the FGLR a is
related to the exponent $x$ of the mass-luminosity relationship

\begin{equation}
 L \sim M^x
\end{equation}
through the simple equation
\begin{equation}
a = -2.5 {x \over 1 - x}.
\end{equation}

A small exponent $x$ as predicted by stellar evolution theory for very massive stars would result in a steep slope of the FGLR (for 
instance $x$ = 1.5 would lead to $a$ = 7.5), whereas a larger
exponent x as encountered for stars of lower mass would flatten the slope  of the FGLR ($x$ = 4 would result in $a$ = 3.3). 

However, in reality the situation is more complicated. Massive stars
after having left the main sequence phase do not exactly evolve at constant luminosity and their masses, when they enter the blue 
supergiant stage, depend on the history of mass-loss. 
\citet{meynet2015} have studied these effects in detail using state-of-the-art stellar evolution calculations and confirmed the general 
concept of the FGLR. While this work investigated the
role of metallicity and the question whether blue supergiants are mostly in an evolutionary phase evolving towards or back from the red 
supergiant stage, it did not focus on the slope of the 
FGLR at its high luminosity end. However, a careful inspection of their population synthesis results based on stellar evolution, in particular 
their Figure 9 (right hand side), indicates that 
the FGLR should indeed significantly steepen for flux-weighted gravities lower than 1.30 dex. A detailed extension of this work for the metallicity of the LMC (their Figure 9 is for solar
 metallicity) concentrating on the high luminosity end of the FGLR and also re-investigating the role of mass-loss on the main sequence and during the blue supergiant stage would be a crucial 
step to understand whether stellar evolution theory reproduces the observed FGLR of our LMC sample of blue supergiants.

\section{Conclusions and future work}

The results obtained in this study of LMC blue supergiant stars confirm the existence of a tight relationship between absolute bolometric magnitude and flux-weighted gravity. However, from the
much larger sample investigated relative to previous calibration attempts it is now evident that the FGLR changes its slope at a flux weighted gravity of \lgf $\,\approx 1.29$\,dex. For lower gravities
the relationship becomes significantly steeper and shows a somewhat larger scatter. As noted in the previous section, it is not clear at this point whether stellar evolution theory is able to 
reproduce and explain the observations presented here. An investigation of this question will require a detailed study combining population synthesis and stellar evolution in future work.

The regression fit applied to the observed FGLR data of our LMC sample provides an accurate new calibration of the relationship, which 
can be used for distance determinations. An obvious next step 
for future work will be the application of this new calibration for improved distance determinations to the set of galaxies for which blue 
supergiant observations have already been carried out or 
will be obtained in the near future. This application will include a careful investigation how the uncertainty of the slope at the upper end 
affects the accuracy of the distances determined. Since
the FGLR data in the galaxies studied so far usually cover a wide range in luminosity and flux-weighted gravity, we are optimistic that the 
new calibration will yield accurate distances. The 
important advantage of the FGLR-method for distance determinations is that reddening and extinction, including variations of the 
reddening law, can be determined individually for each 
stellar target by the combination of spectroscopy and photometry as demonstrated in this work. This is of particular importance  in all 
situations, where \rv, the ratio of visual extinction to
reddening, varies over a wide range in a similar way as we have found for the LMC in this work. We note that extragalactic distance 
determinations using Cepheid stars usually assume a standard
reddening law with \rv = 3.1 to either calculate \textquotedblleft reddening-free\textquotedblright~magnitudes or to apply a reddening 
correction to apparent distance moduli obtained in different filter passbands. Even in cases
where the HST photometry of Cepheids is extended to the H-band in the near-infrared, this has the potential to introduce systematic errors 
of a few percent, when the reddening is a few tenths of a 
magnitude and the deviations from the standard reddening law are substantial. In this sense it will be interesting to compare distances 
obtained with the FGLR with distances using Cepheids and
also with other methods such as the tip of the red giant branch. In view of the ambitious goal to determine the Hubble constant with a 
precision of one percent \citep{riess2011} the FGLR-method with
the new calibration obtained in this work can contribute to investigate the role of systematic effects for the determination of extragalactic 
distances.

\acknowledgments

We would like to thank our colleagues Phillip Dufton and Chris Evans for making the VFTS spectra available to us. 
This research was supported by the Munich Institute for Astro- and Particle Physics (MIAPP) of the DFG 
cluster of excellence \textquotedblleft Origin and Structure of the Universe\textquotedblright. 
RPK and FB acknowledge support by the National Science Foundation under grant AST-1008789.
WG and GP gratefully acknowledge financial support for this work received from the BASAL Centro de Astrof\'{\i}sica y 
Tecnolog\'{\i}as Afines (CATA), PFB-06/2007. WG also acknowledges support from the Millenium Institute of Astrophysics (MAS)
of the Iniciativa Cient\'{\i}fica Milenio del Ministerio de Econom\'{\i}a, Fomento y Turismo de Chile, grant IC120009.
Support from the Ideas Plus grant of the Polish Ministry of Science and Higher Education and TEAM subsidies of the Foundation for 
Polish Science (FNP) is also acknowledged.
This research has made use of the VizieR catalogue access tool, CDS, Strasbourg, France. The original description of the VizieR service 
was published in 
A\&AS 143, 23. This research has made use of NASA's Astrophysics Data
System. This paper includes data gathered with the 6.5 meter Magellan
Telescopes located at the Las Campanas Observatory, Chile. This
publication makes use of data products from the Two Micron All Sky
Survey, which is a joint project of the University of Massachusetts
and the Infrared Processing and Analysis Center/California Institute
of Technology, funded by the National Aeronautics and Space
Administration and the National Science Foundation. Finally, we wish to thank our 
anonymous referee for a very engaged, careful, constructive and helpful review. 

\vspace{5mm}
\facilities{Magellan:Clay (MagE), VLT:Kueyen (FLAMES), ESO:2.2m (FEROS)}


\clearpage

\clearpage

\clearpage

\clearpage

\clearpage

\clearpage

\clearpage

\clearpage

\clearpage

\clearpage

\clearpage

\clearpage


\clearpage

\clearpage



\clearpage 
\appendix
\section{Photometric data used in this investigation}
\begin{deluxetable}{l r r r l }
\tablecaption{LMC Supergiants -- Optical photometry. \label{table_photometry_optical}}
\tablehead{
\colhead{Star} & \colhead{V} & \colhead{B-V}  & \colhead{U-B}  & \colhead{Reference} \\
\colhead{ }      & \colhead{ }   & \colhead{  }      & \colhead{  }       & \colhead{  }   
}
\tablewidth{0pt}
\startdata
N11 24                 & 13.45    &   -0.05    &   -0.87    & \cite{brunet1975}        \\   
N11 36                 & 13.72    &   -0.15    &   \nodata    & \cite{evans2006}         \\   
N11 54                 & 14.10    &   -0.06    &   \nodata    & \cite{evans2006}         \\   
NGC\,2004 8                 & 12.43    &   -0.03    &   \nodata    & \cite{evans2007}         \\   
NGC\,2004 12            & 13.39    &  -0.20    &  \nodata    & \cite{evans2006}         \\   
NGC\,2004 21            & 13.67    &  -0.14    &   \nodata   & \cite{evans2006}         \\   
NGC\,2004 22            & 13.77    &  -0.17    &   \nodata    & \cite{evans2006}         \\   
Sk-65 67                & 11.44    &   0.05    &   -0.31    & \cite{ardeberg1972}      \\   
Sk-66 1                 & 11.61    &   -0.06    &   -0.86    & \cite{isserstedt1982}    \\   
Sk-66 5                 & 10.73    &   -0.03    &   -0.78    & \cite{ardeberg1972}      \\   
Sk-66 15                & 12.81    &   -0.12    &   -0.95    & \cite{isserstedt1979}    \\   
Sk-66 23                & 13.09    &  0.08      &  -0.64     & \cite{isserstedt1979} \\ 
Sk-66 26                & 12.91    &   -0.05    &   -0.81    & \cite{isserstedt1975}    \\   
Sk-66 27                & 11.82    &   0.02    &   -0.74    & \cite{isserstedt1975}    \\   
Sk-66 35                & 11.60    &   -0.07    &   -0.89    & \cite{nicolet1978}       \\   
Sk-66 36                & 11.35    &   0.07    &   -0.76    & \cite{isserstedt1975}    \\   
Sk-66 37                & 12.98    &   -0.09    &   -0.89    & \cite{isserstedt1979}    \\   
Sk-66 50                & 10.63    &   0.02    &   -0.67    & \cite{ardeberg1972}      \\   
Sk-66 106               & 11.72    &   -0.08    &   -0.91    & \cite{isserstedt1975}    \\   
Sk-66 118               & 11.81    &   -0.05    &   -0.86    & \cite{nicolet1978}       \\   
Sk-66 166               & 11.71    &   -0.04    &   -0.79    & \cite{ardeberg1972}      \\   
Sk-67 2                 & 11.26    &   0.08    &   -0.77    & \cite{ardeberg1972}      \\   
Sk-67 14                & 11.52    &   -0.10    &   -0.91    & \cite{ardeberg1972}      \\   
Sk-67 19                & 11.19    &   0.14    &   -0.10    & \cite{ardeberg1972}      \\   
Sk-67 28                & 12.28    &   -0.14    &   -0.97    & \cite{isserstedt1982}    \\   
Sk-67 36                & 12.01    &   -0.08    &   -0.81    & \cite{isserstedt1975}    \\   
Sk-67 78                & 11.26    &   -0.04    &   -0.73    & \cite{ardeberg1972}      \\   
Sk-67 90                & 11.29    &   -0.09    &   -0.90    & \cite{ardeberg1972}      \\   
Sk-67 112               & 11.90    &   -0.13    &   -0.98    & \cite{ardeberg1972}      \\   
Sk-67 133               & 12.59    &   -0.05    &   -0.75    & \cite{isserstedt1975}    \\   
Sk-67 137               & 11.93    &   0.04    &   -0.58    & \cite{ardeberg1972}      \\   
Sk-67 140               & 12.42    &   0.07    &   -0.23    & \cite{ardeberg1972}      \\   
Sk-67 141               & 12.01    &   0.07    &   -0.03    & \cite{ardeberg1972}      \\   
Sk-67 143               & 11.46    &   0.05    &   -0.42    & \cite{ardeberg1972}      \\   
Sk-67 150               & 12.24    &   -0.11    &   -0.92    & \cite{isserstedt1975}    \\   
Sk-67 154               & 12.61    &   -0.06    &   -0.87    & \cite{isserstedt1979}    \\   
Sk-67 155               & 11.60    &   0.09    &   -0.18    & \cite{ardeberg1972}      \\   
Sk-67 157               & 11.95    &   0.00    &   -0.51    & \cite{ardeberg1972}      \\   
Sk-67 169               & 12.18    &   -0.12    &   -0.90    & \cite{isserstedt1975}    \\   
Sk-67 170               & 12.48    &   0.05    &   -0.22    & \cite{ardeberg1972}      \\   
Sk-67 171               & 12.04    &   -0.03    &   -0.57    & \cite{ardeberg1972}      \\   
Sk-67 172               & 11.88    &   -0.07    &   -0.81    & \cite{ardeberg1972}      \\   
Sk-67 173               & 12.04    &   -0.12    &   -0.96    & \cite{isserstedt1975}    \\   
Sk-67 201               &  9.87    &   0.07    &   \nodata    & \cite{feast1960}         \\   
Sk-67 204               & 10.87    &   0.04    &   -0.53    & \cite{ardeberg1972}      \\   
Sk-67 206               & 12.00    &   -0.11    &   -0.95    & \cite{isserstedt1975}    \\   
Sk-67 228               & 11.49    &   -0.05    &   -0.82    & \cite{ardeberg1972}      \\   
Sk-67 256               & 11.90    &   -0.08    &   -0.89    & \cite{isserstedt1975}    \\   
Sk-68 26                & 11.63    &   0.12    &  -0.78    & \cite{gordon2003}        \\   
Sk-68 40                & 11.71    &   -0.07    &   -0.79    & \cite{ardeberg1972}      \\   
Sk-68 41                & 12.01    &   -0.14    &   -0.97    & \cite{isserstedt1982}    \\   
Sk-68 45                & 12.03    &   -0.11    &   -0.94    & \cite{ardeberg1972}      \\   
Sk-68 92                & 11.71    &   -0.07    &   -0.88    & \cite{ardeberg1972}      \\   
Sk-68 111               & 12.01    &   -0.08    &   -0.89    & \cite{ardeberg1972}      \\   
Sk-68 171               & 12.01    &   -0.09    &   -0.89    & \cite{ardeberg1972}      \\   
Sk-69 2A                & 12.47    &   0.23    &   0.20    & \cite{ardeberg1972}      \\   
Sk-69 24                & 12.52    &   0.10    &   -0.41    & \cite{ardeberg1972}      \\   
Sk-69 39A               & 12.35    &   0.10    &   -0.17    & \cite{isserstedt1975}     \\   
Sk-69 43                & 11.94    &   -0.07    &   -0.88    & \cite{ardeberg1972}      \\   
Sk-69 82                & 10.92    &   0.04    &   -0.57    & \cite{ardeberg1972}      \\   
Sk-69 89                & 11.39    &   -0.05    &   -0.82    & \cite{isserstedt1975}    \\   
Sk-69 113               & 10.71    &   0.09    &   -0.42    & \cite{isserstedt1979}     \\   
Sk-69 170               & 10.34    &   0.13    &   -0.52    & \cite{ardeberg1972}      \\   
Sk-69 214               & 12.19    &   0.01    &   -0.84    & \cite{isserstedt1975}    \\   
Sk-69 228               & 12.12    &   0.07    &   -0.76    & \cite{isserstedt1975}    \\   
Sk-69 237               & 12.08    &   -0.03    &   -0.86    & \cite{isserstedt1975}    \\   
Sk-69 270               & 11.27    &   0.14    &   -0.66    & \cite{ardeberg1972}      \\   
Sk-69 274               & 11.21    &   0.04    &   -0.74    & \cite{ardeberg1972}      \\   
Sk-69 299               & 10.24    &   0.23    &   -0.26    & \cite{ardeberg1972}      \\   
Sk-70 45                & 12.53    &   0.01    &   -0.46    & \cite{isserstedt1975}     \\   
Sk-70 78                & 11.29    &   -0.08    &   -0.89    & \cite{ardeberg1972}      \\   
Sk-70 111               & 11.85    &   -0.07    &   -0.87    & \cite{ardeberg1972}      \\   
Sk-70 120               & 11.59    &   -0.06    &   -0.88    & \cite{ardeberg1972}      \\   
Sk-71 14                & 10.62    &   0.09    &   -0.45    & \cite{ardeberg1972}      \\   
Sk-71 42                & 11.17    &   -0.04    &   -0.88    & \cite{isserstedt1982}    \\   
VFTS 3     &    11.59 &     0.11 &    -0.73 & \cite{feitzinger1983}    \\ 
VFTS 28   &    13.40 &     0.31 &    -0.71 & \cite{brunet1975}        \\ 
VFTS 69   &    13.59 &     0.17 &   \nodata       & \cite{evans2011}         \\ 
VFTS 82   &    13.61 &    -0.06 &   \nodata       & \cite{evans2011}         \\ 
VFTS 232 &    14.52 &     0.50 &   \nodata       & \cite{evans2011}         \\ 
VFTS 270 &    14.35 &    -0.07 &   \nodata       & \cite{evans2011}         \\ 
VFTS 302 &    15.53 &     0.32 &    \nodata      & \cite{evans2011}         \\ 
VFTS 315 &    14.81 &     0.03 &    \nodata      & \cite{evans2011}         \\ 
VFTS 431 &    12.05 &     0.11 &     \nodata     & \cite{selman1999}        \\ 
VFTS 533 &    11.82 &     0.29 &     \nodata     & \cite{selman1999}        \\ 
VFTS 590 &    12.49 &     0.22 &    \nodata      & \cite{selman1999}        \\ 
VFTS 696 &    12.74 &     0.09 &    -0.79 & \cite{isserstedt1975}    \\ 
VFTS 732 &    13.00 &     0.20 &    -0.66 & \cite{isserstedt1982}    \\ 
VFTS 831 &    13.04 &     0.29 &    \nodata      & \cite{evans2011}         \\ 
VFTS 867 &    14.63 &    0.13 &    \nodata & \cite{evans2011}        \\ 
\enddata
\tablecomments{Unfortunately, individual errors in the photometric measurements for the full sample are not readily available. 
Many of the works do not provide this valuable information. In what follows, we list the uncertainties quoted in the those references listed:
a) \citet{ardeberg1972} : sig(V)=0.017, sig(B-V)=0.013, sig(U-B)=0.017
b) \citet{isserstedt1975, isserstedt1979, isserstedt1982} : sig(V)=0.025, sig(B-V)=0.020, sig(U-B)=0.025
c) \citet{nicolet1978} : sig(V)=0.02, sig(B-V)=0.015, sig(U-B)=0.02 
d) \citet{gordon2003} -- for Sk-68 26: sig(V)=0.003, sig(B-V)=0.002, sig(U-B)=0.001.
In lieu of this information being missing for all the other references, we adopted a flat error of sig(B-V)=0.02 mag and sig(V) = 0.02 mag for all the 
optical photometry used in this work.}

\end{deluxetable}

\begin{deluxetable}{l l r r r }
\tablecaption{LMC Supergiants -- 2MASS photometry \citep{skrutskie2006}. \label{table_photometry_2mass}}
\tablehead{
\colhead{Star} & \colhead{SSTISAGEMC} & \colhead{J}         & \colhead{H}            & \colhead{Ks}     \\
 \colhead{  }     &                                         & \colhead{(mag)}  & \colhead{(mag)}    & \colhead{(mag)}     
}
\tablewidth{0pt}
\startdata
N11 24        &  J045532.92-662527.7   &  13.529 $\pm$  0.022  &   13.602 $\pm$  0.025   &  13.633 $\pm$  0.030    \\ %
N11 36        &  J045740.99-662956.6   &  13.937 $\pm$  0.029  &   13.956 $\pm$  0.037   &  14.007 $\pm$  0.032    \\ %
N11 54        &  J045718.33-662559.7   &  14.080 $\pm$  0.023  &   14.080 $\pm$  0.025   &  14.110 $\pm$  0.032    \\ %
NGC\,2004 8   &  J053040.09-671638.0   &  12.296 $\pm$  0.022  &   12.278 $\pm$  0.027   &  12.263 $\pm$  0.026    \\ %
NGC\,2004 12  &  J053037.48-671653.6   &  13.847 $\pm$  0.026  &   13.928 $\pm$  0.029   &  13.923 $\pm$  0.035    \\ %
NGC\,2004 21  &  J053042.02-672141.7   &  14.087 $\pm$  0.035  &   14.153 $\pm$  0.041   &  14.166 $\pm$  0.048    \\ %
NGC\,2004 22  &  J053047.33-671723.3   &  13.889 $\pm$  0.036  &   13.849 $\pm$  0.050   &  13.820 $\pm$  0.063    \\ %
Sk-65 67      &  J053435.20-653852.8   &  11.168 $\pm$  0.022  &   11.189 $\pm$  0.024   &  11.105 $\pm$  0.023    \\ %
Sk-66 1       &  J045219.10-664353.2   &  11.656 $\pm$  0.023  &   11.627 $\pm$  0.023   &  11.665 $\pm$  0.028    \\ %
Sk-66 5       &  J045330.03-665528.1   &  10.797 $\pm$  0.021  &   10.808 $\pm$  0.024   &  10.810 $\pm$  0.023    \\ %
Sk-66 15      &  J045522.35-662819.0   &  13.020 $\pm$  0.022  &   13.089 $\pm$  0.024   &  13.090 $\pm$  0.026    \\ %
Sk-66 23      &  J045617.53-661818.9   &  12.839 $\pm$  0.024  &   12.805 $\pm$  0.026   &   12.761 $\pm$ 0.032  \\ %
Sk-66 26      &  J045620.57-662713.8   &  13.086 $\pm$  0.022  &   13.124 $\pm$  0.025   &  13.128 $\pm$  0.026    \\ %
Sk-66 27      &  J045623.48-662951.9   &  11.679 $\pm$  0.021  &   11.686 $\pm$  0.024   &  11.691 $\pm$  0.025    \\ %
Sk-66 35      &  J045704.43-663438.5   &  11.722 $\pm$  0.021  &   11.738 $\pm$  0.025   &  11.717 $\pm$  0.026    \\ %
Sk-66 36      &  J045708.85-662325.0   &  11.104 $\pm$  0.023  &   11.075 $\pm$  0.026   &  10.973 $\pm$  0.026    \\ %
Sk-66 37      &  J045722.10-662427.5   &  13.094 $\pm$  0.023  &   13.069 $\pm$  0.024   &  13.106 $\pm$  0.025    \\ %
Sk-66 50      &  J050308.81-665734.9   &  10.480 $\pm$  0.023  &   10.445 $\pm$  0.027   &  10.403 $\pm$  0.025    \\ %
Sk-66 106     &  J052900.98-663827.8   &  11.849 $\pm$  0.026  &   11.921 $\pm$  0.027   &  11.942 $\pm$  0.026    \\ %
Sk-66 118     &  J053051.90-665409.1   &  11.964 $\pm$  0.024  &   11.988 $\pm$  0.022   &  11.985 $\pm$  0.024    \\ %
Sk-66 166     &  J053604.15-661343.6   &  11.726 $\pm$  0.023  &   11.742 $\pm$  0.027   &  11.735 $\pm$  0.026    \\ %
Sk-67 2       &  J044704.44-670653.1   &  11.019 $\pm$  0.023  &   11.003 $\pm$  0.024   &  10.917 $\pm$  0.019    \\ %
Sk-67 14      &  J045431.87-671524.6   &  11.784 $\pm$  0.023  &   11.779 $\pm$  0.024   &  11.799 $\pm$  0.025    \\ %
Sk-67 19      &  J045521.60-672611.3   &  10.790 $\pm$  0.023  &   10.734 $\pm$  0.025   &  10.658 $\pm$  0.023    \\ %
Sk-67 28      &  J045839.21-671118.7   &  12.561 $\pm$  0.024  &   12.600 $\pm$  0.030   &  12.632 $\pm$  0.033    \\ %
Sk-67 36      &  J050122.56-672009.9   &  12.134 $\pm$  0.024  &   12.146 $\pm$  0.030   &  12.193 $\pm$  0.028    \\ %
Sk-67 78      &  J052019.07-671805.6   &  11.423 $\pm$  0.026  &   11.452 $\pm$  0.035   &  11.378 $\pm$  0.024    \\ %
Sk-67 90      &  J052300.66-671122.0   &  11.499 $\pm$  0.023  &   11.479 $\pm$  0.024   &  11.510 $\pm$  0.023    \\ %
Sk-67 112     &  J052656.46-673935.1   &  12.282 $\pm$  0.038  &   12.352 $\pm$  0.040   &  12.338 $\pm$  0.038    \\ %
Sk-67 133     &  J052921.70-672011.2   &  12.754 $\pm$  0.024  &   12.758 $\pm$  0.023   &  12.835 $\pm$  0.029    \\ %
Sk-67 137     &  J052942.60-672047.7   &  11.983 $\pm$  0.033  &   11.964 $\pm$  0.049   &  11.965 $\pm$  0.042    \\ %
Sk-67 140     &  J052956.50-672730.8   &  12.152 $\pm$  0.024  &   12.102 $\pm$  0.021   &  12.111 $\pm$  0.024    \\ %
Sk-67 141     &  J053001.23-671436.9   &  11.712 $\pm$  0.027  &   11.676 $\pm$  0.025   &  11.636 $\pm$  0.027    \\ %
Sk-67 143     &  J053007.07-671543.2   &  11.283 $\pm$  0.026  &   11.253 $\pm$  0.025   &  11.217 $\pm$  0.023    \\ %
Sk-67 150     &  J053031.69-670053.3   &  12.438 $\pm$  0.022  &   12.509 $\pm$  0.028   &  12.478 $\pm$  0.029    \\ %
Sk-67 154     &  J053103.75-672120.5   &       \nodata                  &   13.069 $\pm$  0.026   &  13.073 $\pm$  0.029    \\ %
Sk-67 155     &  J053112.81-671508.0   &  11.367 $\pm$  0.022  &   11.317 $\pm$  0.025   &  11.296 $\pm$  0.021    \\ %
Sk-67 157     &  J053127.92-672444.2   &  11.845 $\pm$  0.023  &   11.919 $\pm$  0.027   &  11.834 $\pm$  0.026    \\ %
Sk-67 169     &  J053151.57-670222.2   &  12.377 $\pm$  0.026  &   12.446 $\pm$  0.027   &  12.444 $\pm$  0.026    \\ %
Sk-67 170     &  J053152.97-671215.3   &  12.332 $\pm$  0.026  &   12.320 $\pm$  0.028   &  12.317 $\pm$  0.032    \\ %
Sk-67 171     &  J053200.76-672023.0   &  12.004 $\pm$  0.021  &   12.023 $\pm$  0.024   &  11.998 $\pm$  0.027    \\ %
Sk-67 172     &  J053207.30-672914.0   &  12.027 $\pm$  0.024  &   12.113 $\pm$  0.028   &  12.054 $\pm$  0.028    \\ %
Sk-67 173     &  J053210.72-674025.1   &  12.282 $\pm$  0.024  &   12.378 $\pm$  0.030   &  12.369 $\pm$  0.034    \\ %
Sk-67 201     &  J053422.45-670123.4   &   9.604 $\pm$  0.023  &    9.540 $\pm$  0.022   &   9.474 $\pm$  0.024    \\ %
Sk-67 204     &  J053450.17-672112.4   &  10.700 $\pm$  0.024  &   10.702 $\pm$  0.023   &  10.657 $\pm$  0.023    \\ %
Sk-67 206     &  J053455.11-670237.2   &  12.276 $\pm$  0.024  &   12.333 $\pm$  0.021   &  12.397 $\pm$  0.029    \\ %
Sk-67 228     &  J053740.99-674316.5   &  11.554 $\pm$  0.024  &   11.515 $\pm$  0.026   &  11.492 $\pm$  0.025    \\ %
Sk-67 256     &  J054425.03-671349.4   &  11.934 $\pm$  0.023  &   12.008 $\pm$  0.025   &  11.982 $\pm$  0.023    \\ %
Sk-68 26      &  J050132.23-681043.0   &  11.343 $\pm$  0.023  &   11.265 $\pm$  0.024   &  11.251 $\pm$  0.026    \\ %
Sk-68 40      &  J050515.18-680214.2   &  11.744 $\pm$  0.022  &   11.760 $\pm$  0.024   &  11.785 $\pm$  0.024    \\ %
Sk-68 41      &  J050527.09-681002.6   &  12.233 $\pm$  0.022  &   12.298 $\pm$  0.022   &  12.239 $\pm$  0.027    \\ %
Sk-68 45      &  J050607.26-680706.2   &  12.360 $\pm$  0.027  &   12.419 $\pm$  0.035   &  12.440 $\pm$  0.027    \\ %
Sk-68 92      &  J052816.15-685145.6   &  11.860 $\pm$  0.021  &   11.867 $\pm$  0.022   &  11.900 $\pm$  0.026    \\ %
Sk-68 111     &  J053100.82-685357.1   &  12.161 $\pm$  0.021  &   12.254 $\pm$  0.025   &  12.225 $\pm$  0.028    \\ %
Sk-68 171     &  J055022.99-681124.7   &  12.180 $\pm$  0.026  &   12.221 $\pm$  0.026   &  12.219 $\pm$  0.032    \\ %
Sk-69 2A      &  J044733.21-691432.9   &  11.799 $\pm$  0.022  &   11.724 $\pm$  0.024   &  11.642 $\pm$  0.021    \\ %
Sk-69 24      &  J045359.50-692242.6   &  12.231 $\pm$  0.021  &   12.211 $\pm$  0.022   &  12.199 $\pm$  0.026    \\ %
Sk-69 39A     & J045540.43-692640.8    & 11.873 $\pm$  0.023  &  11.854 $\pm$  0.025   &  11.812 $\pm$  0.025    \\ %
Sk-69 43      &  J045610.43-691538.3   &  12.110 $\pm$  0.023  &   12.126 $\pm$  0.027   &  12.179 $\pm$  0.030    \\ %
Sk-69 82      &  J051431.55-691353.5   &  10.724 $\pm$  0.022  &   10.714 $\pm$  0.028   &  10.668 $\pm$  0.025    \\ %
Sk-69 89      &  J051717.52-694644.1   &  11.431 $\pm$  0.025  &   11.429 $\pm$  0.025   &  11.415 $\pm$  0.025    \\ %
Sk-69 113     &  J052122.38-692707.9   &  10.435 $\pm$  0.023  &   10.369 $\pm$  0.027   &  10.336 $\pm$  0.026    \\ %
Sk-69 170     &  J053050.06-693129.4   &  10.007 $\pm$  0.024  &    9.882 $\pm$  0.025   &   9.779 $\pm$  0.023    \\ %
Sk-69 214     &  J053616.41-693126.8   &  12.178 $\pm$  0.023  &   12.217 $\pm$  0.026   &  12.173 $\pm$  0.026    \\ %
Sk-69 228     &  J053709.18-692019.4   &  12.001 $\pm$  0.027  &   11.947 $\pm$  0.031   &  11.956 $\pm$  0.034    \\ %
Sk-69 237     &  J053801.30-692213.9   &  12.186 $\pm$  0.024  &   12.179 $\pm$  0.027   &  12.179 $\pm$  0.026    \\ %
Sk-69 270     &  J054120.37-690507.2   &  10.983 $\pm$  0.024  &   10.954 $\pm$  0.026   &  10.892 $\pm$  0.026    \\ %
Sk-69 274     &  J054127.71-694803.8   &  11.077 $\pm$  0.021  &   11.128 $\pm$  0.024   &  11.035 $\pm$  0.025    \\ %
Sk-69 299     &  J054516.60-685952.0   &   9.690 $\pm$  0.021  &    9.615 $\pm$  0.024   &   9.545 $\pm$  0.023    \\ %
Sk-70 45      &  J050217.78-702656.8   &  12.493 $\pm$  0.026  &   12.461 $\pm$  0.030   &  12.463 $\pm$  0.028    \\ %
Sk-70 78      &  J050616.04-702935.7   &  11.495 $\pm$  0.023  &   11.548 $\pm$  0.025   &  11.563 $\pm$  0.026    \\ %
Sk-70 111     &  J054136.82-700052.7   &  12.034 $\pm$  0.022  &   12.050 $\pm$  0.024   &  12.047 $\pm$  0.027    \\ %
Sk-70 120     &  J055120.76-701709.4   &  11.781 $\pm$  0.021  &   11.801 $\pm$  0.027   &  11.840 $\pm$  0.029    \\ %
Sk-71 14      &  J050938.92-712402.2   &  10.306 $\pm$  0.024  &   10.218 $\pm$  0.026   &  10.155 $\pm$  0.020    \\ %
Sk-71 42      &  J053047.82-710402.4   &  11.163 $\pm$  0.023  &   11.227 $\pm$  0.025   &  11.211 $\pm$  0.026    \\ %
VFTS 3    &  J053655.17-691137.4 &       11.390 $\pm$  0.020 &  11.260 $\pm$  0.010 &   11.210 $\pm$  0.020  \\ 
VFTS 28   &  J053717.86-690946.1 &     12.580 $\pm$  0.023 &   12.339 $\pm$  0.026 &   12.281$\pm$  0.028   \\
VFTS 69   &  J053733.73-690813.0 &     12.801 $\pm$  0.024 &   12.643 $\pm$  0.030 &   12.554 $\pm$  0.030   \\
VFTS 82   &  J053736.08-690645.0 &     13.498 $\pm$  0.023 &   13.496 $\pm$  0.027 &   13.416 $\pm$  0.028   \\
VFTS 232  &  J053804.78-690905.4 &     13.347 $\pm$  0.036 &   13.157 $\pm$  0.044 &   12.988 $\pm$  0.057   \\
VFTS 302  &  J053818.97-691112.5 &     14.498 $\pm$  0.025 &   14.305 $\pm$  0.032 &   14.223 $\pm$  0.041   \\
VFTS 315  &  J053820.57-691537.7 &     14.698 $\pm$  0.023 &   14.715 $\pm$  0.035 &   14.686 $\pm$  0.054   \\
VFTS 431  &  J053836.95-690508.1 &     11.579 $\pm$  0.036 &   11.480 $\pm$  0.038 &   11.384 $\pm$  0.043   \\
VFTS 696  &  J053857.14-685653.0 &     12.491 $\pm$  0.026 &   12.426 $\pm$  0.024 &   12.380 $\pm$  0.031   \\
VFTS 732  &  J053904.76-690409.9 &     12.337 $\pm$  0.029 &   12.180 $\pm$  0.037 &   12.074 $\pm$  0.036   \\
VFTS 831  &  J053939.86-691204.2 &     12.229 $\pm$  0.029 &   12.152 $\pm$  0.038 &   12.027 $\pm$  0.035   \\
VFTS 867  &  J054001.31-690759.4 &     14.273 $\pm$  0.030 &   14.174 $\pm$  0.033 &   14.189 $\pm$  0.039   \\
\enddata
\end{deluxetable}

\begin{deluxetable}{l l r r r r}
\tablecaption{LMC Supergiants -- IRAC photometry \citep{meixner2006}. \label{table_photometry_irac}}
\tablehead{
\colhead{Star} & \colhead{SSTISAGEMC} & \colhead{[3.6]}     & \colhead{[4.5]}      & \colhead{[5.8]}   & \colhead{[8.0]}     \\
 \colhead{  }     &                                         & \colhead{(mag)}  & \colhead{(mag)}    & \colhead{(mag)} & \colhead{(mag)}     
}
\tablewidth{0pt}
\startdata
N11 24        &  J045532.92-662527.7   &  13.579 $\pm$  0.043  &   13.571 $\pm$  0.037  &   13.353 $\pm$  0.102  &   \nodata                \\ %
N11 36        &  J045740.99-662956.6   &  14.082 $\pm$  0.031  &   14.091 $\pm$  0.039  &   14.022 $\pm$  0.114  &   \nodata                \\ %
N11 54        &  J045718.33-662559.7   &  14.071 $\pm$  0.030  &   14.043 $\pm$  0.039  &   14.114 $\pm$  0.093  &   \nodata                \\ %
NGC\,2004 8   &  J053040.09-671638.0   &  12.149 $\pm$  0.036  &   12.163 $\pm$  0.025  &   12.190 $\pm$  0.045  &   12.265 $\pm$  0.046 \\ 
NGC\,2004 12  &  J053037.48-671653.6   &  13.868 $\pm$  0.052  &   13.874 $\pm$  0.044  &   14.018 $\pm$  0.071  &   \nodata                \\ %
NGC\,2004 21  &  J053042.02-672141.7   &  14.220 $\pm$  0.031  &   14.286 $\pm$  0.044  &   14.276 $\pm$  0.070  &   \nodata               \\ %
NGC\,2004 22  &  J053047.33-671723.3   &   \nodata                &   14.444 $\pm$  0.052  &   14.247 $\pm$  0.107  &     \nodata              \\ %
Sk-65 67      &  J053435.20-653852.8   &  11.068 $\pm$  0.023  &   11.030 $\pm$  0.026  &   11.028 $\pm$  0.031  &   11.010 $\pm$  0.036 \\ %
Sk-66 1       &  J045219.10-664353.2   &  11.521 $\pm$  0.027  &   11.503 $\pm$  0.026  &   11.437 $\pm$  0.034  &   11.423 $\pm$  0.040 \\ %
Sk-66 5       &  J045330.03-665528.1   &  10.725 $\pm$  0.020  &   10.685 $\pm$  0.027  &   10.628 $\pm$  0.031  &   10.624 $\pm$  0.030 \\ %
Sk-66 15      &  J045522.35-662819.0   &  13.132 $\pm$  0.030  &   13.073 $\pm$  0.026  &   13.024 $\pm$  0.039  &   12.772 $\pm$  0.089 \\ %
Sk-66 23      & J045617.53-661818.9    &  12.708 $\pm$  0.034	 &   12.661 $\pm$  0.024	&   12.620 $\pm$  0.052 & 12.556 $\pm$  0.127 \\ %
Sk-66 26      &  J045620.57-662713.8   &  13.071 $\pm$  0.053  &   13.073 $\pm$  0.051  &   12.983 $\pm$  0.061  &    \nodata               \\ %
Sk-66 27      &  J045623.48-662951.9   &  11.633 $\pm$  0.029  &   11.581 $\pm$  0.022  &   11.553 $\pm$  0.035  &   11.521 $\pm$  0.065 \\ %
Sk-66 35      &  J045704.43-663438.5   &  11.620 $\pm$  0.027  &   11.571 $\pm$  0.024  &   11.534 $\pm$  0.035  &   11.467 $\pm$  0.045 \\ %
Sk-66 36      &  J045708.85-662325.0   &  10.923 $\pm$  0.042  &   10.801 $\pm$  0.070  &   10.683 $\pm$  0.143  &    \nodata               \\ %
Sk-66 37      &  J045722.10-662427.5   &  13.111 $\pm$  0.044  &   13.063 $\pm$  0.048  &   12.953 $\pm$  0.084  &     \nodata              \\ %
Sk-66 50      &  J050308.81-665734.9   &  10.255 $\pm$  0.023  &   10.240 $\pm$  0.023  &   10.221 $\pm$  0.023  &   10.173 $\pm$  0.027 \\ %
Sk-66 106     &  J052900.98-663827.8   &  11.780 $\pm$  0.028  &   11.811 $\pm$  0.026  &   11.766 $\pm$  0.032  &   11.684 $\pm$  0.037 \\ %
Sk-66 118     &  J053051.90-665409.1   &  11.844 $\pm$  0.031  &   11.887 $\pm$  0.025  &   11.841 $\pm$  0.038  &   11.793 $\pm$  0.040 \\ %
Sk-66 166     &  J053604.15-661343.6   &  11.628 $\pm$  0.036  &   11.589 $\pm$  0.023  &   11.526 $\pm$  0.033  &   11.549 $\pm$  0.034 \\ %
Sk-67 2       &  J044704.44-670653.1   &  10.804 $\pm$  0.022  &   10.744 $\pm$  0.026  &   10.712 $\pm$  0.025  &   10.648 $\pm$  0.054 \\ %
Sk-67 14      &  J045431.87-671524.6   &  11.772 $\pm$  0.032  &   11.713 $\pm$  0.018  &   11.661 $\pm$  0.042  &   11.622 $\pm$  0.030 \\ %
Sk-67 19      &  J045521.60-672611.3   &  10.539 $\pm$  0.024  &   10.494 $\pm$  0.017  &   10.454 $\pm$  0.021  &   10.452 $\pm$  0.023 \\ %
Sk-67 28      &  J045839.21-671118.7   &  12.531 $\pm$  0.031  &   12.529 $\pm$  0.029  &   12.509 $\pm$  0.046  &   12.523 $\pm$  0.058 \\ %
Sk-67 36      &  J050122.56-672009.9   &  12.181 $\pm$  0.031  &   12.124 $\pm$  0.026  &   12.082 $\pm$  0.044  &   12.107 $\pm$  0.042 \\ %
Sk-67 78      &  J052019.07-671805.6   &  11.307 $\pm$  0.028  &   11.266 $\pm$  0.022  &   11.237 $\pm$  0.026  &   11.217 $\pm$  0.030 \\ %
Sk-67 90      &  J052300.66-671122.0   &  11.502 $\pm$  0.027  &   11.472 $\pm$  0.020  &   11.420 $\pm$  0.027  &   11.410 $\pm$  0.029 \\ %
Sk-67 112     &  J052656.46-673935.1   &  12.359 $\pm$  0.035  &   12.331 $\pm$  0.022  &   12.302 $\pm$  0.044  &   12.318 $\pm$  0.045 \\ %
Sk-67 133     &  J052921.70-672011.2   &  12.790 $\pm$  0.024  &   12.771 $\pm$  0.027  &   12.731 $\pm$  0.045  &   12.772 $\pm$  0.084 \\ %
Sk-67 137     &  J052942.60-672047.7   &  11.939 $\pm$  0.025  &   11.915 $\pm$  0.027  &   11.878 $\pm$  0.033  &   11.838 $\pm$  0.041 \\ %
Sk-67 140     &  J052956.50-672730.8   &  11.988 $\pm$  0.028  &   11.966 $\pm$  0.021  &   11.949 $\pm$  0.033  &   11.914 $\pm$  0.032 \\ %
Sk-67 141     &  J053001.23-671436.9   &  11.568 $\pm$  0.033  &   11.518 $\pm$  0.024  &   11.496 $\pm$  0.032  &   11.514 $\pm$  0.031 \\ %
Sk-67 143     &  J053007.07-671543.2   &  11.159 $\pm$  0.024  &   11.129 $\pm$  0.027  &   11.101 $\pm$  0.027  &   11.063 $\pm$  0.033 \\ %
Sk-67 150     &  J053031.69-670053.3   &  12.458 $\pm$  0.031  &   12.406 $\pm$  0.029  &   12.472 $\pm$  0.047  &   12.356 $\pm$  0.042 \\ %
Sk-67 154     &  J053103.75-672120.5   &  13.091 $\pm$  0.029  &   13.043 $\pm$  0.026  &   12.999 $\pm$  0.036  &   12.996 $\pm$  0.049 \\ %
Sk-67 155     &  J053112.81-671508.0   &  11.210 $\pm$  0.031  &   11.198 $\pm$  0.027  &   11.136 $\pm$  0.029  &   11.142 $\pm$  0.034 \\ %
Sk-67 157     &  J053127.92-672444.2   &  11.772 $\pm$  0.021  &   11.770 $\pm$  0.027  &   11.715 $\pm$  0.032  &   11.716 $\pm$  0.035 \\ %
Sk-67 169     &  J053151.57-670222.2   &  12.454 $\pm$  0.028  &   12.439 $\pm$  0.027  &   12.409 $\pm$  0.038  &   12.402 $\pm$  0.063 \\ %
Sk-67 170     &  J053152.97-671215.3   &  12.152 $\pm$  0.032  &   12.181 $\pm$  0.027  &   12.176 $\pm$  0.034  &   12.274 $\pm$  0.051 \\ %
Sk-67 171     &  J053200.76-672023.0   &  11.964 $\pm$  0.025  &   11.980 $\pm$  0.025  &   11.927 $\pm$  0.035  &   11.965 $\pm$  0.045 \\ %
Sk-67 172     &  J053207.30-672914.0   &  12.003 $\pm$  0.026  &   12.011 $\pm$  0.018  &   11.972 $\pm$  0.034  &   11.965 $\pm$  0.031 \\ %
Sk-67 173     &  J053210.72-674025.1   &  12.418 $\pm$  0.035  &   12.341 $\pm$  0.025  &   12.342 $\pm$  0.049  &   \nodata               \\ %
Sk-67 201     &  J053422.45-670123.4   &   9.272 $\pm$  0.030  &    9.246 $\pm$  0.018  &    9.197 $\pm$  0.020  &    9.152 $\pm$  0.026 \\ %
Sk-67 204     &  J053450.17-672112.4   &  10.558 $\pm$  0.020  &   10.533 $\pm$  0.021  &   10.513 $\pm$  0.021  &   10.496 $\pm$  0.026 \\ %
Sk-67 206     &  J053455.11-670237.2   &  12.293 $\pm$  0.046  &   12.266 $\pm$  0.028  &   12.226 $\pm$  0.045  &   12.188 $\pm$  0.044 \\ %
Sk-67 228     &  J053740.99-674316.5   &  11.393 $\pm$  0.019  &   11.285 $\pm$  0.022  &   11.237 $\pm$  0.026  &   11.152 $\pm$  0.028 \\ %
Sk-67 256     &  J054425.03-671349.4   &  11.967 $\pm$  0.026  &   11.973 $\pm$  0.030  &   11.899 $\pm$  0.033  &   11.829 $\pm$  0.040 \\ %
Sk-68 26      &  J050132.23-681043.0   &  11.111 $\pm$  0.022  &   11.004 $\pm$  0.027  &   11.004 $\pm$  0.025  &   10.869 $\pm$  0.033 \\ %
Sk-68 40      &  J050515.18-680214.2   &  11.653 $\pm$  0.027  &   11.631 $\pm$  0.029  &   11.565 $\pm$  0.038  &   11.556 $\pm$  0.046 \\ %
Sk-68 41      &  J050527.09-681002.6   &  12.271 $\pm$  0.034  &   12.270 $\pm$  0.031  &   12.233 $\pm$  0.046  &   12.261 $\pm$  0.072 \\ %
Sk-68 45      &  J050607.26-680706.2   &  12.372 $\pm$  0.034  &   12.352 $\pm$  0.036  &   12.359 $\pm$  0.045  &   12.391 $\pm$  0.055 \\ %
Sk-68 92      &  J052816.15-685145.6   &  11.836 $\pm$  0.034  &   11.764 $\pm$  0.024  &   11.743 $\pm$  0.031  &   11.660 $\pm$  0.036 \\ %
Sk-68 111     &  J053100.82-685357.1   &  12.087 $\pm$  0.030  &   12.120 $\pm$  0.026  &   12.038 $\pm$  0.034  &   11.975 $\pm$  0.044 \\ %
Sk-68 171     &  J055022.99-681124.7   &  12.189 $\pm$  0.039  &   12.124 $\pm$  0.029  &   12.141 $\pm$  0.038  &   12.007 $\pm$  0.043 \\ %
Sk-69 2A      &  J044733.21-691432.9   &  11.551 $\pm$  0.025  &   11.535 $\pm$  0.025  &   11.500 $\pm$  0.036  &   11.487 $\pm$  0.037 \\ %
Sk-69 24      &  J045359.50-692242.6   &  12.053 $\pm$  0.034  &   12.066 $\pm$  0.028  &   12.088 $\pm$  0.040  &    \nodata              \\ %
Sk-69 39A     &  J045540.43-692642.8   &  \nodata  &   \nodata  &   \nodata  &    \nodata              \\ %
Sk-69 43      &  J045610.43-691538.3   &  12.152 $\pm$  0.027  &   12.058 $\pm$  0.025  &   12.044 $\pm$  0.036  &   12.007 $\pm$  0.055 \\ %
Sk-69 82      &  J051431.55-691353.5   &  10.611 $\pm$  0.025  &   10.594 $\pm$  0.023  &   10.583 $\pm$  0.027  &   10.583 $\pm$  0.043 \\ %
Sk-69 89      &  J051717.52-694644.1   &  11.343 $\pm$  0.024  &   11.334 $\pm$  0.023  &   11.288 $\pm$  0.029  &   11.233 $\pm$  0.033 \\ %
Sk-69 113     &  J052122.38-692707.9   &  10.208 $\pm$  0.023  &   10.175 $\pm$  0.020  &   10.152 $\pm$  0.025  &   10.100 $\pm$  0.022 \\ %
Sk-69 170     &  J053050.06-693129.4   &   9.666 $\pm$  0.027  &    9.629 $\pm$  0.021  &    9.569 $\pm$  0.026  &    9.507 $\pm$  0.028 \\ %
Sk-69 214     &  J053616.41-693126.8   &  12.122 $\pm$  0.037  &   12.120 $\pm$  0.030  &   12.095 $\pm$  0.043  &   12.059 $\pm$  0.071 \\ %
Sk-69 228     &  J053709.18-692019.4   &  11.807 $\pm$  0.057  &   11.801 $\pm$  0.027  &   11.697 $\pm$  0.034  &   11.676 $\pm$  0.055 \\ %
Sk-69 237     &  J053801.30-692213.9   &  11.974 $\pm$  0.035  &   11.969 $\pm$  0.025  &   11.883 $\pm$  0.038  &   11.954 $\pm$  0.048 \\ %
Sk-69 270     &  J054120.37-690507.2   &  10.804 $\pm$  0.026  &   10.745 $\pm$  0.026  &   10.705 $\pm$  0.031  &   10.731 $\pm$  0.040 \\ %
Sk-69 274     &  J054127.71-694803.8   &  10.970 $\pm$  0.029  &   10.909 $\pm$  0.025  &   10.862 $\pm$  0.035  &   10.748 $\pm$  0.035 \\ %
Sk-69 299     &  J054516.60-685952.0   &   9.315 $\pm$  0.041  &    9.264 $\pm$  0.020  &    9.201 $\pm$  0.024  &    9.168 $\pm$  0.030 \\ %
Sk-70 45      &  J050217.78-702656.8   &  12.396 $\pm$  0.032  &   12.368 $\pm$  0.024  &   12.384 $\pm$  0.048  &   12.374 $\pm$  0.035 \\ %
Sk-70 78      &  J050616.04-702935.7   &  11.435 $\pm$  0.024  &   11.389 $\pm$  0.024  &   11.339 $\pm$  0.046  &   11.362 $\pm$  0.039 \\ %
Sk-70 111     &  J054136.82-700052.7   &  11.930 $\pm$  0.040  &   11.995 $\pm$  0.038  &   11.894 $\pm$  0.068  &   11.793 $\pm$  0.061 \\ %
Sk-70 120     &  J055120.76-701709.4   &  11.707 $\pm$  0.036  &   11.653 $\pm$  0.021  &   11.610 $\pm$  0.035  &   11.599 $\pm$  0.036 \\ %
Sk-71 14      &  J050938.92-712402.2   &  10.076 $\pm$  0.021  &   10.010 $\pm$  0.021  &   10.000 $\pm$  0.024  &    9.975 $\pm$  0.027 \\ %
Sk-71 42      &  J053047.82-710402.4   &  11.089 $\pm$  0.025  &   11.025 $\pm$  0.021  &   11.058 $\pm$  0.031  &   11.014 $\pm$  0.037 \\ %
VFTS 3    &  J053655.17-691137.4  &   10.951 $\pm$  0.032 &   10.896 $\pm$  0.029 &   10.853 $\pm$  0.039 &   10.807 $\pm$ 0.078 \\
VFTS 28   &  J053717.86-690946.1  &   12.084 $\pm$  0.039 &   11.980 $\pm$  0.025 &   12.038 $\pm$  0.060 &   \nodata              \\
VFTS 69   &  J053733.73-690813.0  &   12.414 $\pm$  0.036 &   12.357 $\pm$  0.024 &   12.271$\pm$  0.065 &   \nodata              \\
VFTS 82   &  J053736.08-690645.0  &   13.388 $\pm$  0.069 &   13.294 $\pm$  0.047 &   12.654 $\pm$  0.117 &   \nodata              \\
VFTS 232  &  J053804.78-690905.4  &   12.821 $\pm$  0.057 &   12.787 $\pm$  0.041 &   11.932 $\pm$  0.149 &  \nodata               \\
VFTS 302  &  J053818.97-691112.5  &   14.034 $\pm$  0.067 &   13.999 $\pm$  0.055 &   13.262 $\pm$  0.163 &   \nodata              \\
VFTS 315  &  J053820.57-691537.7  &   14.655 $\pm$  0.040 &   14.646 $\pm$  0.052 &   14.601$\pm$  0.121 &   \nodata              \\
VFTS 431  &  J053836.95-690508.1  &   11.259 $\pm$  0.055 &    \nodata                  &    \nodata                  &      \nodata           \\
VFTS 696  &  J053857.14-685653.0  &   12.267 $\pm$  0.035 &   12.218 $\pm$  0.029 &   12.128 $\pm$  0.046 &  \nodata               \\
VFTS 732  &  J053904.76-690409.9  &   11.846 $\pm$  0.050 &   11.731 $\pm$  0.071 &   11.298 $\pm$  0.113 &   \nodata              \\
VFTS 831  &  J053939.86-691204.2  &   12.013 $\pm$  0.058 &   12.011 $\pm$  0.042 &   11.978 $\pm$  0.042 &   11.986 $\pm$ 0.066 \\
VFTS 867  &  J054001.31-690759.4  &   14.120 $\pm$  0.045 &   14.086 $\pm$  0.044 &   13.945 $\pm$  0.084 &  \nodata               \\
\enddata
\end{deluxetable}


\end{document}